\documentclass[annual]{acmsiggraph}

\TOGonlineid{0041}

\TOGvolume{30}
\TOGnumber{6}

\TOGarticleDOI{2024156.2024191}

\TOGprojectURL{http://kevinkarsch.com/publications/sa11.html}
\TOGvideoURL{}
\TOGdataURL{}
\TOGcodeURL{}

\title{Rendering Synthetic Objects into Legacy Photographs}

\author{Kevin Karsch \hspace{.5in} Varsha Hedau \hspace{.5in} David Forsyth  \hspace{.5in} Derek Hoiem  \vspace{2mm} \\
University of Illinois at Urbana-Champaign \\
\{karsch1,vhedau2,daf,dhoiem\}@uiuc.edu
\vspace{-2mm}
}

\pdfauthor{Kevin Karsch, Varsha Hedau, David Forsyth, Derek Hoiem}

\keywords{image-based rendering, computational photography, light estimation, photo editing}

\usepackage{amssymb}

\usepackage{fancyhdr}
\usepackage{amsmath}    
\usepackage{graphicx}   
\usepackage{multirow}
\usepackage{amsthm}
\usepackage{algorithm2e}
\usepackage{algorithmic}
\usepackage{url}
\usepackage{float}

\usepackage{color}

\def\argmin{\mathop{\rm argmin}}

\newcommand{\boldhead}[1]{\vspace{0.05in}\noindent\textbf{#1.}}

\definecolor{mycolor}{RGB}{128,200,255}
\definecolor{dblue}{RGB}{67,88,178}
\definecolor{dyellow}{RGB}{178,146,50}

\newcommand{\ignore}[1]{}

\newcommand{\ea}{{\em et al.~}}

\usepackage[scaled=.92]{helvet}
\usepackage{times}
\usepackage{paralist}

\usepackage[scaled=.92]{helvet}
\usepackage{times}

\usepackage{graphicx}

\usepackage{parskip}

\usepackage[labelfont=bf,textfont=it]{caption}

\begin{document}
\teaser{
\centerline{
\includegraphics[width=43mm]{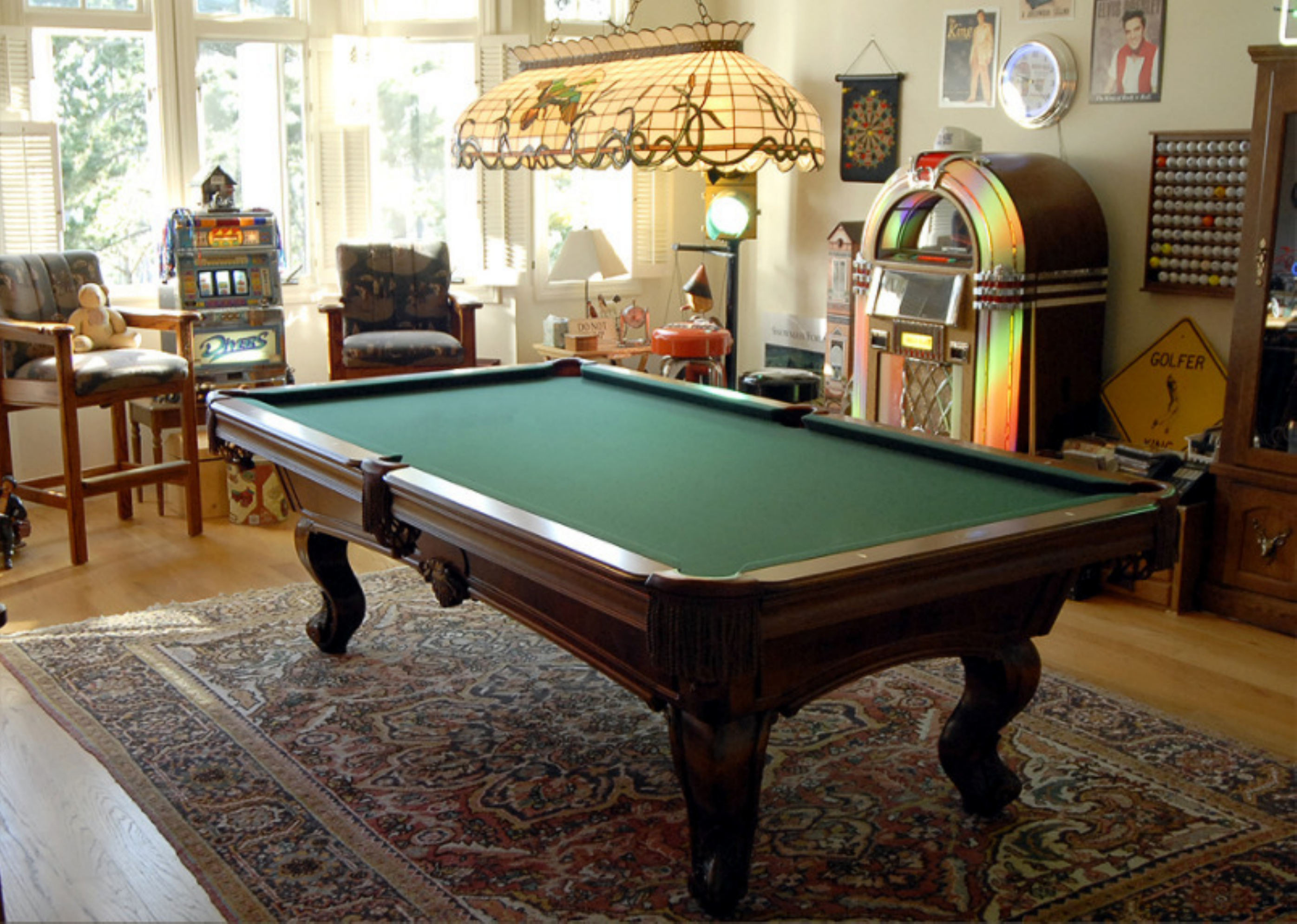}
\includegraphics[width=43mm]{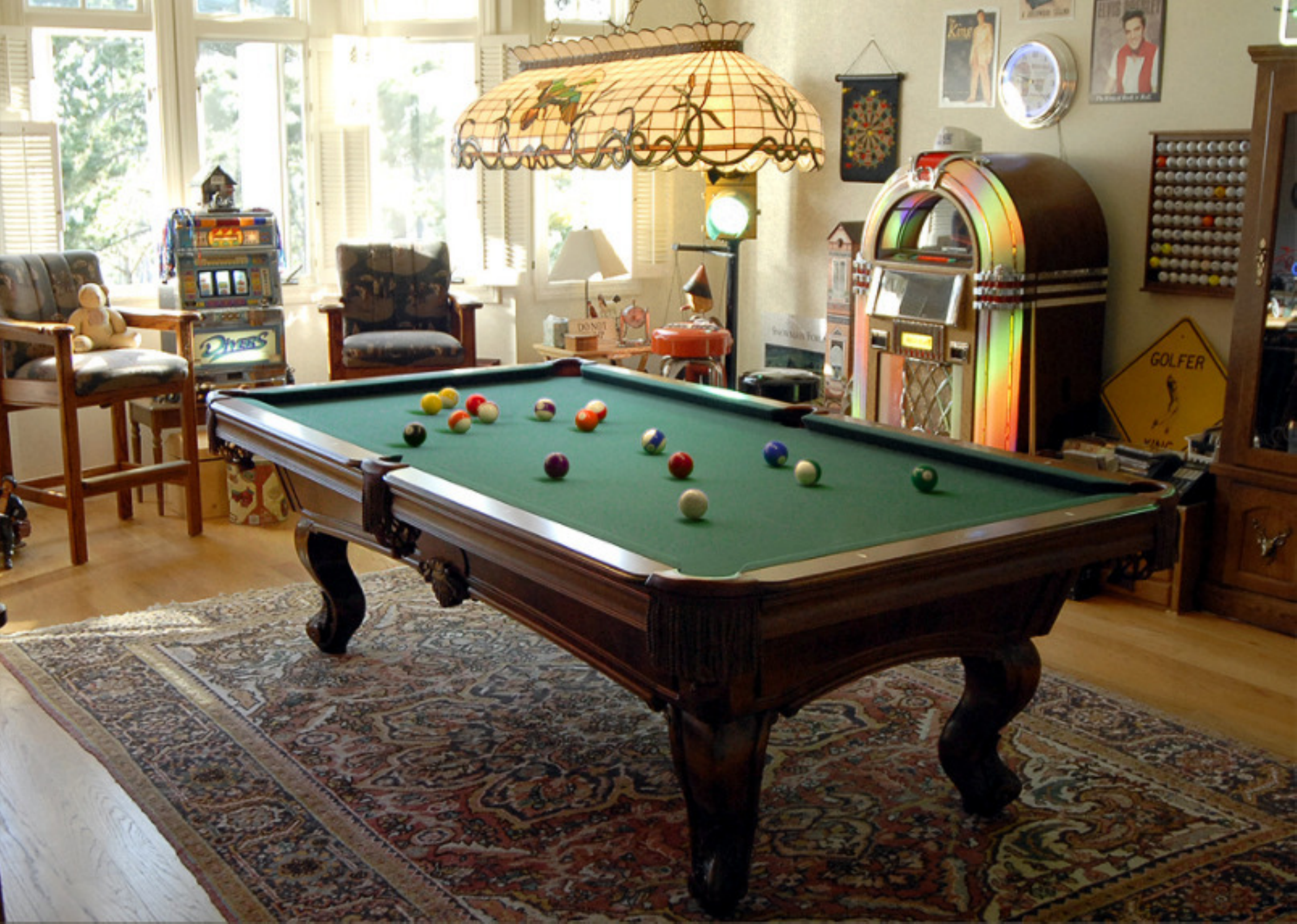}
\hspace{1mm}
\includegraphics[width=43mm]{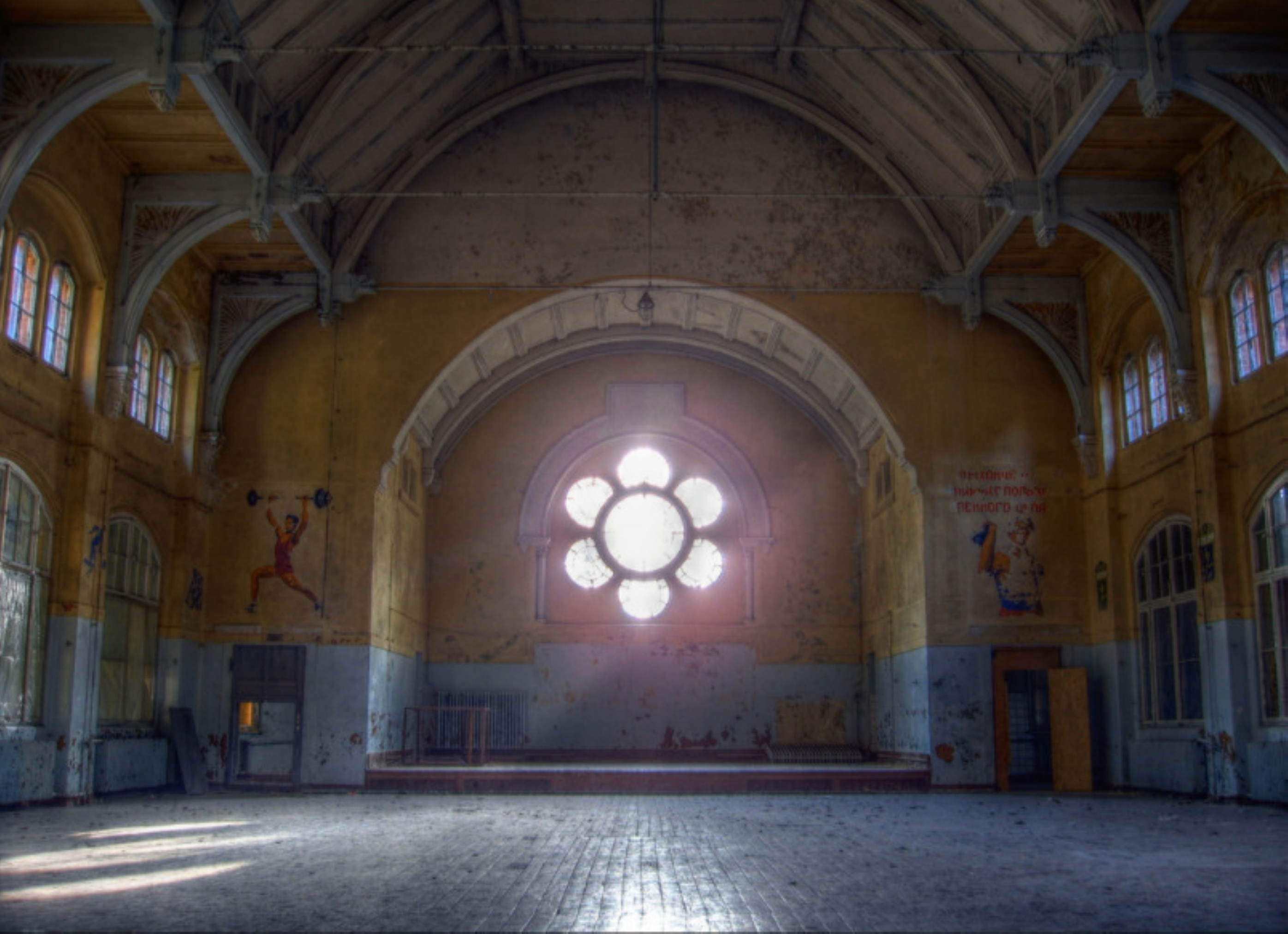}
\includegraphics[width=43mm]{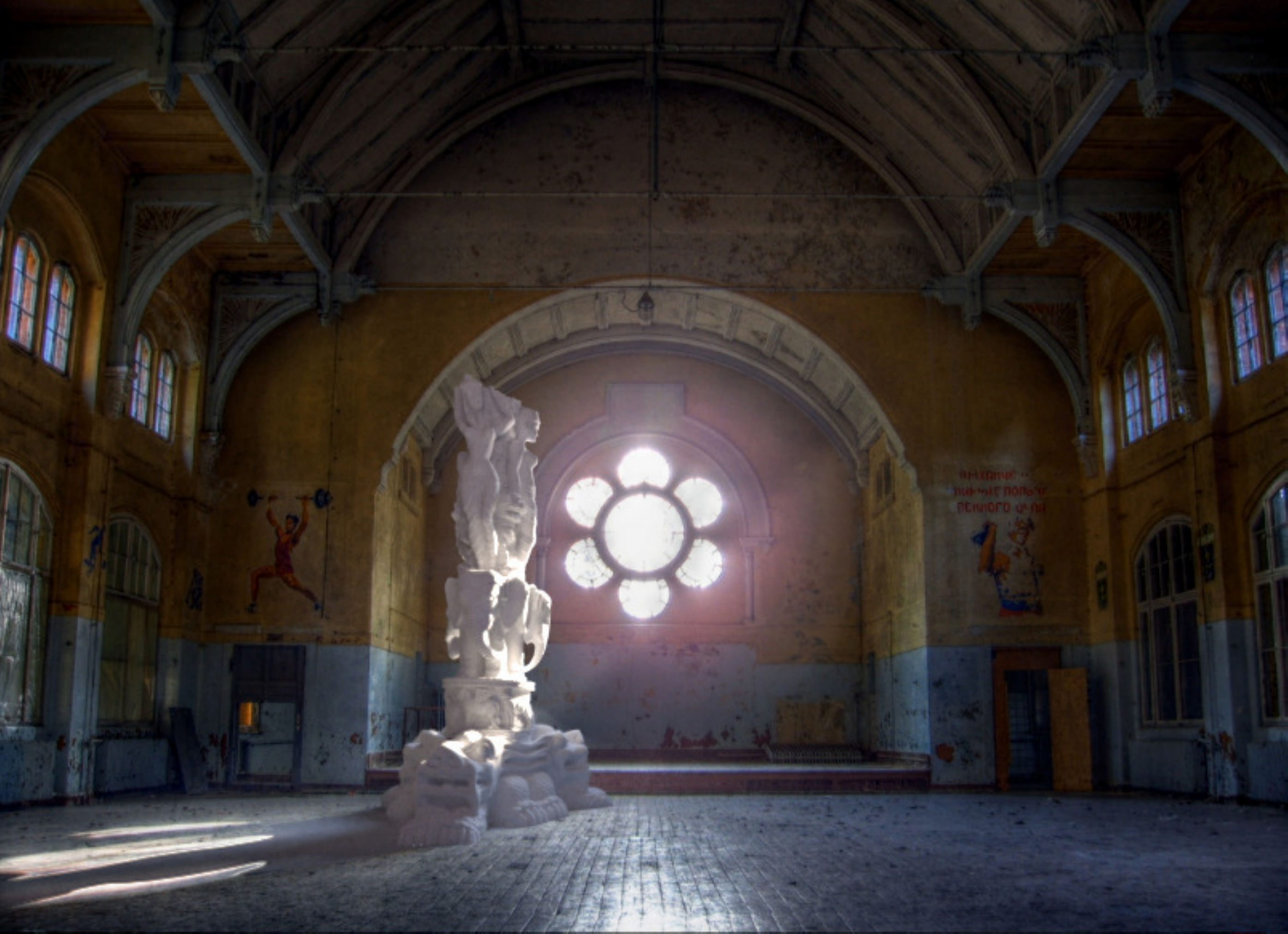}
}
\caption{With only a small amount of user interaction, our system allows objects to be inserted into legacy images so that perspective, occlusion, and lighting of inserted objects adhere to the physical properties of the scene. Our method works with only a single LDR photograph, and no access to the scene is required.
}
\label{fig:teaser}
}


\maketitle


\begin{abstract}
We propose a method to realistically insert synthetic objects into existing photographs without requiring access to the scene or any additional scene measurements.  With a single image and a small amount of annotation, our method creates a physical model of the scene that is suitable for realistically rendering synthetic objects with diffuse, specular, and even glowing materials while accounting for lighting interactions between the objects and the scene. We demonstrate in a user study that synthetic images produced by our method are confusable with real scenes, even for people who believe they are good at telling the difference. Further, our study shows that our method is competitive with other insertion methods while requiring less scene information. We also collected new illumination and reflectance datasets; renderings produced by our system compare well to ground truth. Our system has applications in the movie and gaming industry, as well as home decorating and user content creation, among others.
\end{abstract}


\begin{CRcatlist}
\CRcat{I.2.10}{Computing Methodologies}%
{Artificial Intelligence}{Vision and Scene Understanding};
\CRcat{I.3.6}{Computing Methodologies}{Computer Graphics}{Methodology and Techniques}
\end{CRcatlist}

\keywordlist

\copyrightspace


\TOGlinkslist

\section{Introduction}
Many applications require a user to insert 3D meshed characters, props, or other synthetic objects into images and videos. Currently, to insert objects into the scene, some scene geometry must be manually created, and lighting models may be produced by photographing mirrored light probes placed in the scene, taking multiple photographs of the scene, or even modeling the sources manually. Either way, the process is painstaking and requires expertise.

\begin{figure*}[htpb!]
\centerline{
\includegraphics[width=167mm]{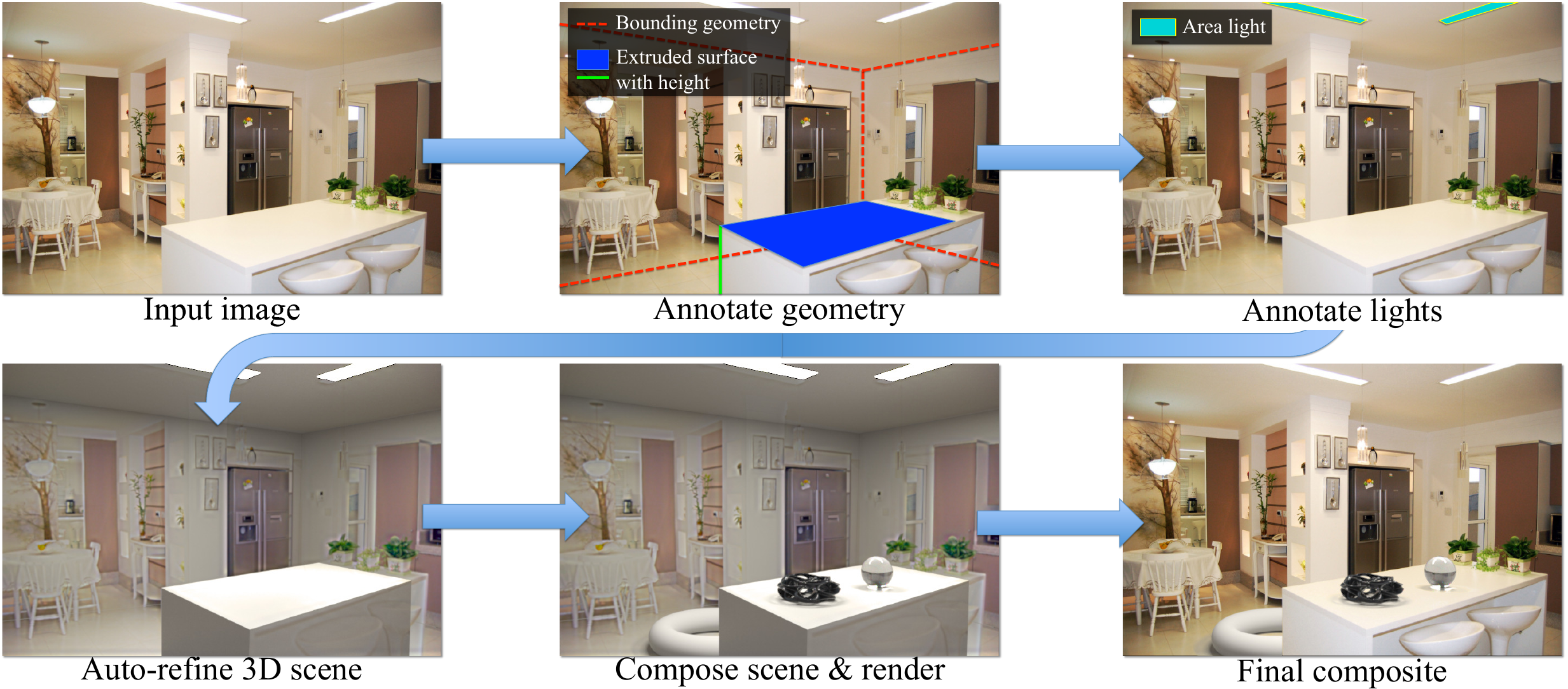}
}
\caption{ Our method for inserting synthetic objects into legacy photographs. From an input image \emph{(top left)}, initial geometry is estimated and a user annotates other necessary geometry \emph{(top middle)} as well as light positions \emph{(top right)}. From this input, our system automatically computes a 3D scene, including a physical light model, surface materials, and camera parameters \emph{(bottom left}). After a user places synthetic objects in the scene \emph{(bottom middle)}, objects are rendered and composited into the original image \emph{(bottom right)}.  Objects appear naturally lit and adhere to the perspective and geometry of the physical scene. From our experience, the markup procedure takes only a minute or two, and the user can begin inserting objects and authoring scenes in a matter of minutes.
}
\label{fig:systemdemo}
\end{figure*}

We propose a method to realistically insert synthetic objects into existing photographs without requiring access to the scene, special equipment, multiple photographs, time lapses, or any other aids.
Our approach, outlined in Figure~\ref{fig:systemdemo}, is to take advantage of small amounts of annotation to recover a simplistic model of geometry and the position, shape, and intensity of light sources.  First, we automatically estimate a rough geometric model of the scene, and ask the user to specify (through image space annotations) any additional geometry that synthetic objects should interact with. Next, the user annotates light sources and light shafts (strongly directed light) in the image. Our system automatically generates a physical model of the scene using these annotations. The models created by our method are suitable for realistically rendering synthetic objects with diffuse, specular, and even glowing materials while accounting for lighting interactions between the objects and the scene.

In addition to our overall system, our primary technical contribution is a semiautomatic algorithm for estimating a physical lighting model from a single image. Our method can generate a full lighting model that is demonstrated to be physically meaningful through a ground truth evaluation. We also introduce a novel image decomposition algorithm that uses geometry to improve lightness estimates, and we show in another evaluation to be state-of-the-art for single image reflectance estimation.
We demonstrate with a user study that the results of our method are confusable with real scenes, even for people who believe they are good at telling the difference. Our study also shows that our method is competitive with other insertion methods while requiring less scene information.
This method has become possible from advances in recent literature. In the past few years, we have learned a great deal about extracting high level information from indoor scenes~\cite{hedau2009iccv,Lee:09,Lee:10}, and that detecting shadows in images is relatively straightforward~\cite{guo_cvpr11}. Grosse \ea~\shortcite{grosse09intrinsic} have also shown that simple lightness assumptions lead to powerful surface estimation algorithms; Retinex remains among the best methods.

\section{Related work}
Debevec's work~\shortcite{Debevecprobe} is most closely related to ours. Debevec shows that a light probe, such as a spherical mirror, can be used to capture a physically accurate radiance map for the position where a synthetic object is to be inserted. This method requires a considerable amount of user input: HDR photographs of the probe, converting these photos into an environment map, and manual modeling of scene geometry and materials.
More robust methods exist at the cost of more setup time (e.g. the plenopter~\cite{Mury:2009vz}). Unlike these methods and others (e.g.~\cite{fournier1992b,alnasser2006,cossairt2008,Lalonde:sa09}), we require no special equipment, measurements, or multiple photographs. Our method can be used with only a single LDR image, e.g. from Flickr, or even historical photos that cannot be recaptured.

\boldhead{Image-based Content Creation}
Like us, Lalonde \ea~\shortcite{lalonde2007} aim to allow a non-expert user populate an image with objects. Objects are segmented from a large database of images, which they automatically sort to present the user with source images that have similar lighting and geometry. Insertion is simplified by automatic blending and shadow transfer, and the object region is resized as the user moves the cursor across the ground.
This method is only suitable if an appropriate exemplar image exists, and even in that case, the object cannot participate in the scene's illumination. Similar methods exist for translucent and refractive objects~\cite{Yeung:2011:MCT:1899404.1899406}, but in either case, inserted objects cannot reflect light onto other objects or cast caustics. Furthermore, these methods do not allow for mesh insertion, because scene illumination is not calculated. We avoid these problems by using synthetic objects (3D textured meshes, now plentiful and mostly free on sites like Google 3D Warehouse and turbosquid.com) and physical lighting models.

\boldhead{Single-view 3D Modeling}
Several user-guided~\cite{Liebowitz99,criminisi2000,Zhang01,TIP,TIP2,Oh01,1409112} or automatic~\cite{hoiem2005siggraph,Make3D} methods are able to perform 3D modeling from a single image.  These works are generally interested in constructing 3D geometric models for novel view synthesis.  Instead, we use the geometry to help infer illumination and to handle perspective and occlusion effects.  Thus, we can use simple box-like models of the scene~\cite{hedau2009iccv} with planar billboard models~\cite{TIP2} of occluding objects.  The geometry of background objects can be safely ignored.  Our ability to appropriately resize 3D objects and place them on supporting surfaces, such as table-tops, is based on the single-view metrology work of Criminisi~\shortcite{criminisi2000}; also described by Hartley and Zisserman~\shortcite{hartley2004}.  We recover focal length and automatically estimate three orthogonal vanishing points, using the method from Hedau \ea~\shortcite{hedau2009iccv}, which is based on Rother's technique~\shortcite{rother2002}.

\begin{figure*}[htp!]
\begin{center}
\begin{minipage}{0.73\linewidth}
\vspace{-3mm}
\centerline{
\includegraphics[width=0.19\linewidth]{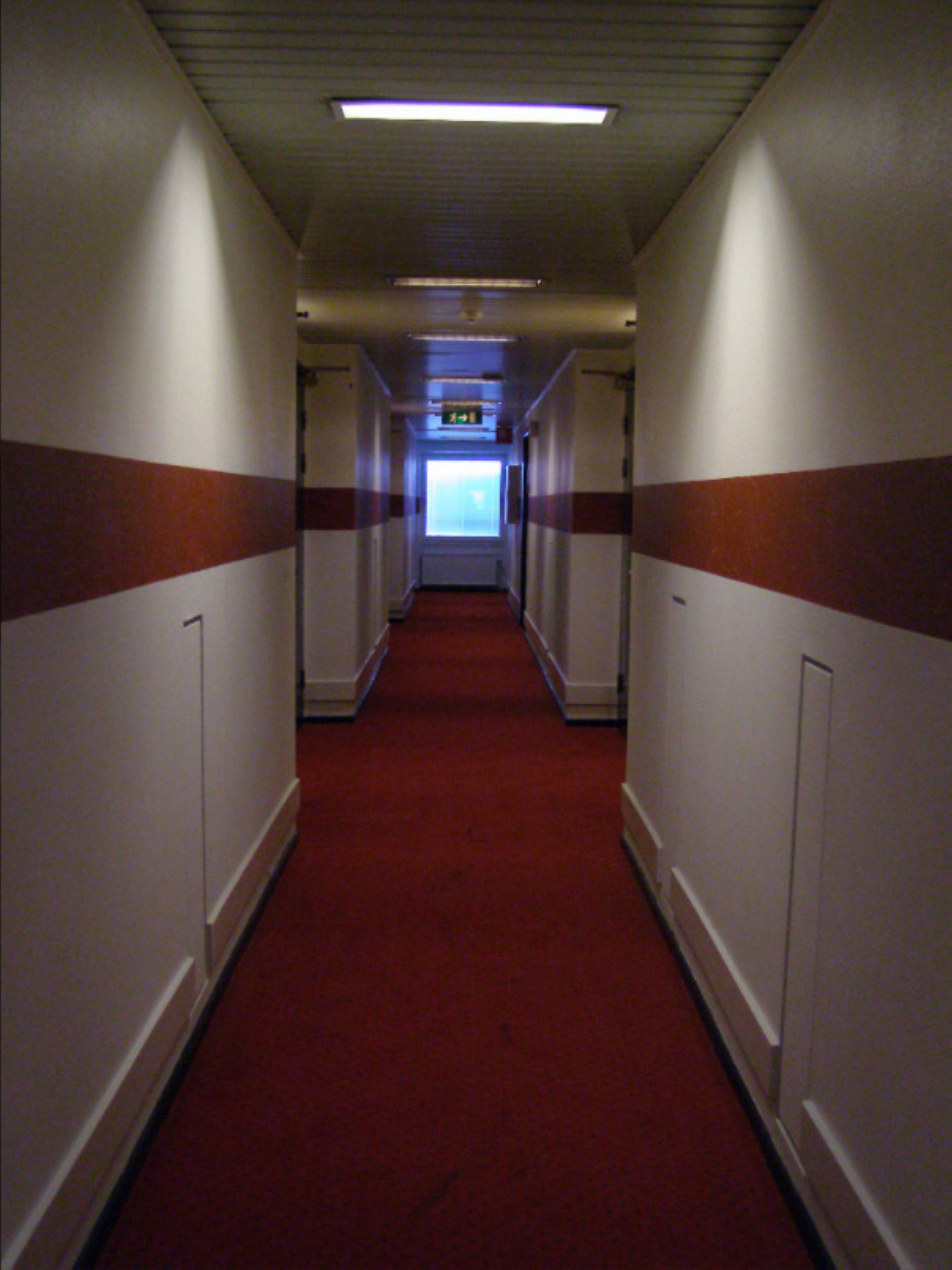} \hfill
\includegraphics[width=0.19\linewidth]{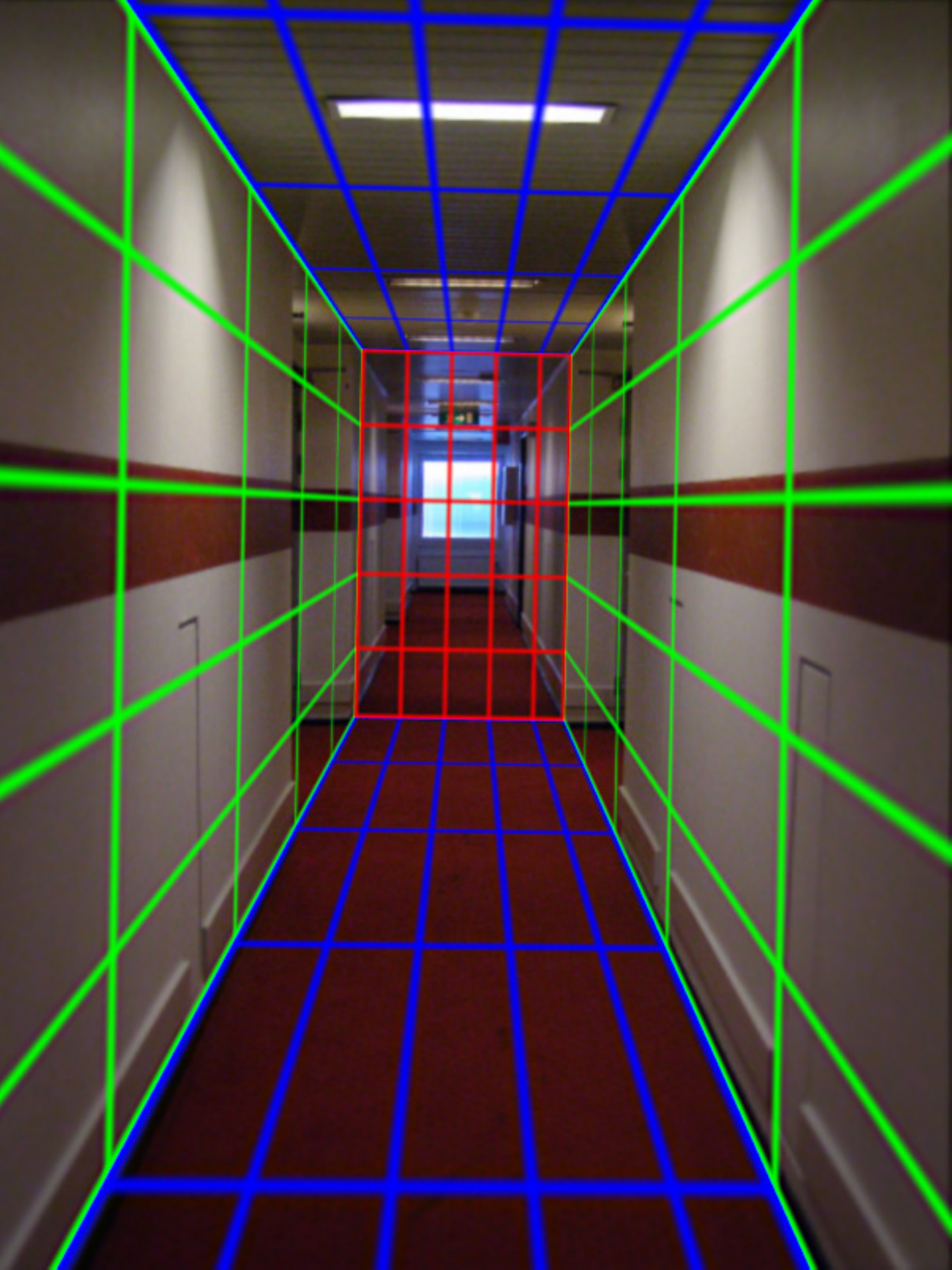} \hfill
\includegraphics[width=0.19\linewidth]{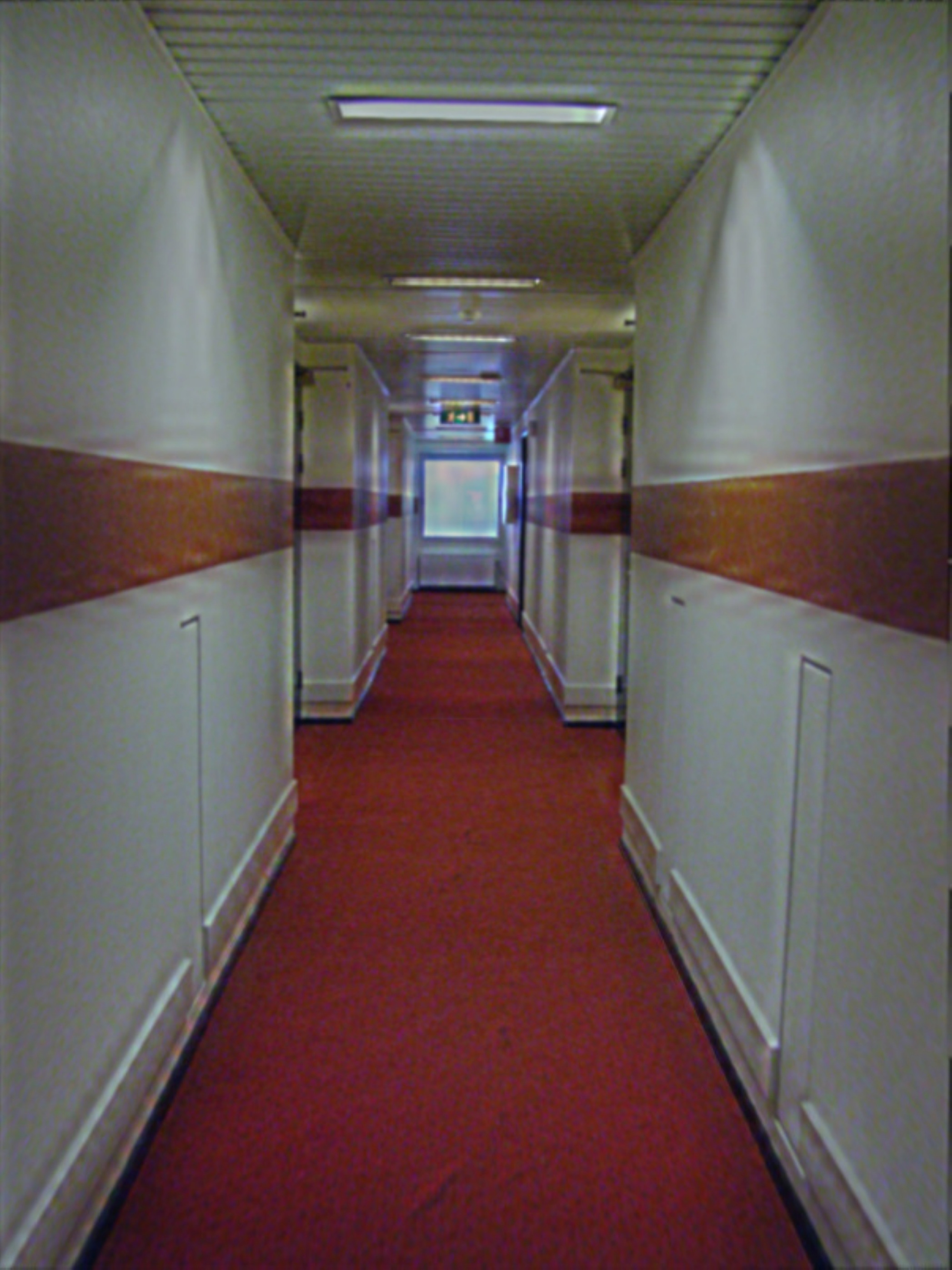}\hfill
\includegraphics[width=0.19\linewidth]{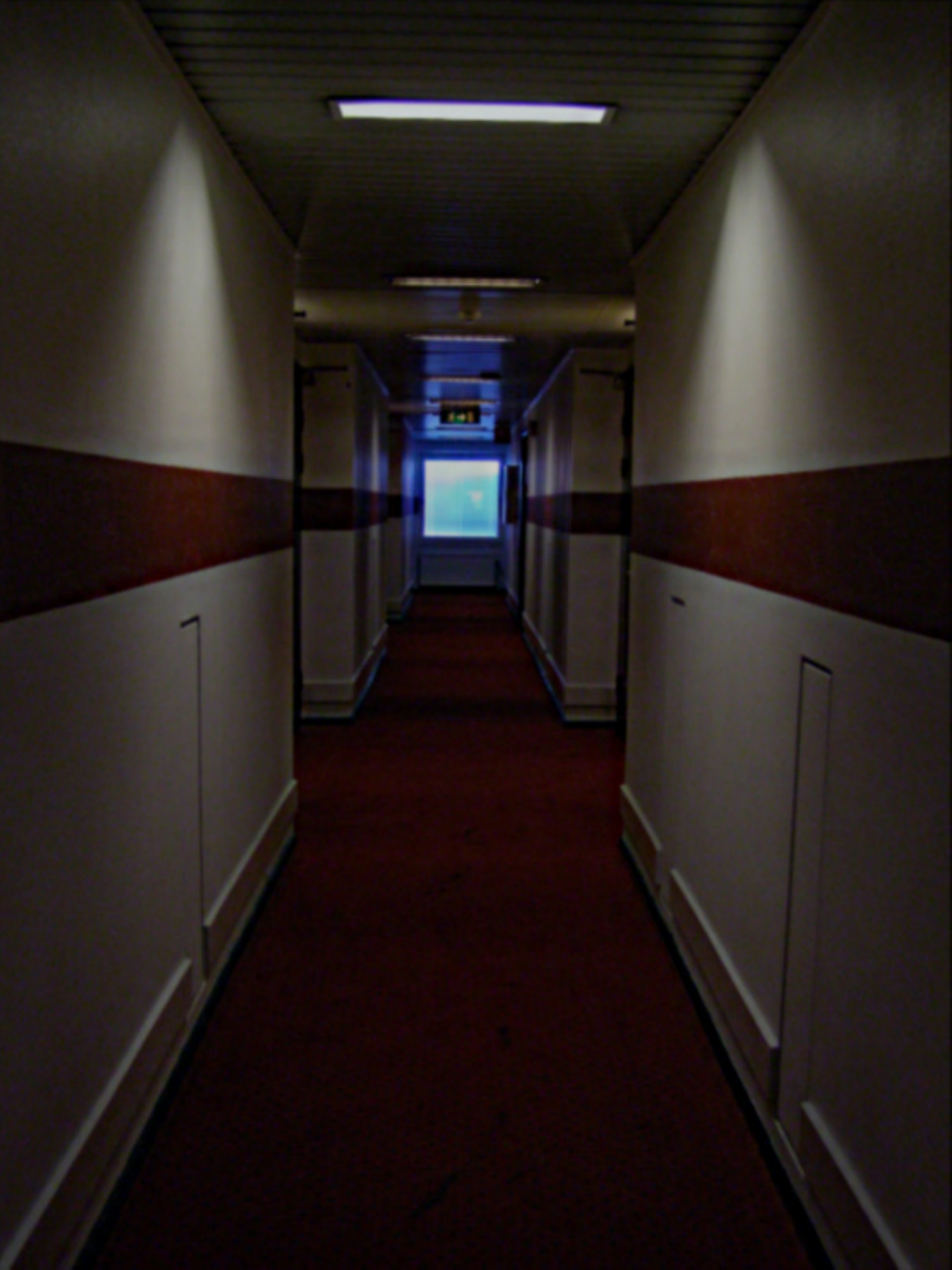}
}
\centerline{\large \ \ \ Input (a) \hspace{0.125\linewidth} Geometry (b)\hspace{0.135\linewidth} Albedo (c) \hspace{0.14\linewidth} Direct (d) \vspace{1.5mm}}
\vspace{-0.5mm}
\centerline{
\includegraphics[width=0.19\linewidth]{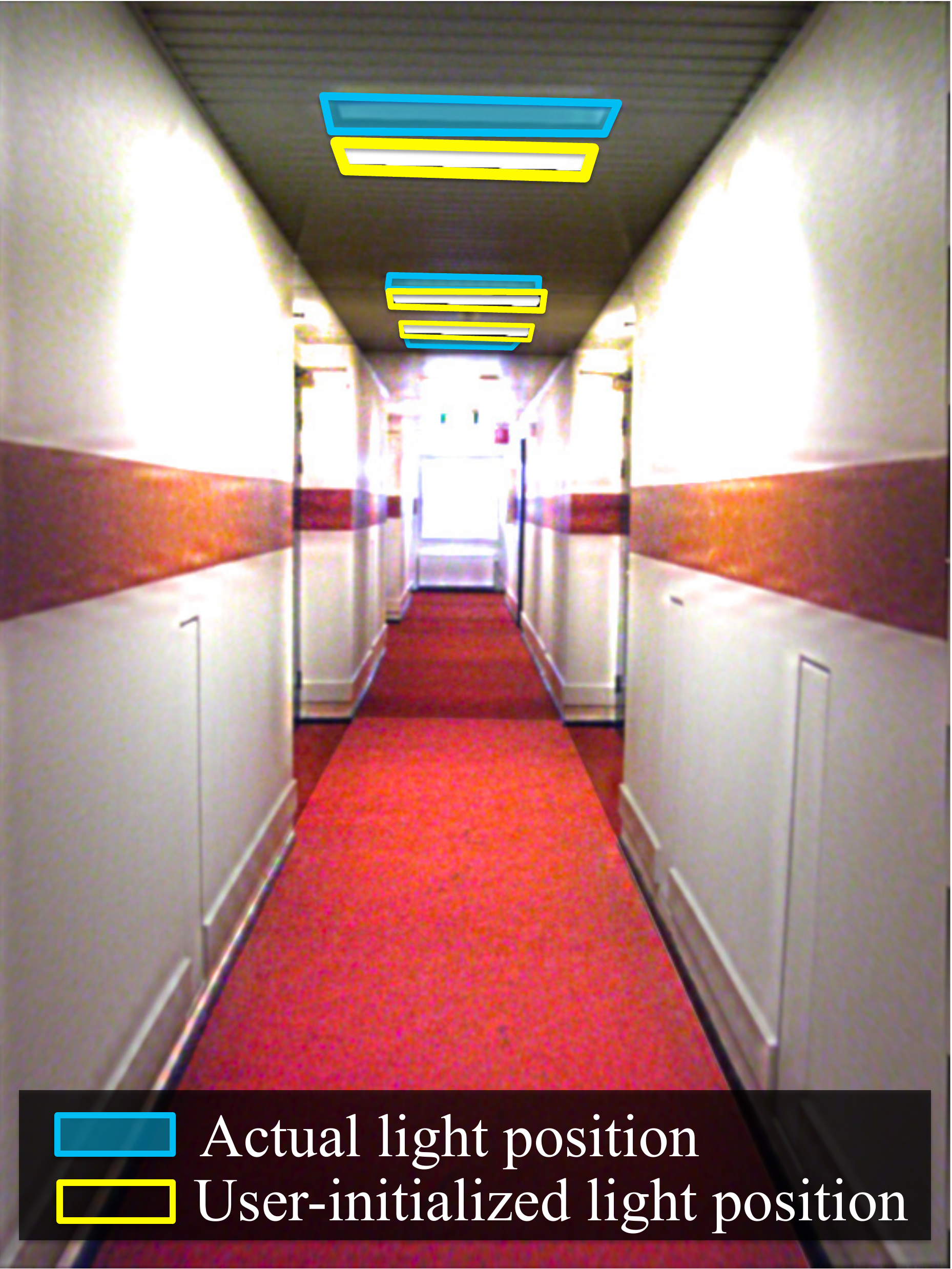} \hfill
\includegraphics[width=0.19\linewidth]{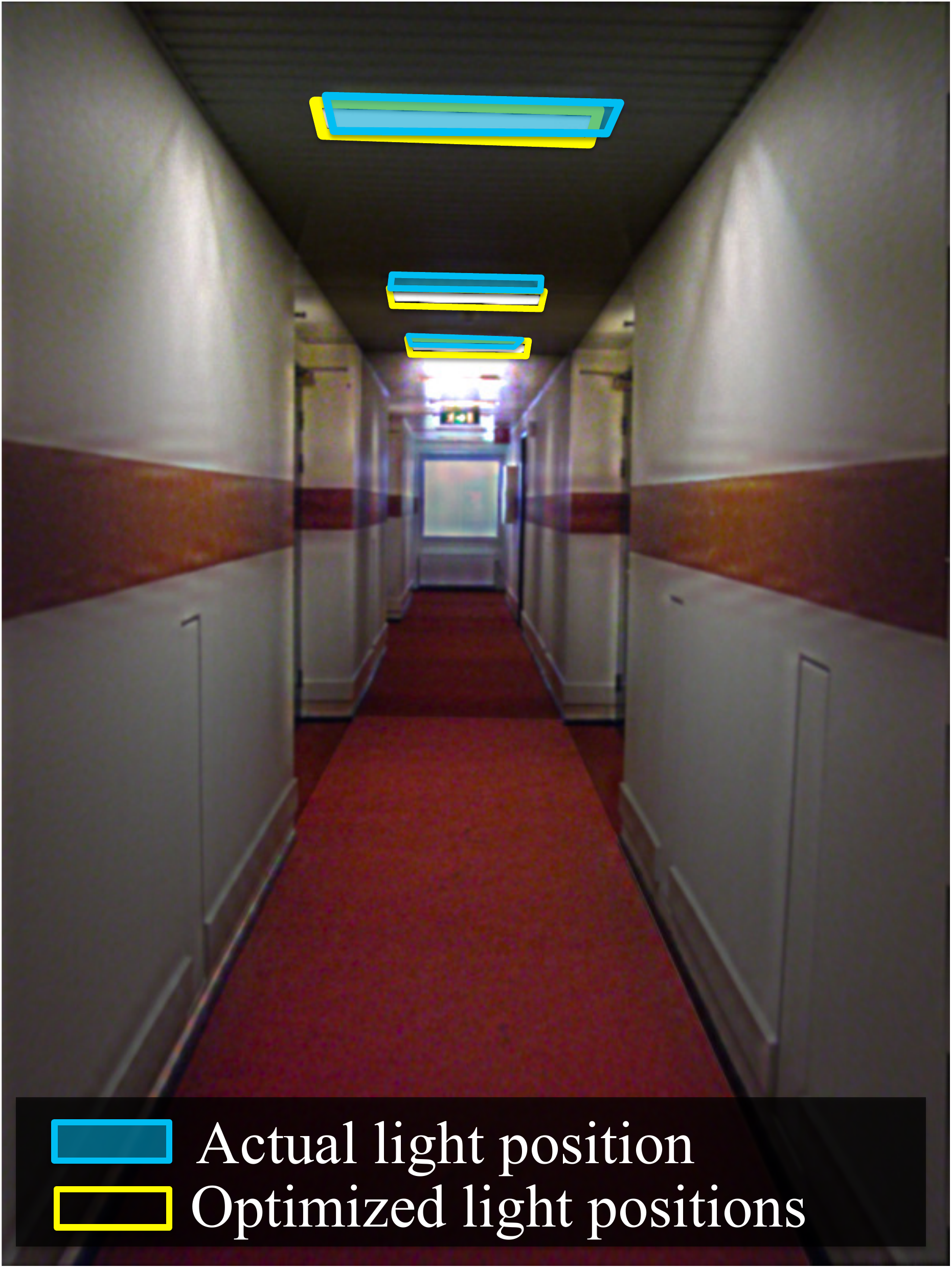} \hfill
\includegraphics[width=0.19\linewidth]{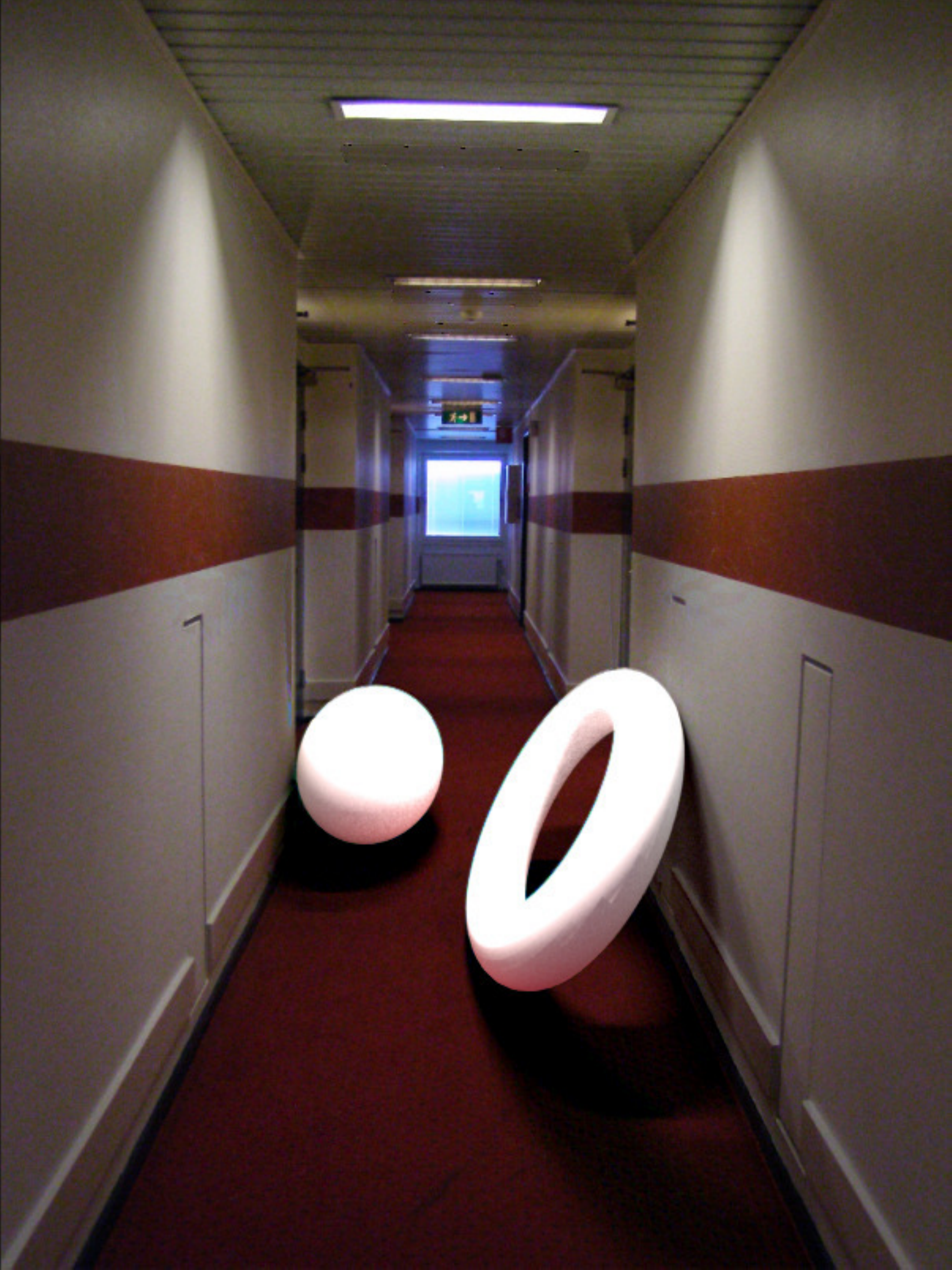}\hfill
\includegraphics[width=0.19\linewidth]{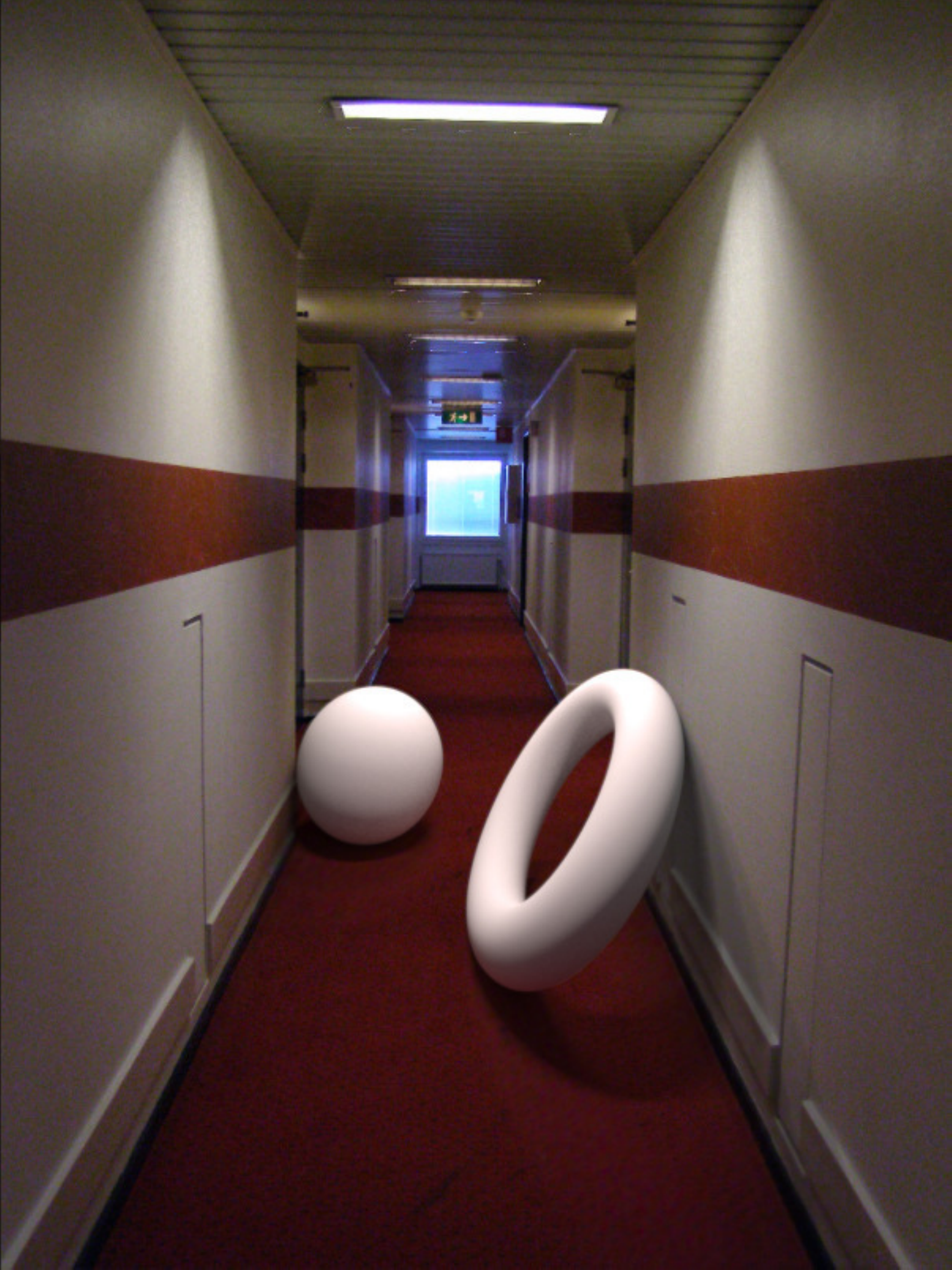}
}
\centerline{\large \ \ Initial lights (e) \hspace{0.08\linewidth} Refined lights (f)\hspace{0.08\linewidth}  Initial result (g) \hspace{0.07\linewidth} Refined result (h)}
\end{minipage}\hfill
\begin{minipage}{0.25\linewidth}
\caption{Overview of our interior lighting algorithm. For an input image \emph{(a)}, we use the modeled geometry (visualization of 3D scene boundaries as colored wireframe mesh, \emph{(b)}) to decompose the image into albedo \emph{(c)} and direct reflected light \emph{(d)}. The user defines initial lighting primitives in the scene \emph{(e)}, and the light parameters are re-estimated \emph{(f)}. The effectiveness of our lighting algorithm is demonstrated by comparing a composited result  \emph{(g)} using the initial light parameters to another composited result \emph{(h)} using the optimized light parameters. Our automatic lighting refinement enhances the realism of inserted objects. Lights are initialized away from the actual sources to demonstrate the effectiveness of our refinement.
}
\label{fig:lightingdemo}
\end{minipage}
\vspace{-4mm}
\end{center}
\end{figure*}

\boldhead{Materials and Illumination}
We use an automatic decomposition of the image into albedo, direct illumination and indirect illumination terms ({\em intrinsic images}~\cite{BarTen78}). Our geometric estimates are used to improve these terms and material estimates, similar to Boivin and Gagalowicz~\shortcite{Boivin:2001:IRD:383259.383270} and Debevec~\shortcite{Debevecprobe}, but our method improves efficiency of our illumination inference algorithm and is sufficient for realistic insertion (as demonstrated in Sections~\ref{sec:Evaluation} and~\ref{sec:study}).
We must work with a single legacy image, and wish to capture a physical light source estimate so that our method can be used in conjunction with any physical rendering software. Such representations as an irradiance volume do not apply~\cite{irvol}. Yu \ea  show that when a comprehensive model of geometry and luminaires is available, scenes can be relit convincingly~\cite{yu99inverse}.  We differ from them in that our estimate of geometry is coarse, and do not require multiple images.  Illumination in a room is not strongly directed, and cannot be encoded with a small set of point light sources, so the methods of Wang and Samaras~\cite{wangsamaras} and Lopez-Moreno \ea~\cite{LopezMoreno2010698} do not apply. As we show in our user study, point light models fail to achieve the realism that physical models do. We also cannot rely on having a known object present~\cite{satoikeuchi}.
In the past, we have seen that people are unable to detect perceptual errors in lighting~\cite{LopezMoreno2010698}. Such observations allow for high level image editing using rough estimates (e.g. materials~\cite{Khan:tog06} and lighting~\cite{kee-farid10a}). Lalonde and Efros~\shortcite{Lalonde_iccv07} consider the color distribution of images to differentiate real and fake images; our user study provides human assessment on this problem as well.


There are standard computational cues for estimating intrinsic images.  Albedo tends to display sharp, localized changes (which result in large image gradients), while shading tends to change slowly.  These rules-of-thumb inform the Retinex method~\cite{Land71} and important variants~\cite{Horn,Blake,BrelstaffBlake}. Sharp changes of shading do occur at shadow boundaries or normal discontinuities, but cues such as chromaticity~\cite{Funtlightness} or differently lit images~\cite{WeissII} can control these difficulties, as can methods that classify edges into albedo or shading~\cite{Tappenpami,Farenzena}. Tappen et al.~\shortcite{TappenCVPR06} assemble example patches of intrinsic image, guided by the real image, and exploiting the constraint that patches join up.  Recent work by Grosse \ea demonstrates that the color variant of Retinex is state-of-the-art for single-image decomposition methods~\cite{grosse09intrinsic}.

\section{Modeling}
\label{sec:technique}
To render synthetic objects realistically into a scene, we need estimates of geometry and lighting. At present, there are no methods for obtaining such information accurately and automatically; we incorporate user guidance to synthesize sufficient models.

Our lighting estimation procedure is the primary technical contribution of our method. With a bit of user markup, we automatically decompose the image with a novel intrinsic image method, refine initial light sources based on this decomposition, and estimate light shafts using a shadow detection method.
Our method can be broken into three phases. The first two phases interactively create models of geometry and lighting respectively, and the final phase renders and composites the synthetic objects into the image. An overview of our method is sketched in Algorithm~\ref{alg:technique}.

\subsection{Estimating geometry and materials}
\label{sec:geometry}
To realistically insert objects into a scene, we only need enough geometry to faithfully model lighting effects. We automatically obtain a coarse geometric representation of the scene using the technique of Hedau \ea\shortcite{hedau2009iccv}, and estimate vanishing points to recover camera pose automatically. Our interface allows a user to correct errors in these estimates, and also create simple geometry (tables and or near-flat surfaces) through image-space annotations. If necessary, other geometry can be added manually, such as complex objects near inserted synthetic objects. However, we have found that in most cases our simple models suffice in creating realistic results; all results in this paper require no additional complex geometry. Refer to Section~\ref{sec:details:geometry} for implementation details.


\subsection{Estimating illumination}
\label{sec:lights}
Estimating physical light sources automatically from a single image is an extremely difficult task. Instead, we describe a method to obtain a physical lighting model that, when rendered, closely resembles the original image.
We wish to reproduce two different types of lighting: {\it interior lighting}, emitters present within the scene, and {\it exterior lighting}, shafts of strongly directed light which lie outside of the immediate scene (e.g. sunlight).

\boldhead{Interior lighting}
Our geometry is generally rough and not canonical, and our lighting model should account for this; lights should be modeled such that renderings of the scene look similar to the original image. This step should be transparent to the user. We ask the user to mark intuitively where light sources should be placed, and then refine the sources so that the rendered image best matches the original image. Also, intensity estimation and color cast can be difficult to estimate, and we correct these automatically (see Fig~\ref{fig:lightingdemo}).

{\it Initializing light sources}.
To begin, the user clicks polygons in the image corresponding to each source. These polygons are projected onto the geometry to define an area light source.  Out-of-view sources are specified with 3D modeling tools.

\floatstyle{boxed}
\newfloat{algorithm}{htp!}{loa}
\floatname{algorithm}{Algorithm}
\begin{algorithm}
\underline{\textsc{legacyInsertion}($img, \textsc{user}$)}
\begin{tabbing}
\emph{Model}\= \emph{ geometry (Sec~\ref{sec:details:geometry}), auto-estimate materials (Sec~\ref{sec:details:materials})} \+ \\
$geometry \gets \textsc{detectBoundaries}(img)$ \\
$geometry \gets \textsc{user}(\text{`Correct boundaries'})$ \\
$geometry \gets \textsc{user}(\text{`Annotate/add additional geometry'})$ \\
$geometry_{mat} \gets \textsc{estMaterials}(img, geometry)$ \emph{[Eq~\ref{eq:decomp}]}\- \\
\emph{Refine initial lights and estimate shafts (Sec~\ref{sec:lights})} \+ \\
$lights \gets \textsc{user}(\text{`Annotate lights/shaft bounding boxes'})$ \\
$lights \gets \textsc{refineLights}(img, geometry)$ \emph{[Eq~\ref{eq:objfunc}]} \\
$lights \gets \textsc{detectShafts}(img)$ \- \\
\emph{Insert objects, render and composite (Sec~\ref{sec:details:insertion})} \+ \\
$scene \gets \textsc{createScene}(geometry, lights)$  \\
$scene \gets \textsc{user}(\text{'Add synthetic objects'})$ \\
return \textsc{composite}($img, \textsc{render}(scene))$  \emph{[Eq~\ref{eq:composite}]} \-
\end{tabbing}
\caption{Our method for rendering objects into legacy images
}
\label{alg:technique}
\end{algorithm}

{\it Improving light parameters}.
Our technique is to choose light parameters to minimize the squared pixel-wise differences between the rendered image (with estimated lighting and geometry) and the target image (e.g. the original image). Denoting $R({\bf L})$ as the rendered image parameterized by the current lighting parameter vector ${\bf L}$, $R^{*}$ as the target image, and ${\bf L}_0$ as the initial lighting parameters, we seek to minimize the objective
\begin{equation}
\label{eq:objfunc}
\begin{array}{c}
\displaystyle \argmin_{\bf L} \sum_{i \in \text{pixels}} \alpha_{i} (R_{i}({\bf L})-R_{i}^{*})^2 + \sum_{j \in \text{params} } w_j ( {\bf L}_j-{\bf L}_{0_j})^2 \vspace{1mm} \\
\hspace{15mm}\text{subject to: } 0 \leq {\bf L}_j \leq 1 \ \forall j
\vspace{-3.5mm}
\end{array}
\vspace{1mm}
\end{equation}

where $w$ is a weight vector that constrains lighting parameters near their initial values, and $\alpha$ is a per-pixel weighting that places less emphasis on pixels near the ground. Our geometry estimates will generally be worse near the bottom of the scene since we may not have geometry for objects near the floor.  In practice, we set $\alpha=1$ for all pixels above the spatial midpoint of the scene (height-wise), and $\alpha$ decreases quadratically from 1 to 0 at floor pixels. Also, in our implementation, ${\bf L}$ contains 6 scalars per light source: RGB intensity, and 3D position. More parameters could also be optimized. For all results, we normalize each light parameter to the range $[0,1]$, and set the corresponding values of $w$ to 10 for spatial parameters and 1 for intensity parameters. A user can also modify these weights depending on the confidence of their manual source estimates.
To render the synthetic scene and determine $R$, we must first estimate materials for all geometry in the scene.  We use our own intrinsic image decomposition algorithm to estimate surface reflectance (albedo), and the albedo is then projected onto the scene geometry as a diffuse texture map, as described in Section~\ref{sec:details:materials}.

\begin{figure}[htp]
\begin{center}
\includegraphics[width=85mm]{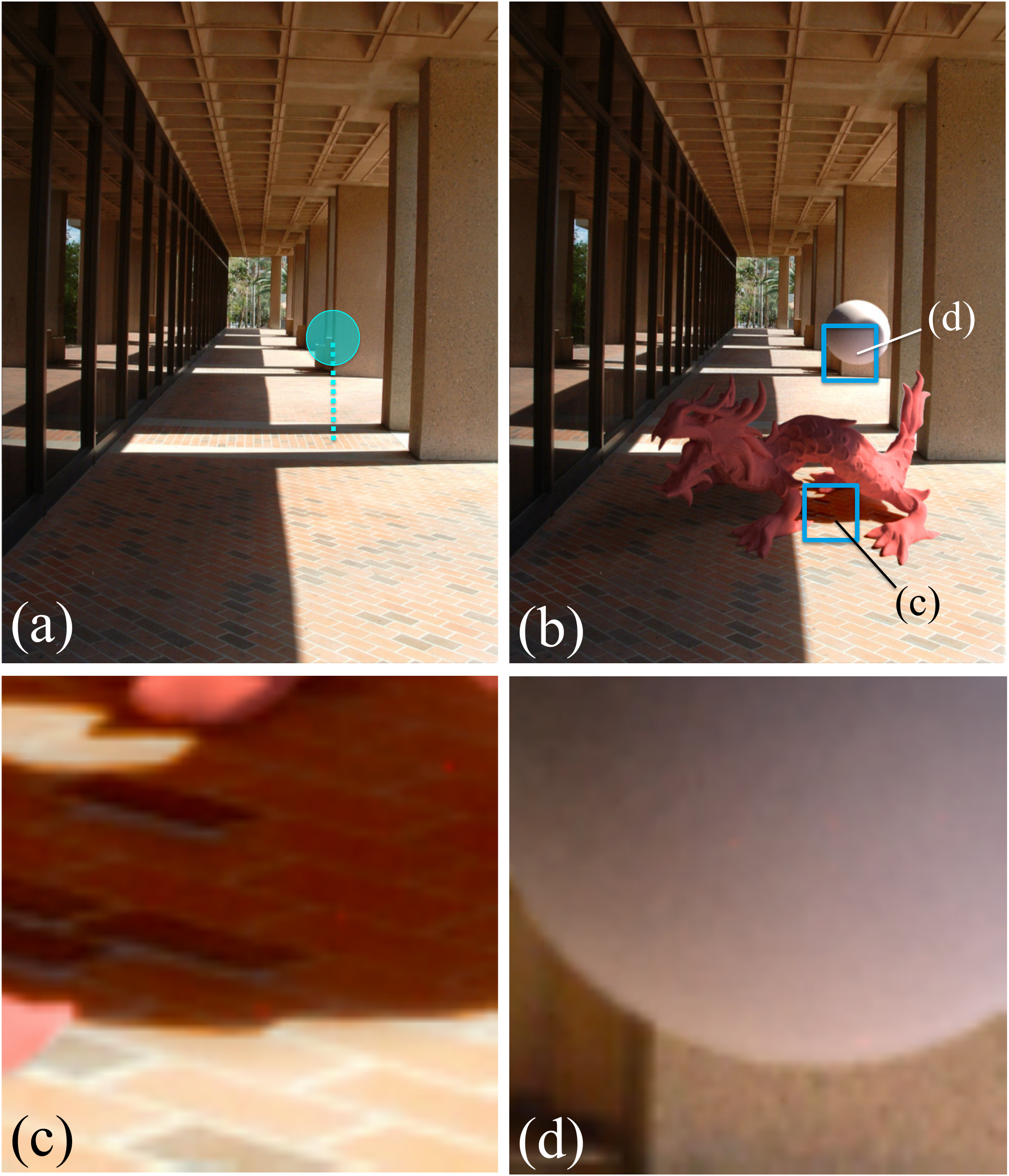}
\end{center}
\caption{Inserted objects fully participate with the scene lighting as if they were naturally a part of the image. Here, an input image \emph{(a)} is augmented with inserted objects and illuminated with a bright shaft of light \emph{(b)}. Interreflected red light from the dragon onto the brick floor is evident in \emph{(c)}, and the underside of the inserted sphere has a slight red tint from light reflecting off of the brick \emph{(d)}. A registration probe in \emph{(a)} displays the scale and location of the sphere in \emph{(b)}. Best viewed on a high resolution, high contrast display.
}
\label{fig:interreflections}
\end{figure}

{\it Intrinsic decomposition}.
Our decomposition method exploits our geometry estimates. First, indirect irradiance is computed by {\it gathering} radiance values at each 3D patch of geometry that a pixel projects onto. The gathered radiance values are obtained by sampling observed pixel values from the original image, which are projected onto geometry along the camera's viewpoint. We denote this indirect irradiance image as $\Gamma$; this term is equivalent to the integral in the radiosity equation. Given the typical Lambertian assumptions, we assume that the original image $B$ can be expressed as the product of albedo $\rho$ and shading $S$ as well as the sum of reflected direct light $D$ and reflected indirect light $I$. Furthermore, reflected gathered irradiance is equivalent to reflected indirect lighting under these assumptions. This leads to the equations
\begin{equation}
B=\rho S,\ \ \ B=D+I,\ \ \ I=\rho\Gamma,\ \ \ B = D + \rho \Gamma.
\label{eq:intr_assumptions}
\end{equation}
We use the last equation as constraints in our optimization below.

We have developed an objective function to decompose an image $B$ into albedo $\rho$ and direct light $D$ by solving
\begin{equation}
\label{eq:decomp}
\begin{split}
\argmin_{\rho, D} &\sum_{i \in \text{pixels}} |\Delta \rho|_i + \gamma_1 m_i (\nabla \rho)_i^2 + \gamma_2 (D_i-D_{0_i})^2 + \gamma_3 (\nabla D)_i^2 \\
&\text{subject to } \ \ \ B = D + \rho \Gamma, \ \ \ 0 \leq \rho \leq 1, \ \ \ 0 \leq D,
\end{split}
\vspace{-3mm}
\end{equation}
where $\gamma_1,\gamma_2,\gamma_3$ are weights, $m$ is a scalar mask taking large values where $B$ has small gradients, and small values otherwise, and $D_0$ is the initial direct lighting estimate. We define $m$ as a sigmoid applied to the gradient magnitude of $B$:  $m_i = 1-1/(1+e^{-s(||\nabla B||_i^2-c)})$, setting $s = 10.0$, $c=0.15$ in our implementation.

\begin{figure*}[htp]
\centerline{
\includegraphics[width=44mm]{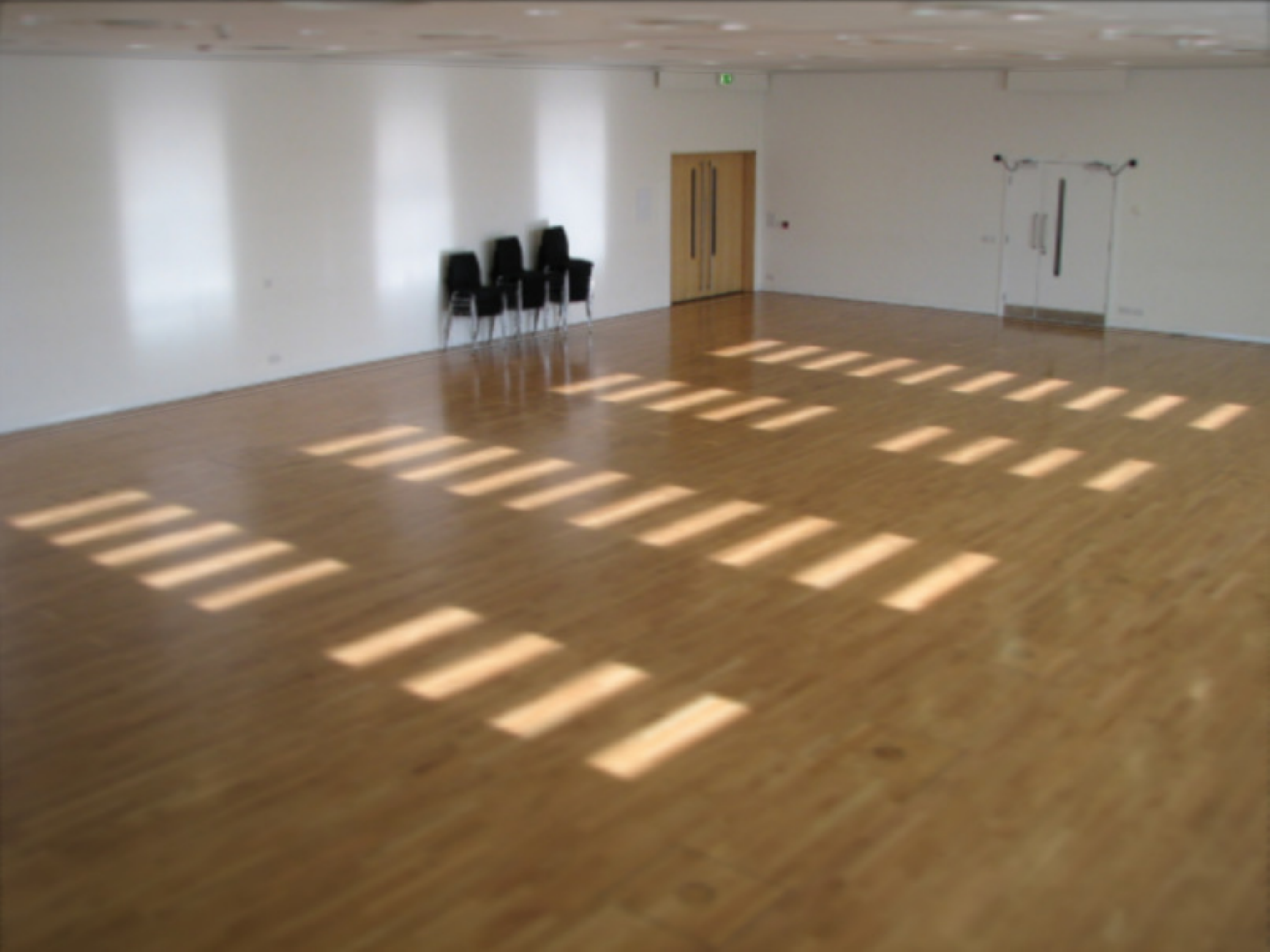}
\includegraphics[width=45mm]{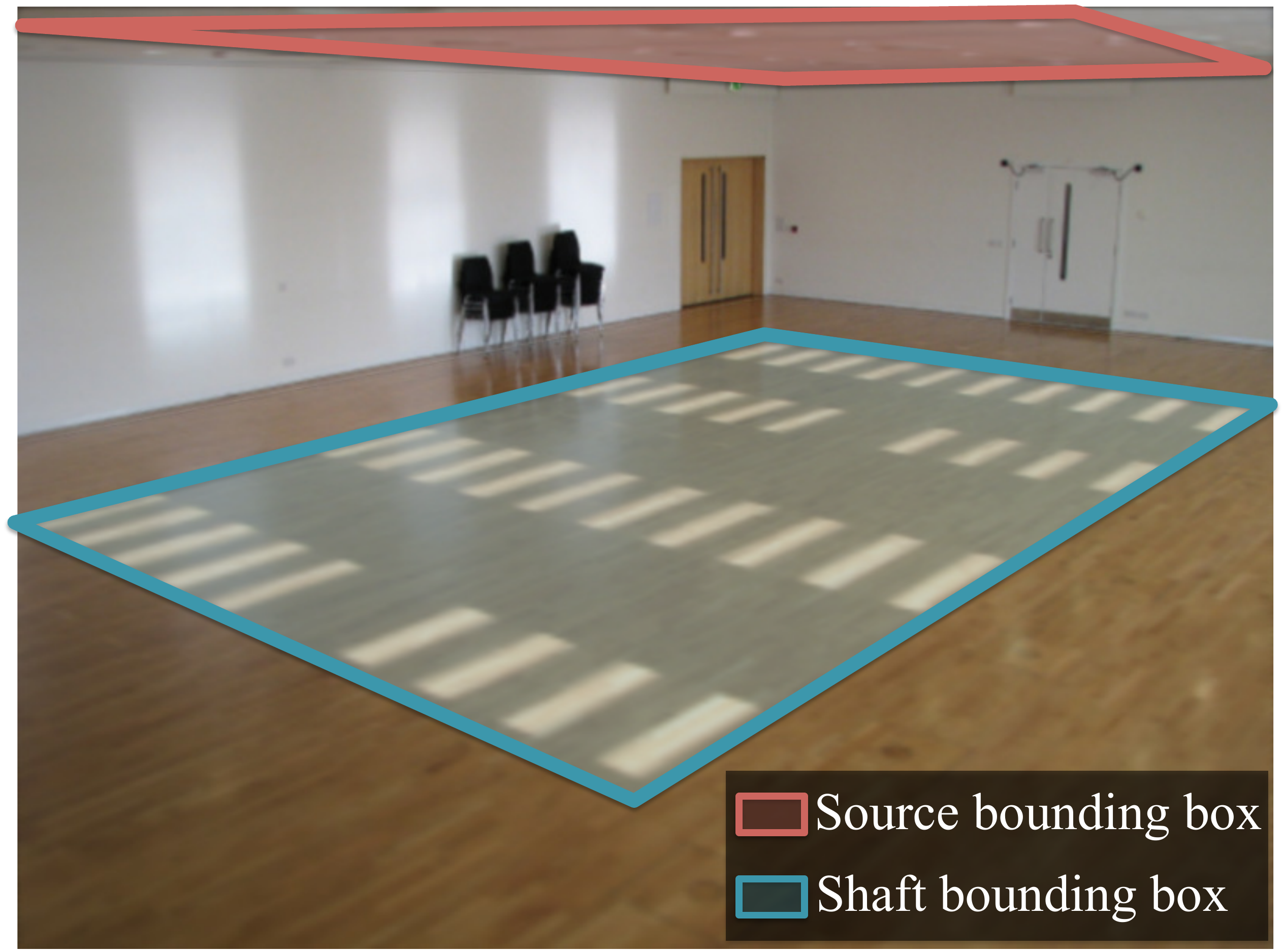}
\includegraphics[width=44mm]{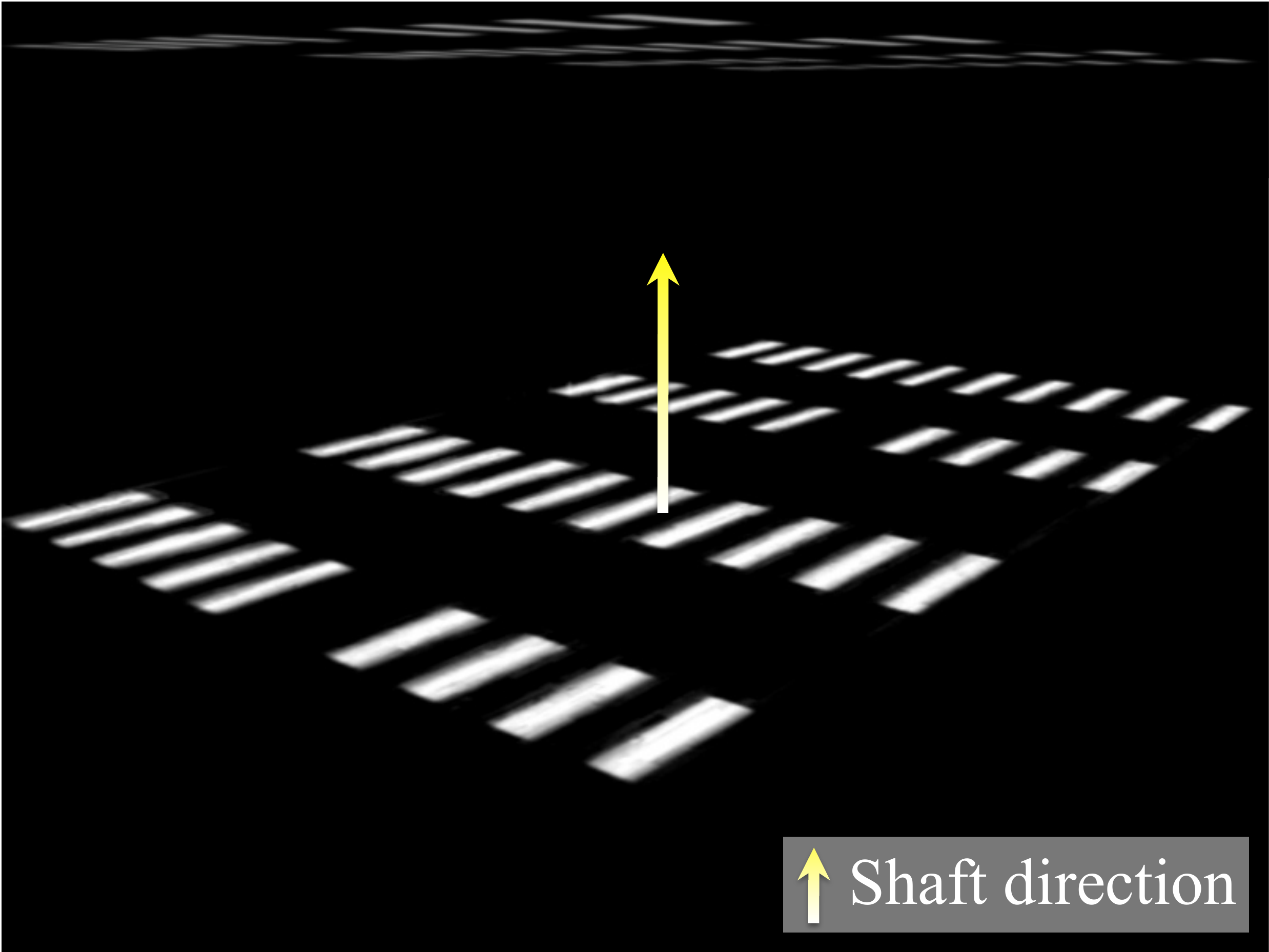}
\includegraphics[width=44mm]{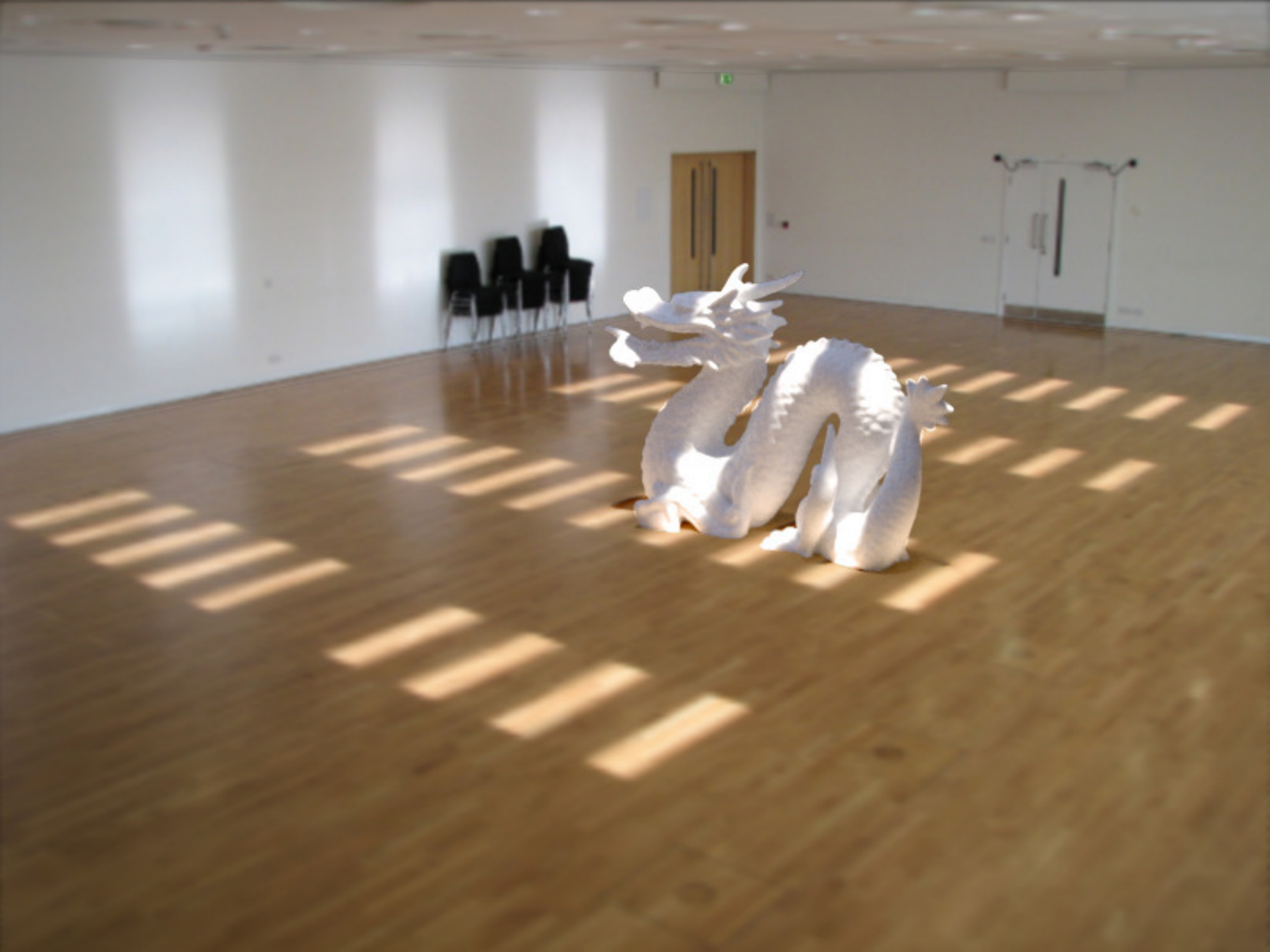}
}
\centerline{\Large (a) \hspace{39mm} (b)\hspace{39mm}  (c) \hspace{39mm} (d)}
\caption{Our algorithm for estimating exterior lighting (light shafts). Given an input image \emph{(a)}, the user specifies bounding boxes around the shafts and their sources \emph{(b)}. The shafts are detected automatically, and the shaft direction is estimated using the centroid of the the bounding boxes in 3D \emph{(c)}. A physical lighting model (e.g. a masked, infinitely far spotlight) is created from this information, and objects can be rendered inserted realistically into the scene \emph{(d)}.
}
\label{fig:shafts}
\end{figure*}

Our objective function is grounded in widespread intrinsic image assumptions \cite{Land71,Blake,BrelstaffBlake}, namely that shading is spatially slow and albedo consists of piecewise constant patches with potentially sharp boundaries. The first two terms in the objective coerce $\rho$ to be piecewise constant. The first term enforces an L1 sparsity penalty on edges in $\rho$, and the second term smoothes albedo only where $B$'s gradients are small. The final two terms smooth $D$ while ensuring it stays near the initial estimate $D_0$. We set the objective weights to $\gamma_1 = 0.2$, $\gamma_2 = 0.9$, and $\gamma_3 = 0.1$. We initialize $\rho$ using the color variant of Retinex as described by Grosse \ea\shortcite{grosse09intrinsic}, and initialize $D$ as $D_0 = B-\rho\Gamma$ (by Eq.~\ref{eq:intr_assumptions}). This optimization problem can be solved in a variety of ways; we use an interior point method (implemented with MATLAB's optimization toolbox).
In our implementation, to improve performance of our lighting optimization (Eq.~\ref{eq:objfunc}), we set the target image as our estimate of the direct term, and render our scene only with direct lighting (which greatly reduces the time in recalculating the rendered image). We choose our method as it utilizes the estimated scene geometry to obtain better albedo estimates, and reduces the computation cost of solving Eq.~\ref{eq:objfunc}, but any decomposition method could be used (e.g. Retinex).

\begin{figure}[tp]
\begin{center}
\includegraphics[width=41mm]{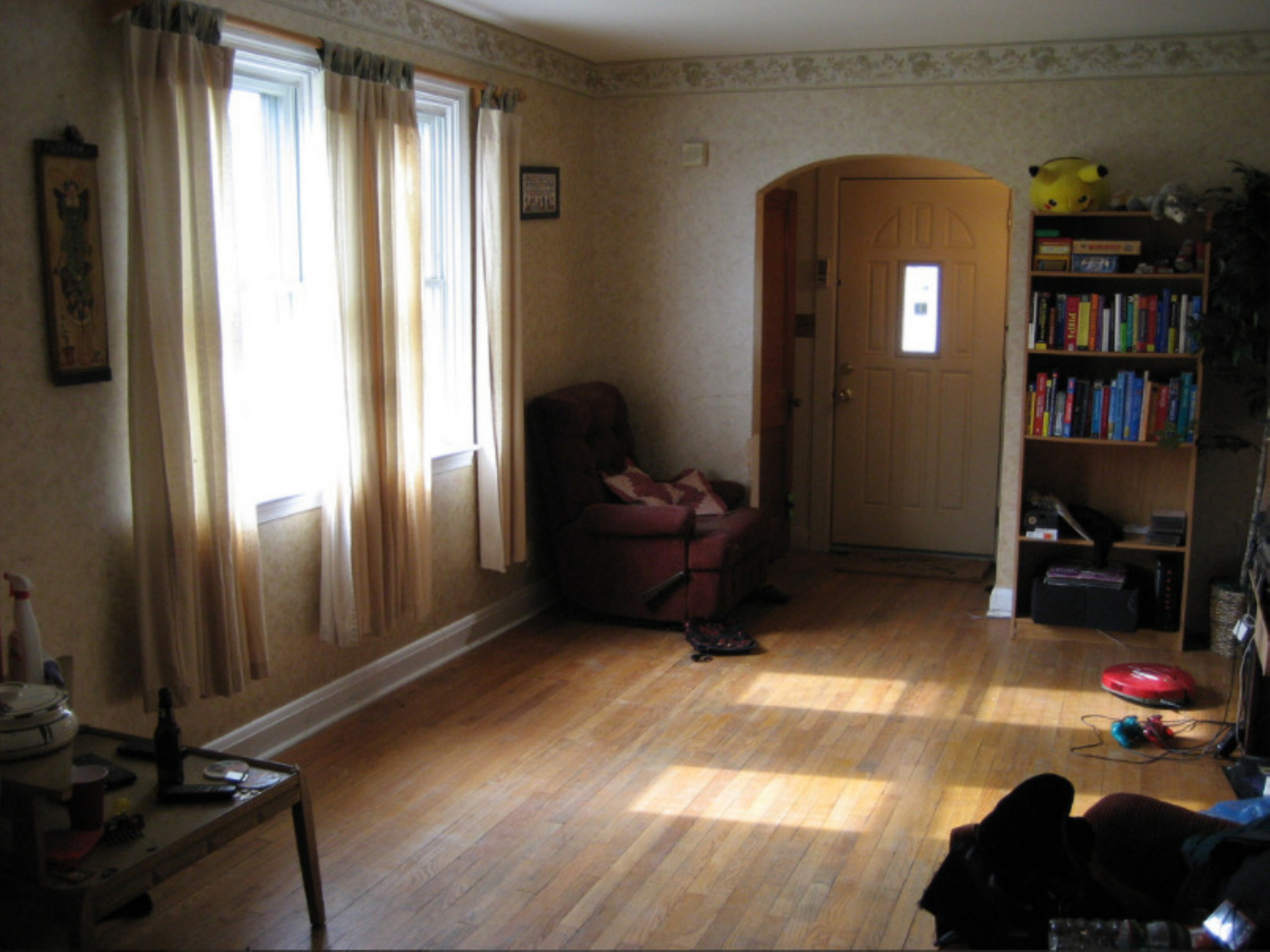}
\includegraphics[width=41mm]{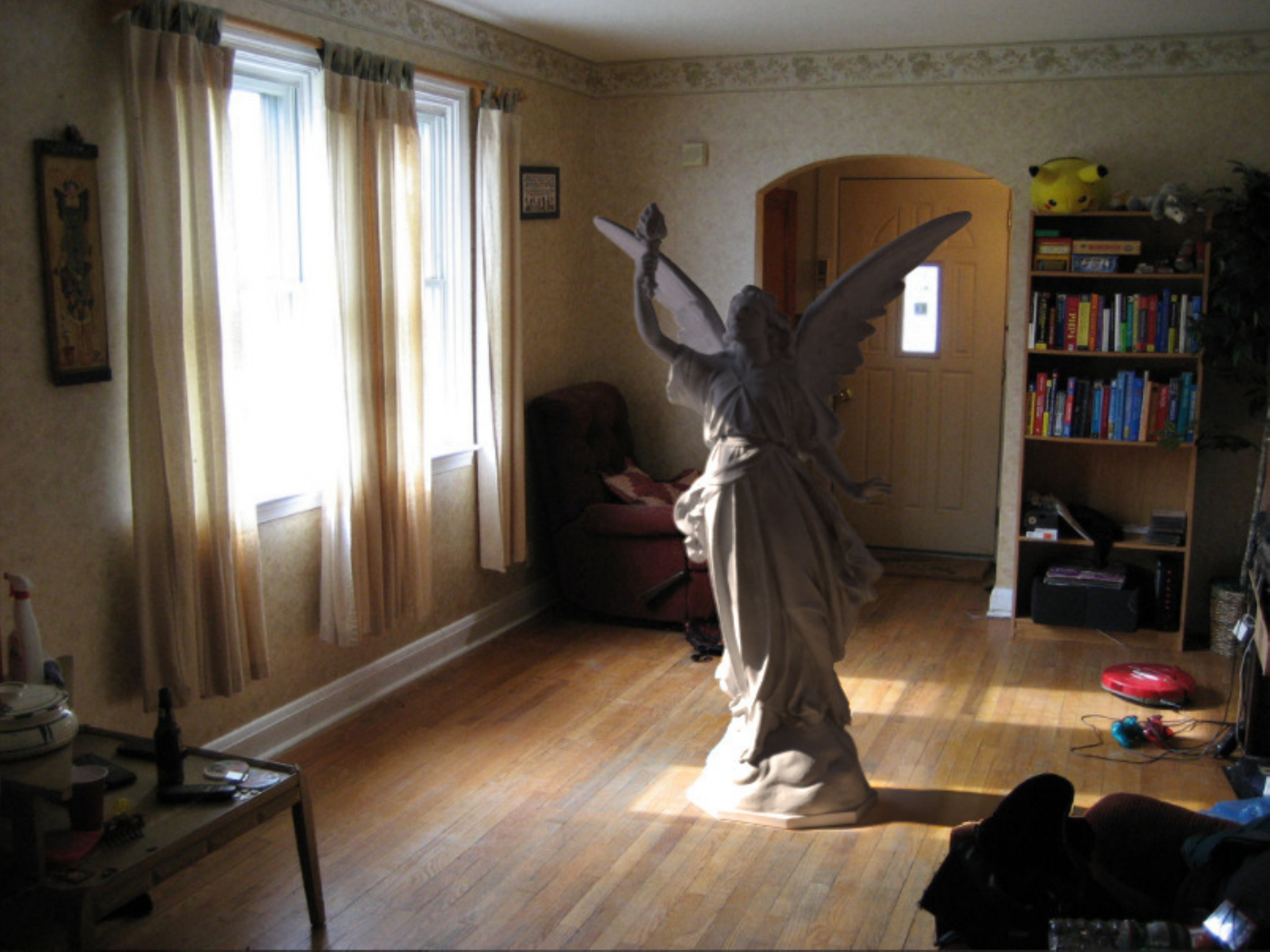}
\end{center}
\caption{A difficult image for detecting light shafts. Many pixels near the window are saturated, and some shaft patterns on the floor are occluded, as in the image on the left. However, an average of the matte produced for the floor and wall provides an acceptable estimate (used to relight the statue on the right).
}
\label{fig:shaftissues}
\end{figure}

\boldhead{Exterior lighting (light shafts)}
Light shafts are usually produced by the sun, or some other extremely far away source. Thus, the type of light we wish to model can be thought of as purely directional, and each shaft in a scene will have the same direction.

We define a light shaft with a 2D polygonal projection of the shaft  and a direction vector. In Figure~\ref{fig:shafts}, the left image shows a scene with many light shafts penetrating the ceiling and projecting onto the floor. Our idea is to detect either the source or the projections of shafts in an image and recover the shaft direction. The user first draws a bounding box encompassing shafts visible in the scene, as well as a bounding box containing shaft sources (windows, etc.).  We then use the shadow detection algorithm of Guo \ea~\shortcite{guo_cvpr11} to determine a scalar mask that estimates the confidence that a pixel is {\it not} illuminated by a shaft. This method models region based appearance features along with pairwise relations between regions that have similar surface material and illumination.  A graph cut inference is then performed to identify the regions that have same material and different illumination conditions, resulting in the confidence mask. The detected shadow mask is then used to recover a soft shadow matte using the spectral matting method of Levin \ea~\shortcite{Levin07spectralmatting}. We then use our estimate of scene geometry to recover the direction of the shafts (the direction defined by the two midpoints of the two bounding boxes). However, it may be the case that either the shaft source or the shaft projection is not visible in an image. In this case, we ask the user to provide an estimate of the direction, and automatically project the source/shaft accordingly. Figure~\ref{fig:shafts} shows an example of our shaft procedure where the direction vector is calculated automatically from the marked bounding boxes. Shafts are represented as masked spotlights for rendering.

\begin{figure}[htpb!]
\centerline{
\includegraphics[width=0.49\columnwidth]{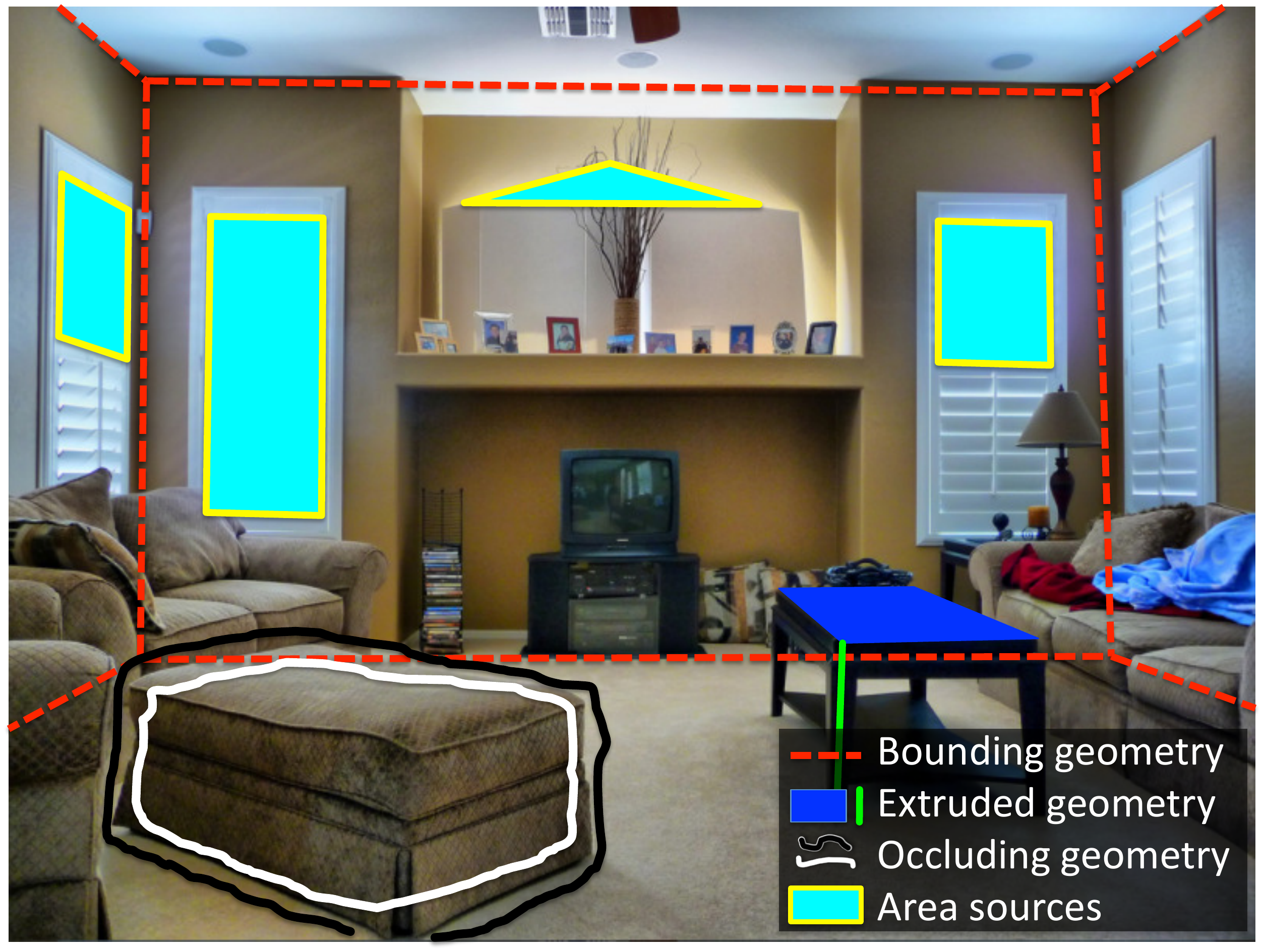}
\includegraphics[width=0.49\columnwidth]{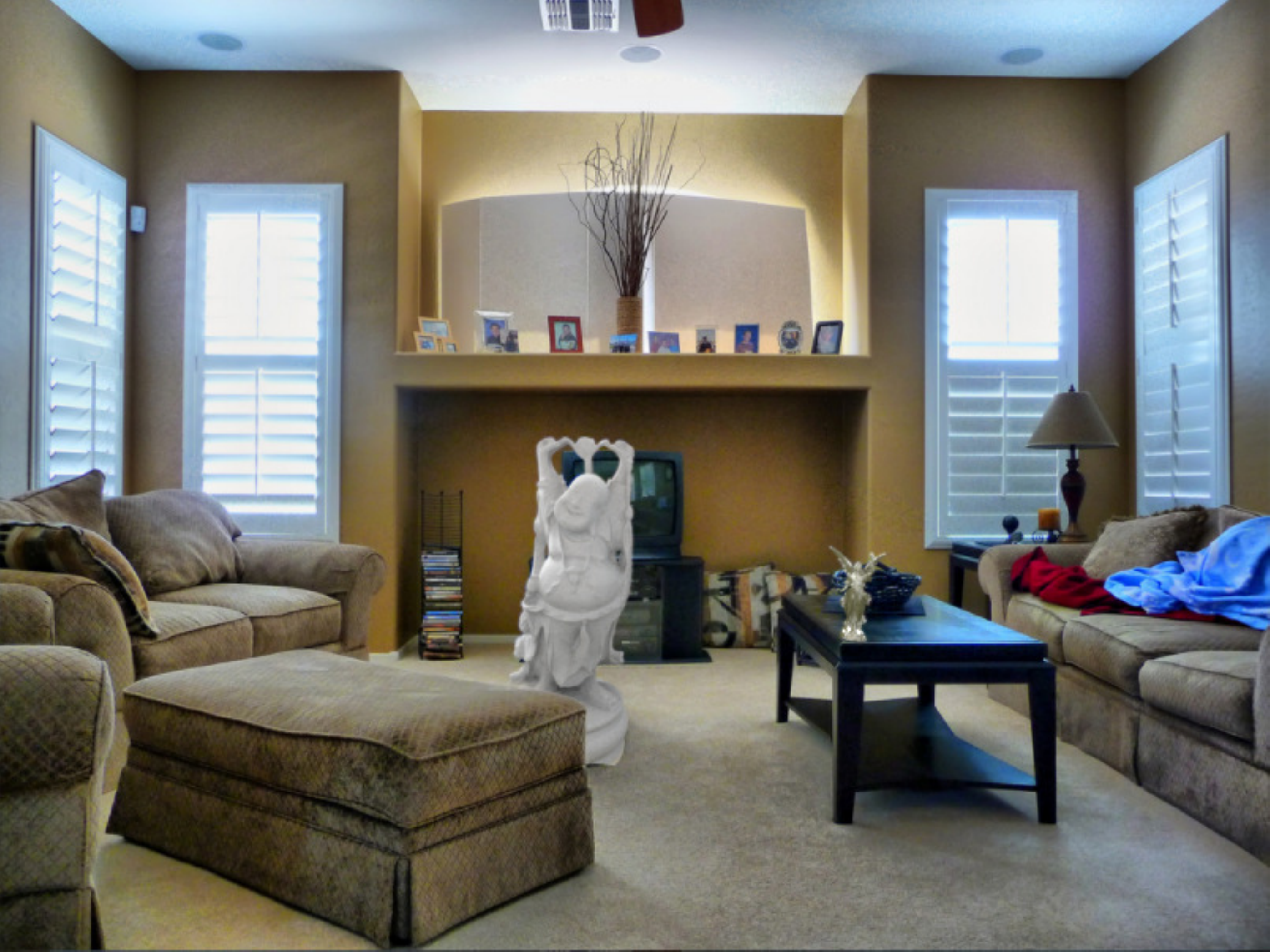}
}
\caption{Our system is intuitive and quick. This result was modeled by a user unfamiliar with our interface (after a short demonstration). From start to finish, this result was created in under 10 minutes (render time not included). User's markup shown on left.
}
\label{fig:noviceresult}
\end{figure}

In some cases, it is  difficult to recover accurate shadow mattes for a window on a wall or a shaft on the floor individually. For instance, it is difficult to detect the window in Figure~\ref{fig:shaftissues}  using only the cues from the wall.  In such cases, we project the recovered mask on the floor along the shaft direction to get the mapping on the wall and average matting results for the wall and floor to improve the results. Similarly, an accurate matte of a window can be used to improve the matte of a shaft  on the floor (as in the right image of Figure~\ref{fig:teaser}).

\subsection{Inserting synthetic objects}
\label{sec:details:insertion}
With the lighting and geometry modeled, a user is now free to insert synthetic 3D geometry
into the scene. Once objects have been inserted, the scene can be rendered with any suitable rendering software.\footnote{For our results, we use LuxRender (http://www.luxrender.net)}~Rendering is trivial, as all of the information required by the renderer has been estimated (lights, geometry, materials, etc).

To complete the insertion process, we composite the rendered objects back into the original photograph using the additive differential rendering method~\cite{Debevecprobe}. This method renders two images: one containing synthetic objects $\mathcal{I}_{obj}$, and one without synthetic objects $\mathcal{I}_{noobj}$, as well as an object mask $M$ (scalar image that is 0 everywhere where no object is present, and $(0,1]$ otherwise). The final composite image $\mathcal{I}_{final}$ is obtained by
\begin{equation}
\label{eq:composite}
\mathcal{I}_{final} = M \odot \mathcal{I}_{obj} + (1-M) \odot (\mathcal{I}_b + \mathcal{I}_{obj}-\mathcal{I}_{noobj})
\end{equation}
where $\mathcal{I}_b$ is the input image, and $\odot$ is the Hadamard product.

\begin{figure}[htp]
\begin{center}
\includegraphics[width=0.49\columnwidth]{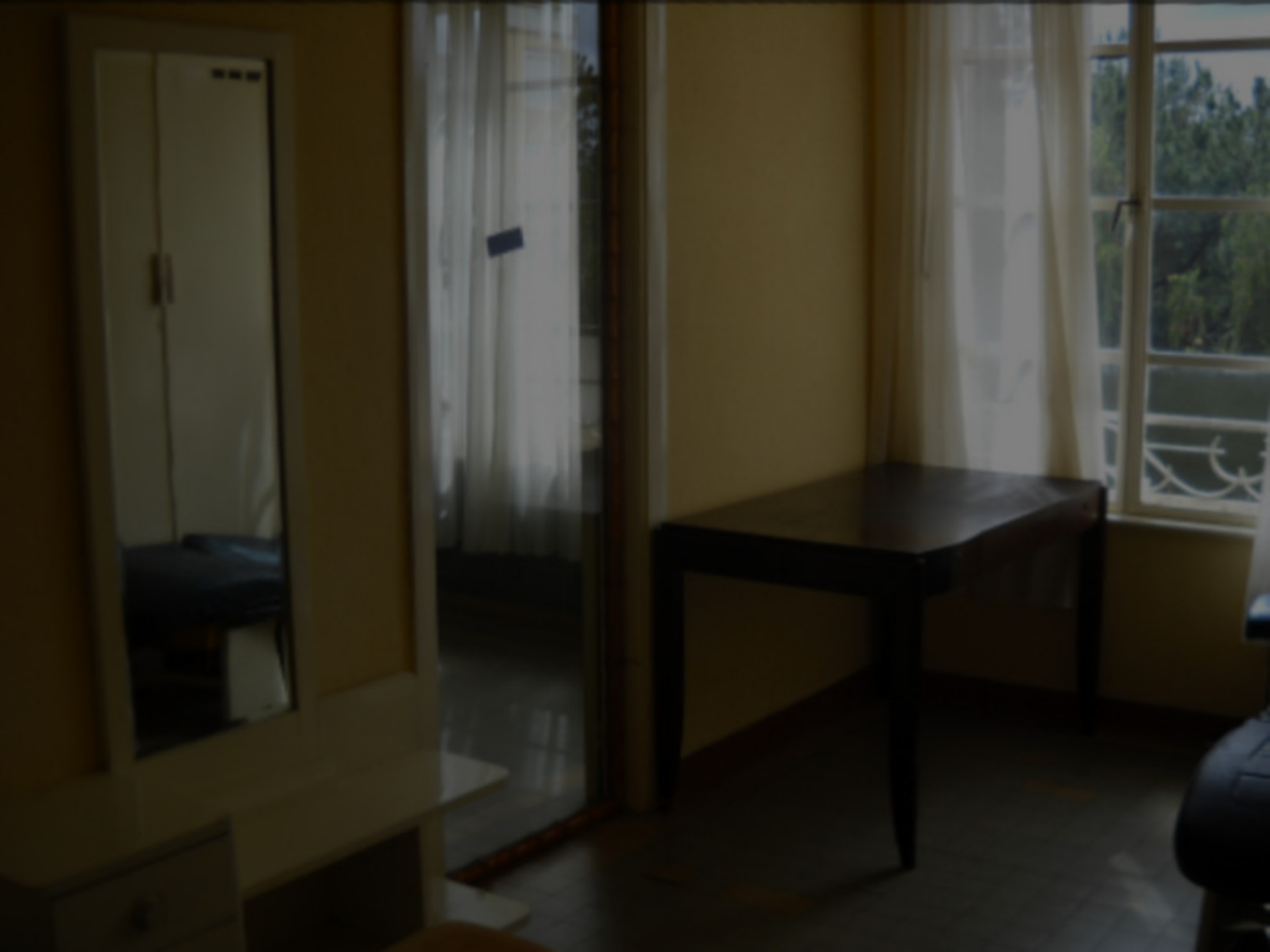}
\includegraphics[width=0.49\columnwidth]{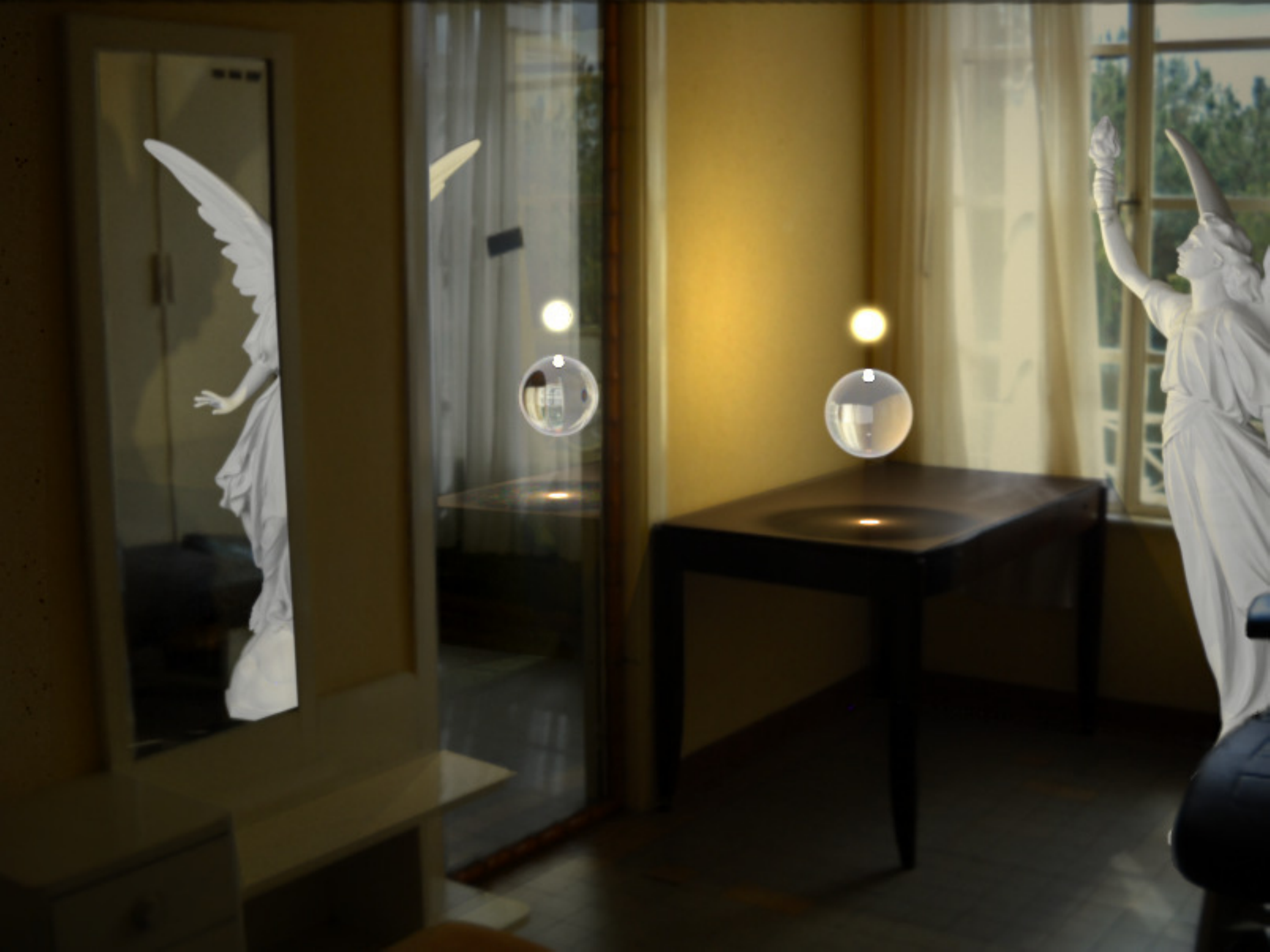}
\end{center}
\caption{Our method allows for light source insertion and easy material reassignment. Here, a glowing ball is inserted above a synthetic glass sphere, casting a caustic on the table. The mirror has been marked as reflective, allowing synthetic objects to realistically interact with the scene.
}
\label{fig:mirror}
\end{figure}

\section{Implementation details}
\label{sec:details}
\subsection{Modeling geometry}
\label{sec:details:geometry}
Rough scene boundaries ({\it bounding geometry}) are estimated first along with the camera pose, and we provide tools for correcting and supplementing these estimates. Our method also assigns materials to this geometry automatically based on our intrinsic decomposition algorithm (Sec.~\ref{sec:lights}).




\boldhead{Bounding geometry}
We model the bounding geometry as a 3D cuboid; essentially the scene is modeled as a box that circumscribes the camera so that up to five faces are visible. Using the technique of Hedau \ea~\shortcite{hedau2009iccv}, we automatically generate an estimate of this box layout for an input image, including camera pose. This method estimates three vanishing points for the scene (which parameterize the box's rotation), as well as a 3D translation to align the box faces with planar faces of the scene (walls, ceiling floor). However, the geometric estimate may be inaccurate, and in that case, we ask the user to manually correct the layout using a simple interface we have developed. The user drags the incorrect vertices of the box to corresponding scene corners, and manipulates vanishing points using a pair of line segments (as in the Google Sketchup\footnote{http://sketchup.google.com} interface) to fully specify the 3D box geometry.

\boldhead{Additional geometry}
We allow the user to easily model {\it extruded geometry}, i.e. geometry defined by a closed 2D curve that is extruded along some 3D vector, such as tables, stairs, and other axis-aligned surfaces. In our interface, a user sketches a 2D curve defining the surface boundary, then clicks a point in the footprint of the object which specifies the 3D height of the object~\cite{criminisi2000}. Previously specified vanishing points and bounding geometry allow for these annotations to be automatically converted to a 3D model.

In our interface, users can also specify {\it occluding surfaces}, complex surfaces which will occlude inserted synthetic objects (if the inserted object is behind the occluding surface). We allow the user to create occlusion boundaries for objects using the interactive spectral matting segmentation approach~\cite{Levin07spectralmatting}. The user defines the interior and exterior of an object by scribbling, and a segmentation matte for the object is computed. These segmentations act as cardboard cutouts in the scene; if an inserted object intersects the segmentation and it is farther from the camera, then it will be occluded by the cutout.  We obtain the depth of an object by assuming the lowermost point on its boundary to be its contact point with the floor. Figures~\ref{fig:systemdemo} and~\ref{fig:noviceresult} show examples of both extruded and occluding geometry.

\begin{figure}[htp]
\begin{minipage}[b]{.65\columnwidth}
\includegraphics[width=.49\columnwidth]{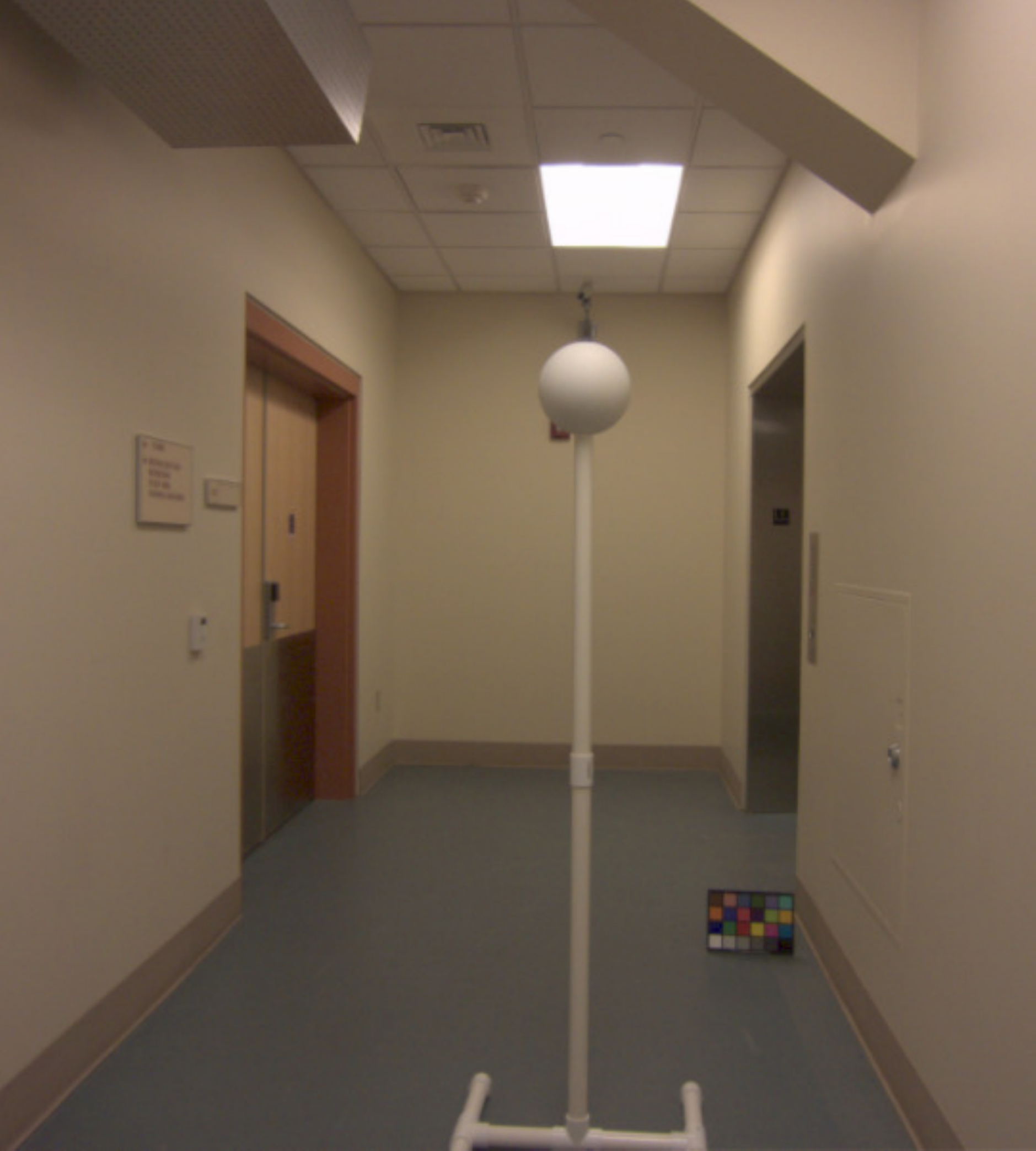}
\includegraphics[width=.49\columnwidth]{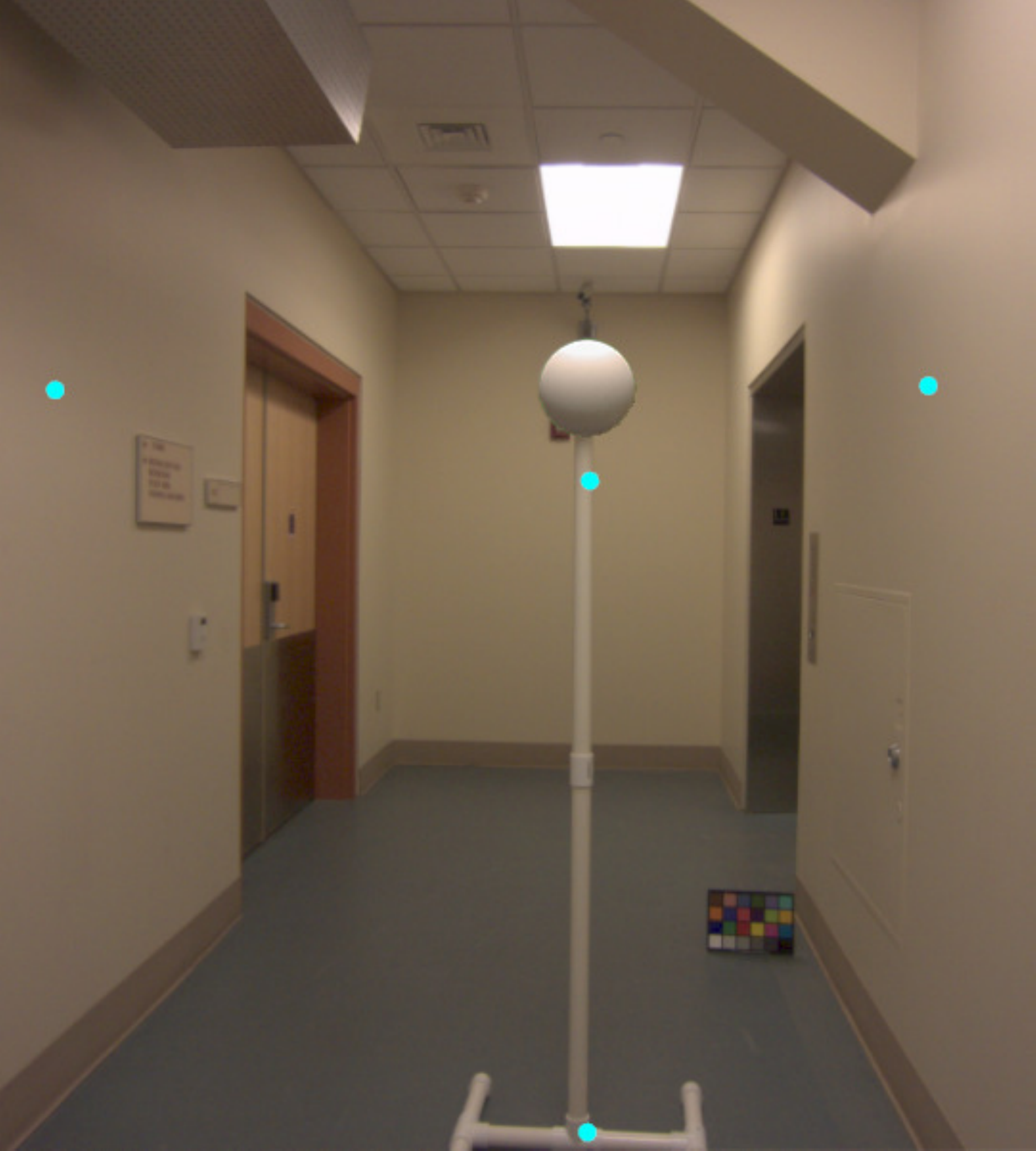}
\caption{The instrument used for collecting ground truth illumination data. The left image shows the apparatus (a white, diffuse ball resting on a plastic, height-adjustable pole). Using knowledge of the physical scene, we can align a rendered sphere over the probe for error measurements \emph{(right}).
}
\label{fig:rig}
\end{minipage}
\hfill
\begin{minipage}[b]{.29\columnwidth}
\includegraphics[width=\columnwidth]{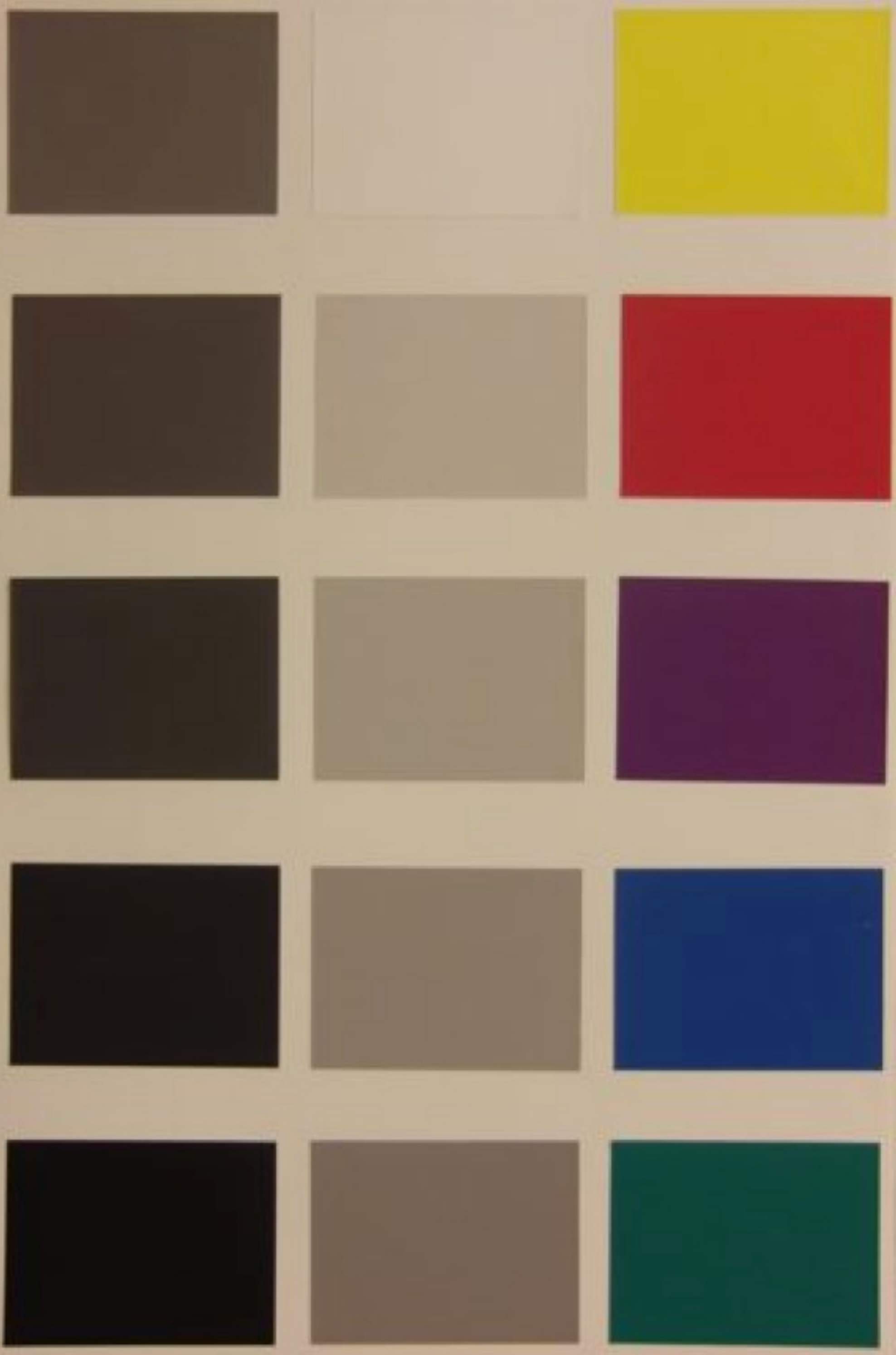}
\caption{The chart used in our ground truth reflectance experiments (Sec.~\ref{sec:eval_intrinsic}).}
\label{fig:colorchecker}
\end{minipage}
\end{figure}

\subsection{Modeling materials}
\label{sec:details:materials}
We assign a material to all estimated geometry based on the albedo estimated during intrinsic image decomposition (Sec~\ref{sec:lights}). We project the estimated albedo along the camera's view vector onto the estimated geometry, and render the objects with a diffuse texture corresponding to projected albedo. This projection applies also to out-of-view geometry (such as the wall behind the camera, or any other hidden geometry). Although unrealistic, this scheme has proven effective for rendering non-diffuse objects (it is generally difficult to tell that out-of-view materials are incorrect; see Fig~\ref{fig:results2}).


\section{Ground truth evaluations}
\label{sec:Evaluation}
Here, we evaluate the physical accuracy of lighting estimates produced by our method as well as our intrinsic decomposition algorithm. We do not strive for physical accuracy (rather, human believability), but we feel that these studies may shed light on how physical accuracy corresponds to people's perception of a real (or synthetic) image. Our studies show that our lighting models are quite accurate, but as we show later in our user study, people are not very good at detecting physical inaccuracies in lighting. Our reflectance estimates are also shown to be more accurate than the color variant of Retinex, which is currently one of the best single-image diffuse reflectance estimators.

\subsection{Lighting evaluation}
\label{sec:eval_lighting}
We have collected a ground truth dataset in which the surface BRDF is known for an object (a white, diffuse ball) in each image.  Using our algorithm, we estimate the lighting for each scene and insert a synthetic sphere. Because we know the rough geometry of the scene, we can place the synthetic sphere at the same spatial location as the sphere in the ground truth image. 

\boldhead{Dataset} Our dataset contains 200 images from 20 indoor scenes illuminated under varying lighting conditions. We use an inflatable ball painted with flat white paint as the object with known BRDF, which was matched and verified using a Macbeth Color Checker. The ball is suspended by a pole that protrudes from the ground and can be positioned at varying heights (see Fig~\ref{fig:rig}).  The images were taken with a Casio EXILIM EX-FH100 using a linear camera response function ($\gamma = 1$).

\begin{figure}[htp]
\begin{center}
\includegraphics[width=80mm]{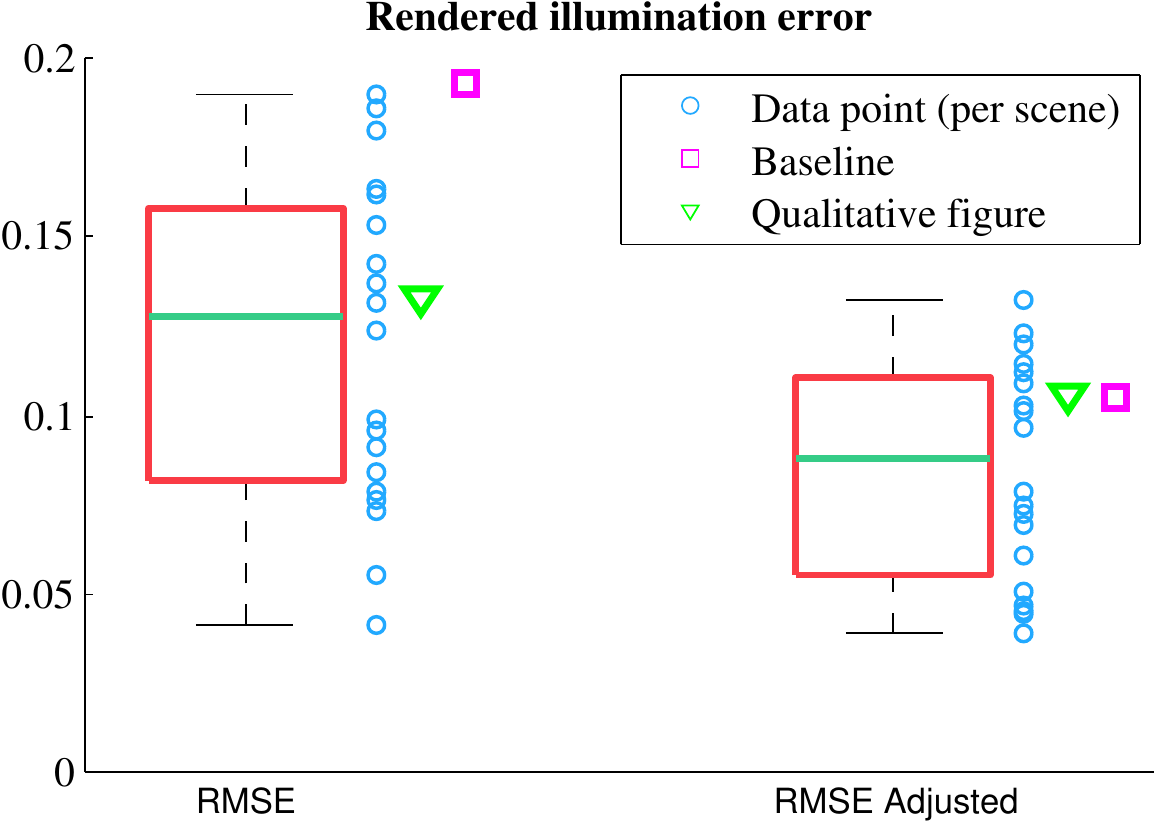}
\end{center}
\caption{We report results for both the root mean squared error (RMSE) and the RMSE after subtracting the mean intensity per sphere (RMSE adjusted). The RMSE metric illustrates how our method compares to the ground truth in an absolute metric, and the RMSE adjusted metric gives a sense of how accurate the lighting pattern is on each of the spheres (indicating whether light size/direction is correct).  For each metric, we show a box plot where the green horizontal line is the median, and the red box extends to the 25th and 75th percentiles. The averaged RMSE per scene (10 spheres are in each scene) is shown as a blue circle. A baseline (purple square) was computed by rendering all spheres with uniform intensity, and set to be the mean intensity of all images in the dataset. The green triangle indicates the error for the qualitative illustration in Fig~\ref{fig:errorvis}. No outliers exist for either metric, and image intensities range from [0,1].
}
\label{fig:errorgraph}
\end{figure}

\begin{figure}[htp]
\centerline{\includegraphics[width=75mm]{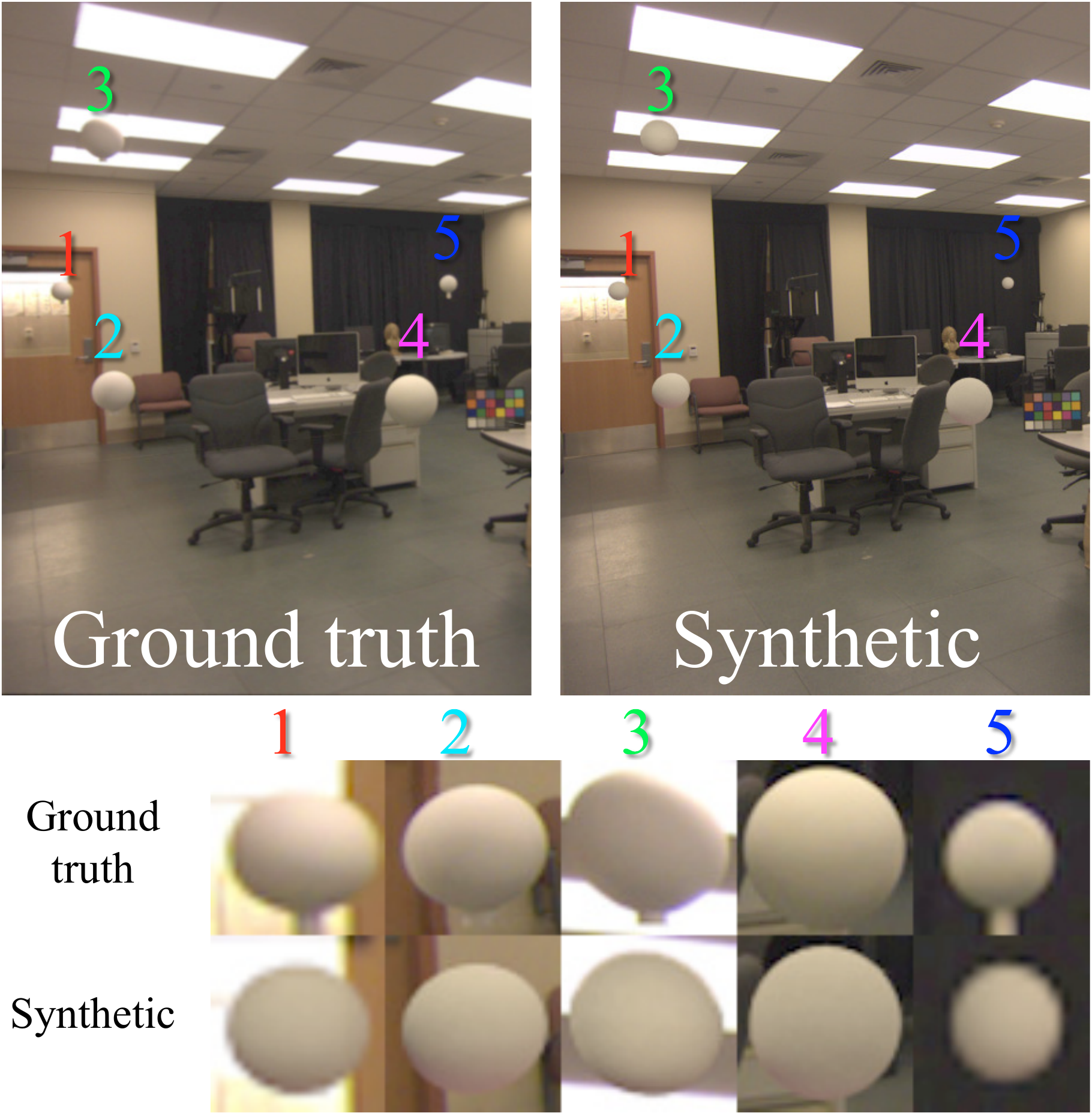} }
\caption{Qualitative comparison of our lighting algorithm to ground truth lighting. The top left image shows a scene containing five real spheres with authentic lighting (poles edited out for visual compactness). We estimate illumination using our algorithm and render the spheres into the scene at the same spatial locations \emph{(top right)}.  The bottom image matrix shows close-up views of the ground truth and rendered spheres.
See Fig~\ref{fig:errorgraph} for quantitative results.
}
\label{fig:errorvis}
\end{figure}

\boldhead{Results} For a pair of corresponding ground truth and rendered images, we measure the error by computing the pixel-wise difference of all pixels that have known BRDF. We measure this error for each image in the dataset, and report the root mean squared error (RMSE).  Overall, we found the RMSE to be $0.12 \pm 0.049$ for images with an intensity range of $[0,1]$. For comparing lighting patterns on the spheres, we also computed the error after subtracting the mean intensity (per sphere) from each sphere. We found that this error to be $0.085\pm 0.03$. Figure~\ref{fig:errorgraph} shows the RMSE for the entire dataset, as well as the RMSE after subtracting the mean intensity (RMSE adjusted), and a baseline for each metric (comparing against a set of uniformly lit spheres with intensity set as the mean of all dataset images). Our method beats the baseline for every example in the RMSE metric, suggesting decent absolute intensity estimates, and about 70\% of our renders beat the adjusted RMSE baseline. A qualitative visualization for five spheres in one scene from the dataset is also displayed in Figure~\ref{fig:errorvis}. In general, baseline renders are not visually pleasing but still do not have tremendous error, suggesting qualitative comparisons may be more useful when evaluating lightness estimation schemes.

\subsection{Intrinsic decomposition evaluation}
\label{sec:eval_intrinsic}
We also collected a ground truth reflectance dataset to compare to the reflectance estimates obtained from our intrinsic decomposition algorithm. We place a chart with known diffuse reflectances (ranging from dark to bright) in each scene, and measure the error in reflectance obtained by our method as well as Retinex. We show that our method achieves more accurate absolute reflectance than Retinex in nearly every scene in the dataset.

\boldhead{Dataset} Our reflectance dataset contains 80 images from different indoor scenes containing our ground truth reflectance chart (shown in Fig~\ref{fig:colorchecker}). We created the chart using 15 Color-aid papers; 10 of which are monochrome patches varying between 3\% reflectance (very dark) and 89\% reflectance (very bright). Reflectances were provided by the manufacturer. Each image in the dataset was captured by with the same camera and response as in Sec.~\ref{sec:eval_lighting}.

\begin{figure}[htp!]
\begin{center}
\includegraphics[width=77mm]{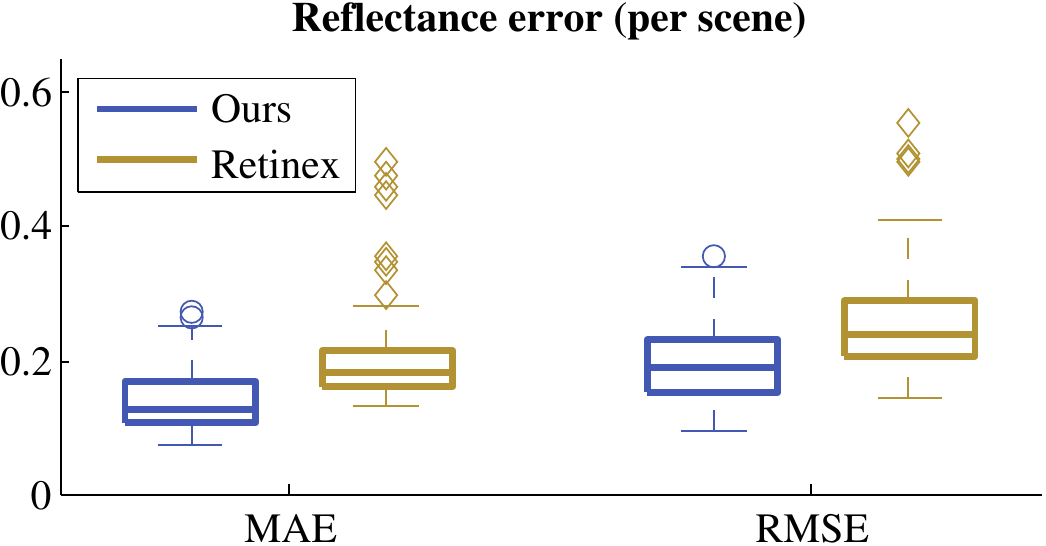}
\end{center}
\setlength{\unitlength}{1mm}
\begin{picture}(0,0)
\color{dblue}
\put(14.5, 35.65){\circle{2}}
\color{dyellow}
\put(13.5, 31.1){$\Diamond$}
\end{picture}
\vspace{-4mm}
\caption{Summary of the reflectance evaluation. Errors are measured per scene using a ground truth reflectance chart and reported in MAE and RMSE.  For each method and metric, a box plot is shown where the center horizontal line indicates the median, and the box extends to the 25th and 75th percentiles. Results from our decomposition method are displayed in blue (outliers as circles); Retinex results are displayed in gold (outliers as diamonds).
}
\label{fig:intrinsiccomp}
\end{figure}

\boldhead{Results} Using our decomposition method described in Sec.~\ref{sec:lights}, we estimate the per-pixel reflectance of each scene in our dataset. We then compute the mean absolute error (MAE) and root mean squared error (RMSE) for each image over all pixels with known reflectance (i.e. only for the pixels inside monochromatic patches). For further comparison, we compute the same error measures using the color variant of Retinex (as described in Grosse \ea~\shortcite{grosse09intrinsic}) as another method for estimating reflectance. Figure~\ref{fig:intrinsiccomp} summarizes these results. Our decomposition method outperforms Retinex for almost a large majority of the scenes in the dataset, and when averaged over the entire dataset, our method produced an MAE and RMSE of .141 and .207 respectively, compared to Retinex's MAE of .205 and RMSE of .272. These results indicate that much improvement can be made to absolute reflectance estimates when the user supplies a small amount of rough geometry, and that our method may improve other user-aided decomposition techniques, such as the method of Carroll~\ea\shortcite{Carroll:sg11}.

\subsection{Physical accuracy of intermediate results}
From these studies, we conclude that our method achieves comparatively accurate illumination and reflection estimates. However, it is important to note that these estimates are heavily influenced by the rough estimates of scene geometry, and optimized to produce a perceptually plausible rendered image (with our method) rather than to achieve physical accuracy. Our method adjusts light positions so that the rendered scenes look most like the original image, and our reflectance estimates are guided by rough scene geometry. Thus, the physical accuracy of the light positions and reflectance bear little correlation on the fidelity of the final result.

\begin{figure}[htp]
\begin{center}
\includegraphics[width=38mm]{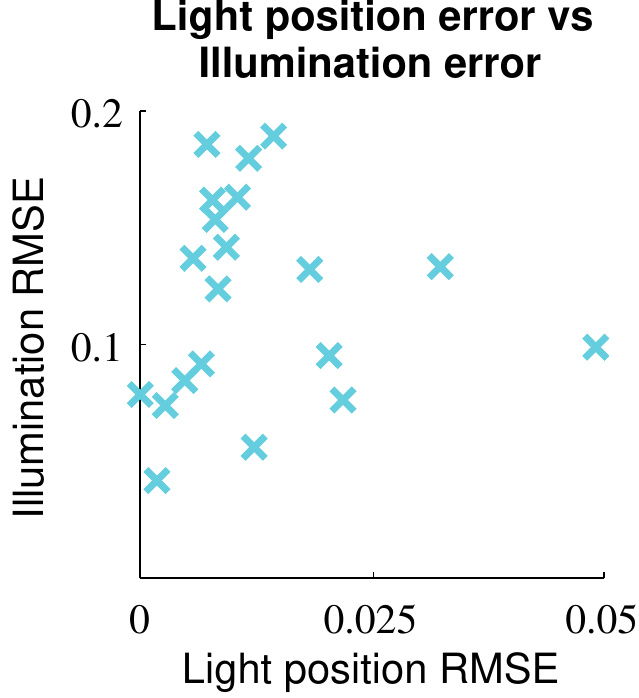}
\hfill
\includegraphics[width=38mm]{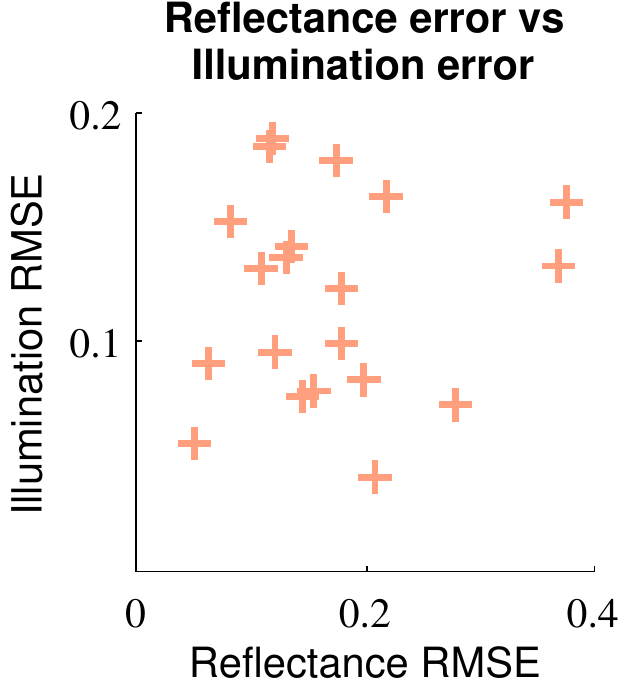}
\end{center}
\caption{The physical accuracy of our light position estimates as well as reflectance have little influence on the accuracy of illumination. This is likely because the light positions are optimized so that the rendered scene looks most like the original image, and the reflectance estimates are biased by our rough geometry estimates.
}
\label{fig:error_correlation}
\end{figure}

To verify this point, for each of the scenes in Sec~\ref{sec:eval_lighting}, we plotted the physical accuracy of our illumination estimates versus the physical accuracy of both our light position and reflectance estimates (Fig~\ref{fig:error_correlation}).  Light positions were marked by hand and a Macbeth ColorChecker was used for ground truth reflectance. We found that the overall Pearson correlation of illumination error and lighting position error was 0.034, and the correlation between illumination error and reflectance error was 0.074. These values and plots indicate a weak relation for both comparisons. Thus, our method is particularly good at achieving the final result, but this comes at the expense of physical inaccuracies along the way.

\begin{figure*}
\centerline{
\includegraphics[width=0.5\columnwidth]{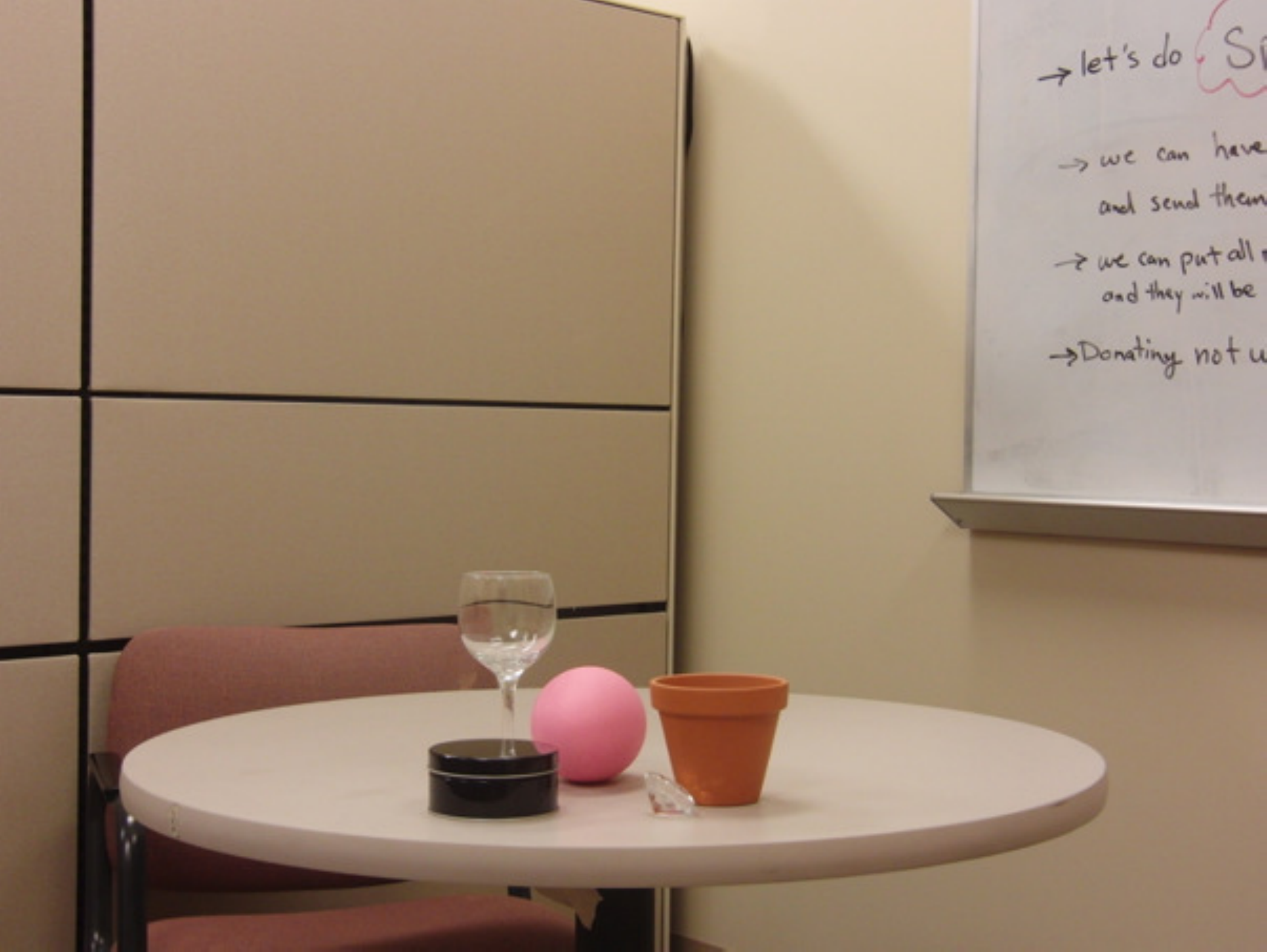}
\includegraphics[width=0.5\columnwidth]{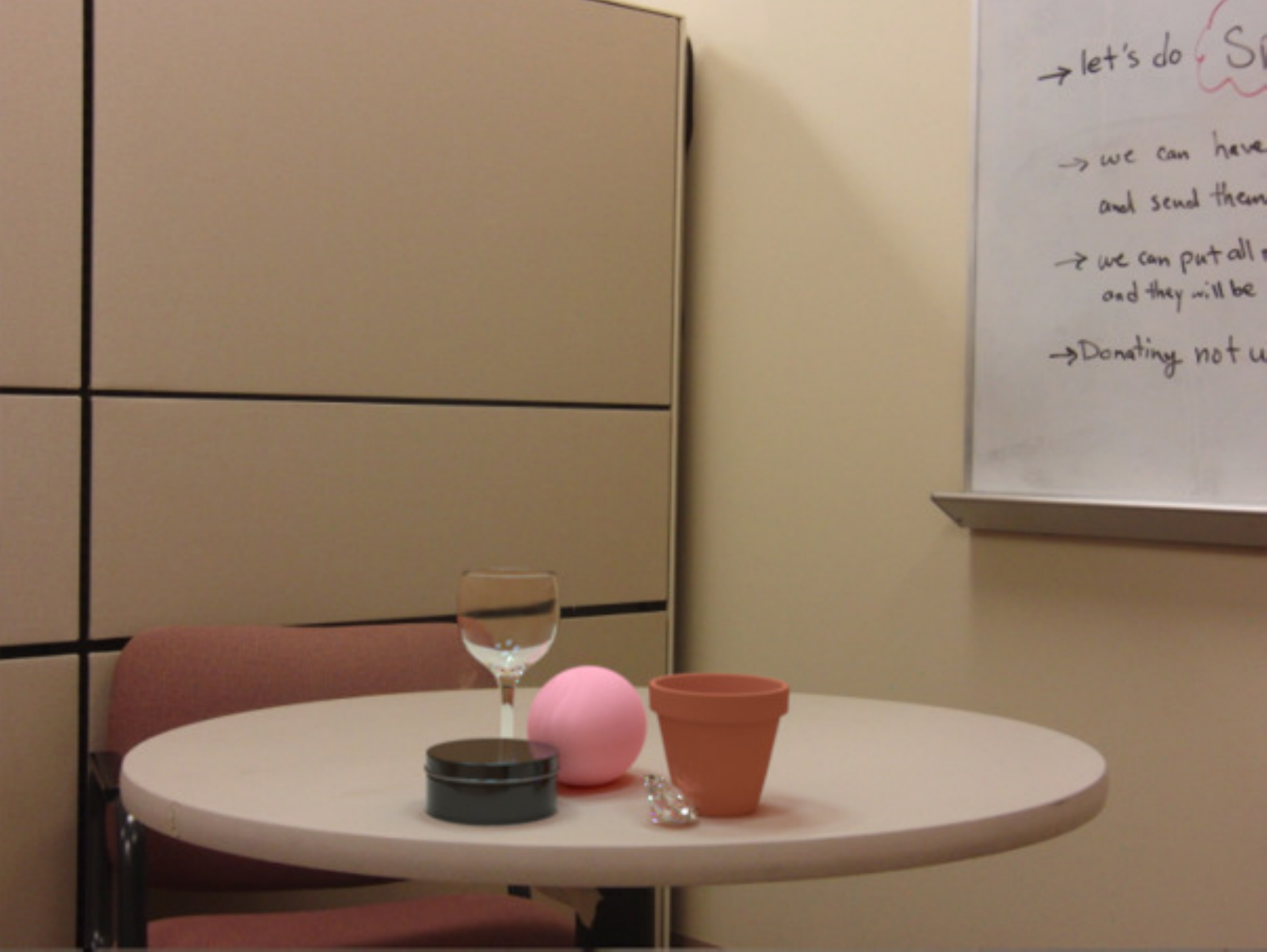}
\includegraphics[width=0.5\columnwidth]{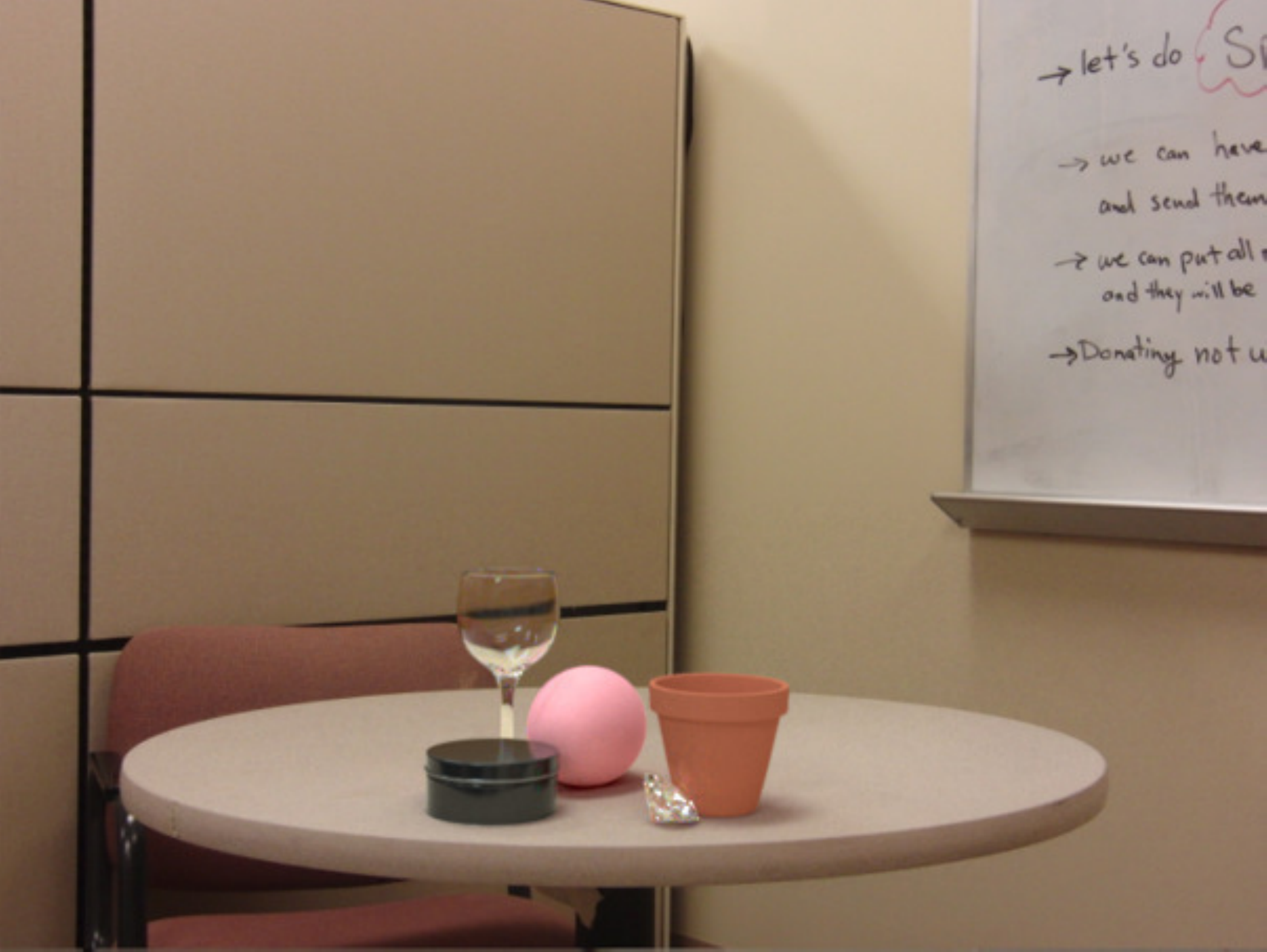}
\includegraphics[width=0.5\columnwidth]{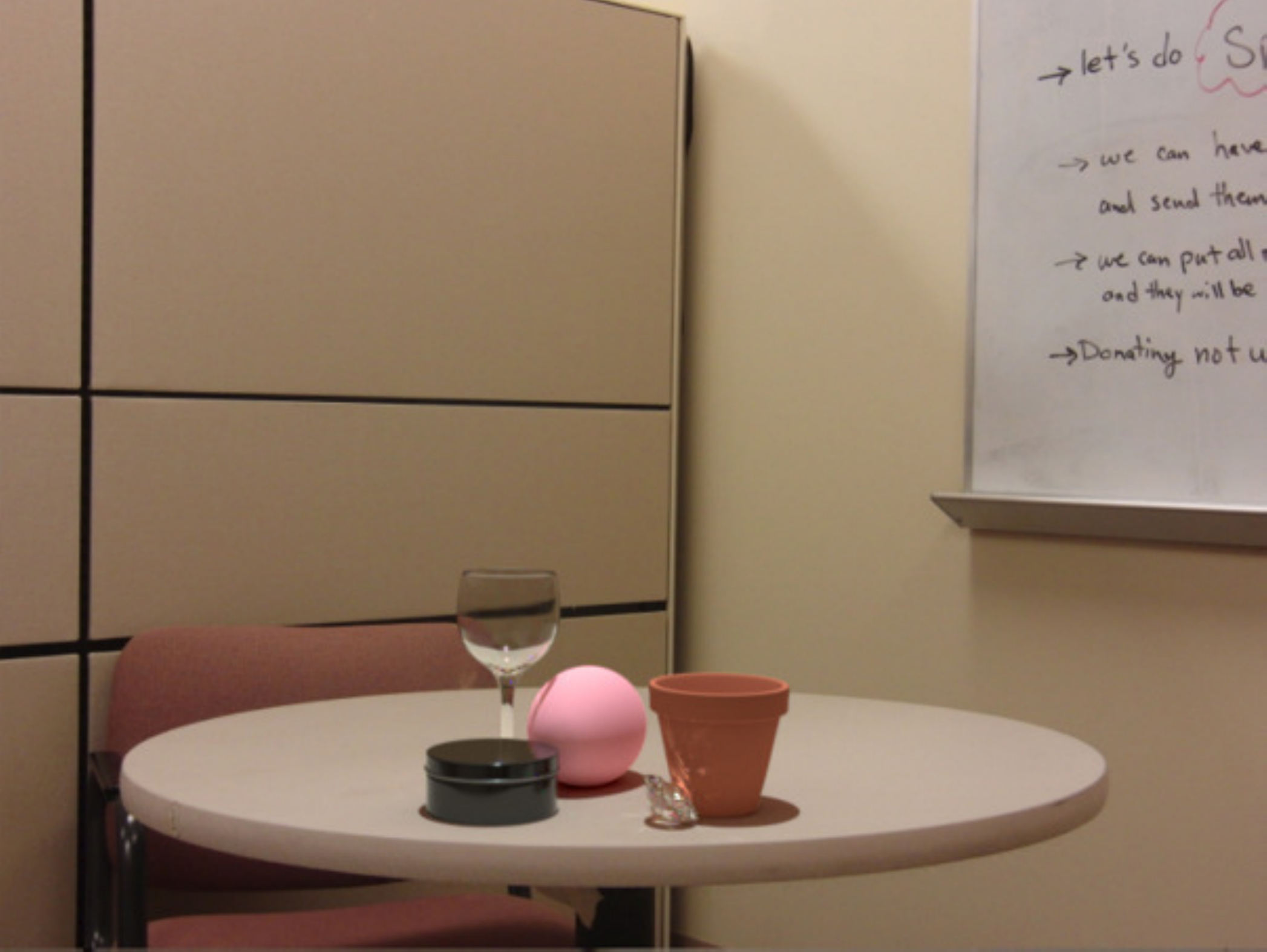} \vspace{-30mm}
}
\centerline{ \color{white} \Huge \hspace{17mm} Real \hspace{25mm} Ours \hspace{16mm}  Light probe  \hspace{10mm} Baseline \hspace{10mm}}
\vspace{23.5mm}
\centerline{
\includegraphics[width=0.5\columnwidth]{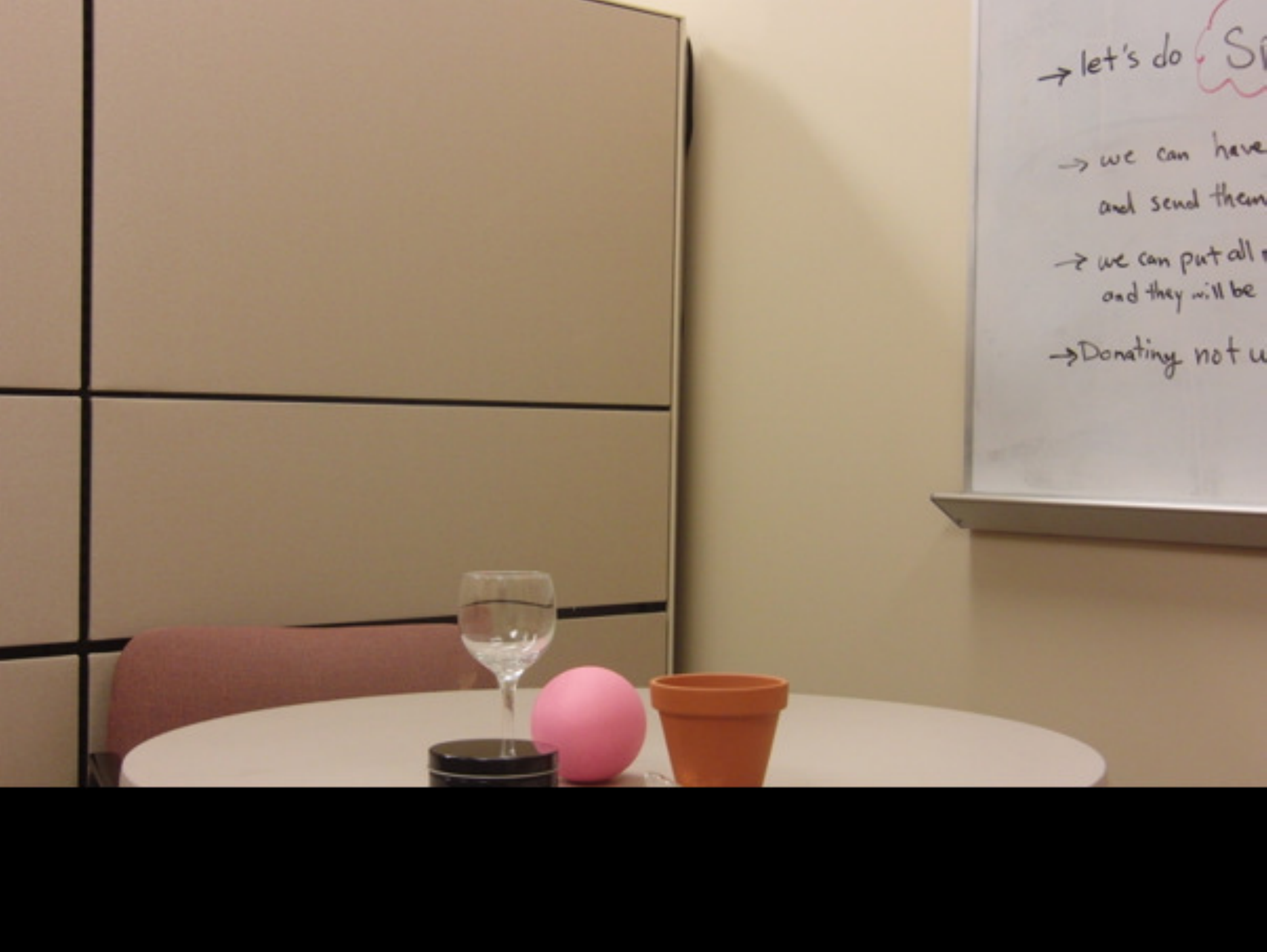}
\includegraphics[width=0.5\columnwidth]{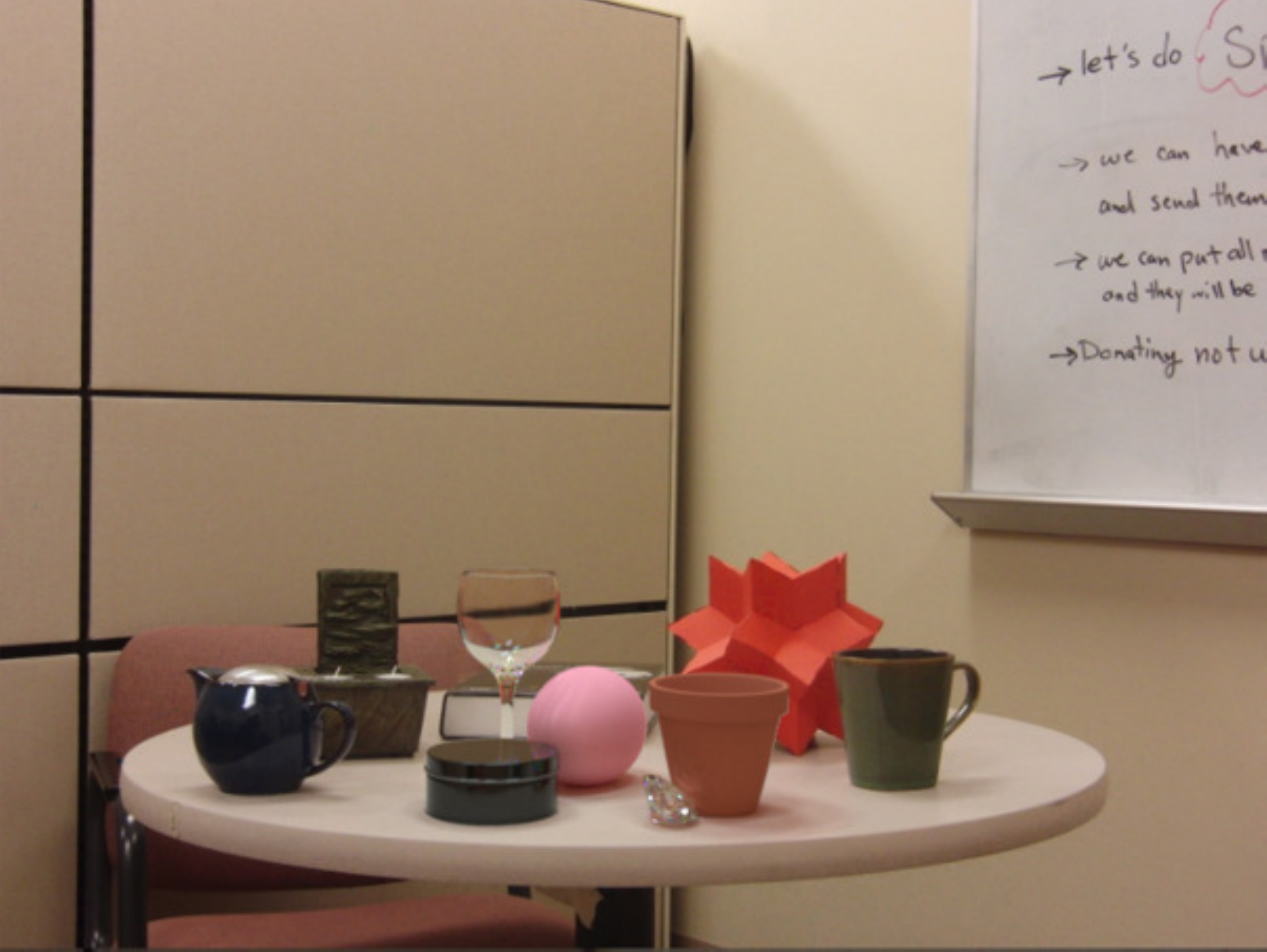}
\includegraphics[width=0.5\columnwidth]{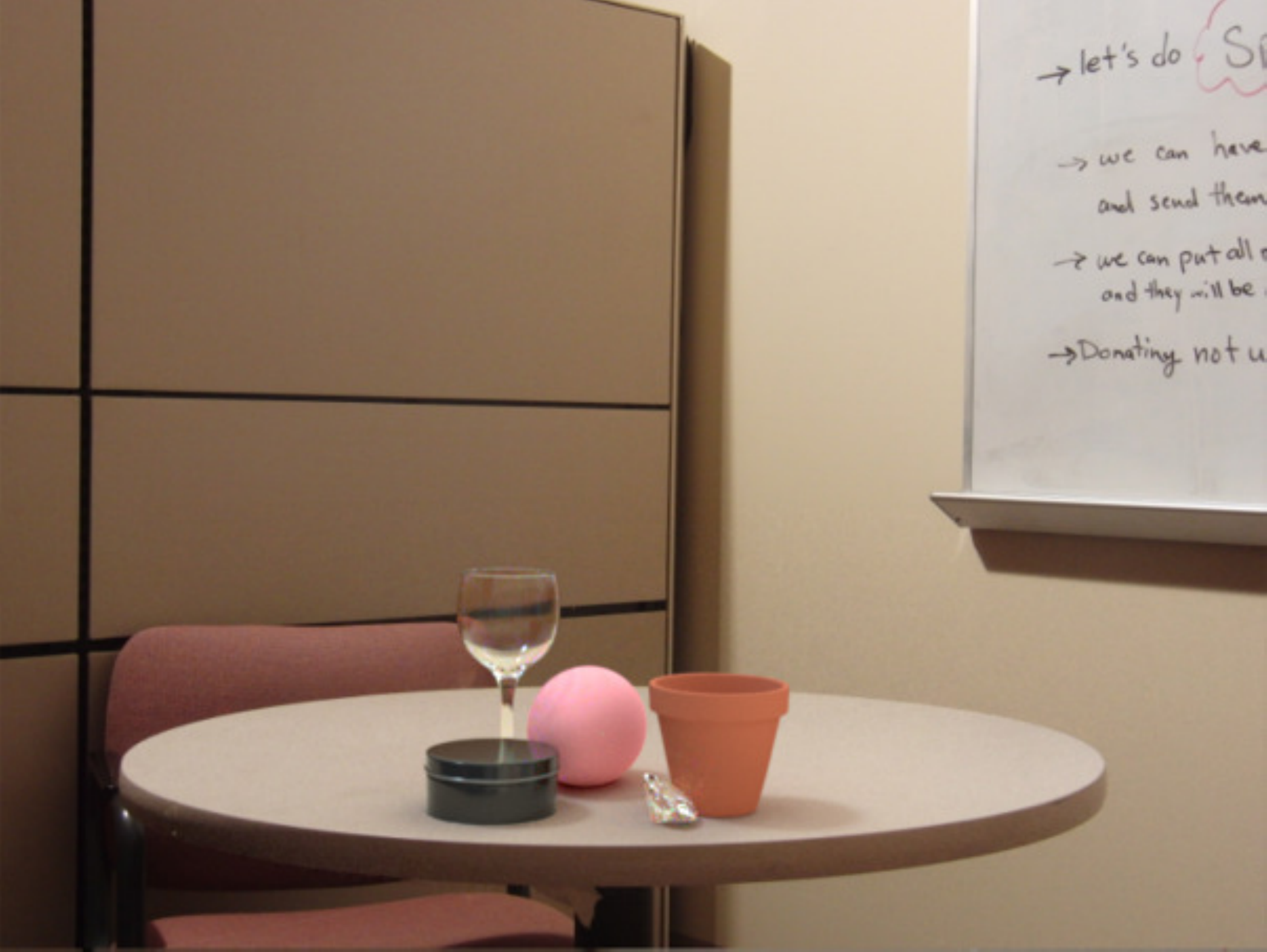}
\includegraphics[width=0.5\columnwidth]{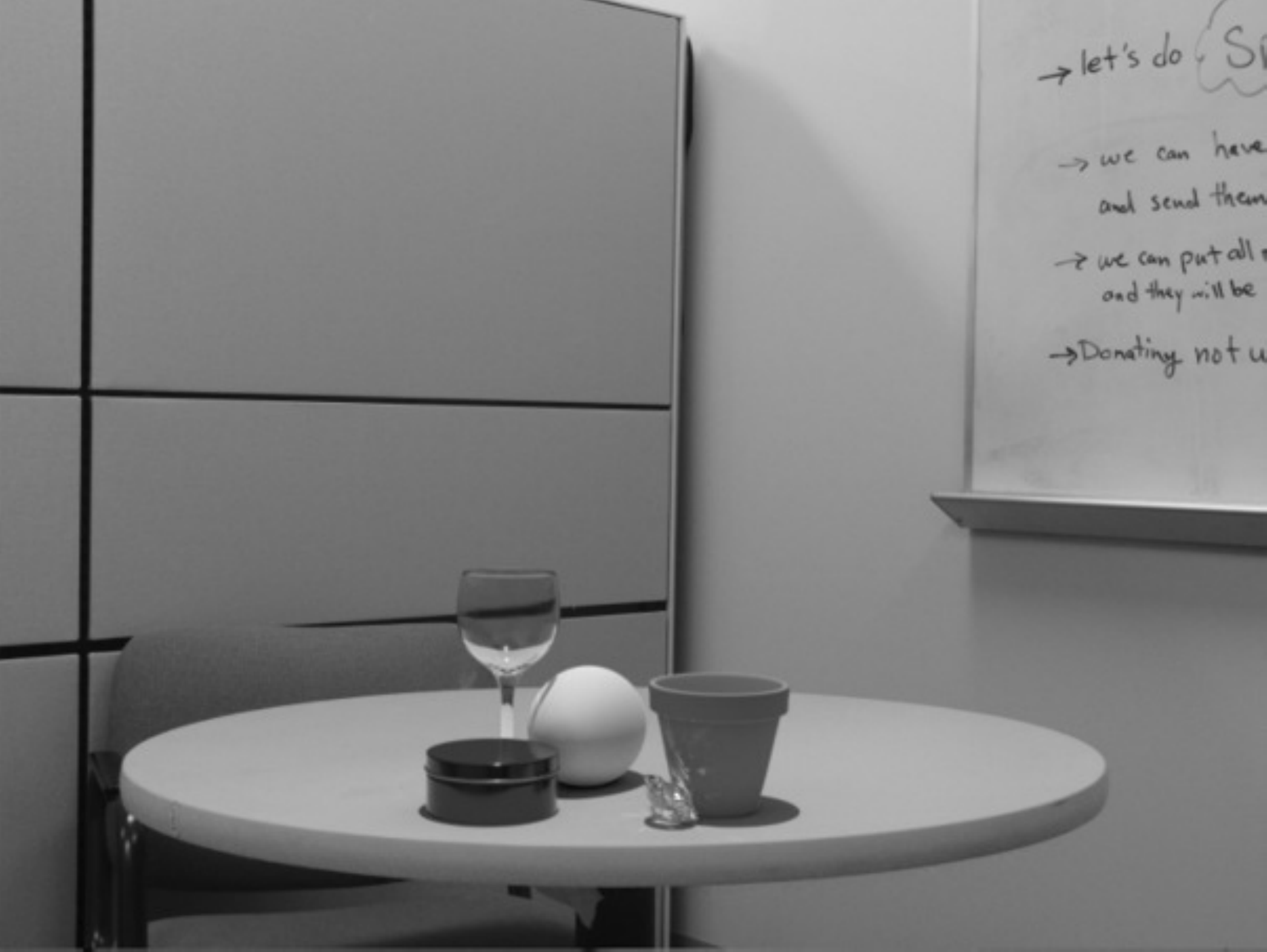}
\vspace{-30mm}
}
\centerline{ \color{white} \LARGE \hspace{10mm} Cropped \hspace{24mm} Clutter \hspace{24mm}  Spotlight  \hspace{17mm} Monochrome \hspace{5mm}}
\vspace{25.5mm}
\caption{Examples of methods and variants for Scene 10 in our user study. In the top row, from left to right, we show the real image, and synthetic images produced by our method, the light probe method, and the baseline method. In the bottom row, the four variants are shown.
}
\label{fig:study_overview}
\end{figure*}

\section{User study}
\label{sec:study}
We also devised a user study to measure how well users can differentiate between real images and synthetic images of the same scene under varying conditions. For the study, we show each participant a sequence of images of the same background scene containing various objects. Some of these images are photographs containing no synthetic objects, and other images contain synthetic objects inserted with one of three methods: our method, a variant of Debevec's light probe method\footnote{We use Debevec's method for estimating illumination through the use of a light probe, coupled with our estimates of geometry and reflectance.}~\shortcite{Debevecprobe}, or a baseline method (our method but with a simplified lighting model). Images are always shown in pairs and each of the paired images contain the exact same objects (although some of these objects may be synthetically inserted). The task presented to each user is a two-alternative forced choice test: for each pair of images, the user must choose (by mouse click) the image which they believe appears {\it most realistic}.

\boldhead{Methods} Our study tests three different methods for inserting synthetic objects into an image.  For the first, we use our method as described in Section~\ref{sec:technique}, which we will call {\bf ours}. We also compare to a method that uses Debevec's light probe method for estimating the illumination, combined with our coarse geometry and reflectance estimates, referred to as {\bf light probe}. To reproduce this method, we capture HDR photographs of a mirrored sphere in the scene from two angles (for removing artifacts/distortion), use these photographs to create a radiance map, model local geometry, and composite the rendered results~\cite{Debevecprobe}. Much more time was spent creating scenes with the light probe method than our own. The third method, denoted as {\bf baseline}, also uses our geometry and reflectances but places a single point light source near the center of the ceiling rather than using our method for estimating light sources. {\it Note that each of these methods use identical reflectance and geometry estimates; the only change is in illumination.}

\boldhead{Variants} We also test four different variations when presenting users with the images to determine whether certain image cues are more or less helpful in completing this task. These variants are {\bf monochrome} (an image pair is converted from RGB to luminance), {\bf cropped} (shadows and regions of surface contact are cropped out of the image), {\bf clutter} (real background objects are added to the scene), and {\bf spotlight} (a strongly directed out of scene light is used rather than diffuse ceiling light). Note that the spotlight variant requires a new lighting estimate using our method, and a new radiance map to be constructed using the light probe method; also, this variant is not applicable to the baseline method. If no variant is applied, we label its variant as {\bf none}.

\boldhead{Study details}
There are 10 total scenes that are used in the study. Each scene contains the same background objects (walls, table, chairs, etc) and has the same camera pose, but the geometry within the scene changes. We use five real objects with varying geometric and material complexity (shown in Fig~\ref{fig:study_overview}), and have recreated synthetic versions of these objects with 3D modeling software. The 10 different scenes correspond to unique combinations and placements of these objects. Each method was rendered using the same software (LuxRender), and the inserted synthetic geometry/materials remained constant for each scene and method. The rendered images were tone mapped with a linear kernel, but the exposure and gamma values differed per method. Tone mapping was performed so that the set of all scenes across a particular method looked most realistic (i.e. our preparation of images was biased towards realistic appearance for a skilled viewer, rather than physical estimates).

We recruited 30 subjects for this task. All subjects had a minimal graphics background, but a majority of the participants were computer scientists and/or graduate students. Each subject sees 24 pairs of images of identical scenes. 14 of these pairs contain one real and one synthetic image. Of these 14 synthetic images, five are created using our method, five are created using the light probe method, and the remaining four are created using the baseline method. Variants are applied to these pairs of images so that each user will see exactly one combination of each method and the applicable variants.  The other 10 pairs of images shown to the subject are all synthetic; one image is created using our method, and the other using the light probe method. No variants are applied to these images.

Users are told that their times are recorded, but no time limit is enforced. We ensure that all scenes, methods, and variants are presented in a randomly permuted order, and that the image placement (left or righthand side) is randomized. In addition to the primary task of choosing the most realistic image in the image pair, users are asked to rate their ability in performing this task both before and after the study using a scale of 1 (poor) to 5 (excellent).

\begin{figure}
\includegraphics[width=.95\columnwidth]{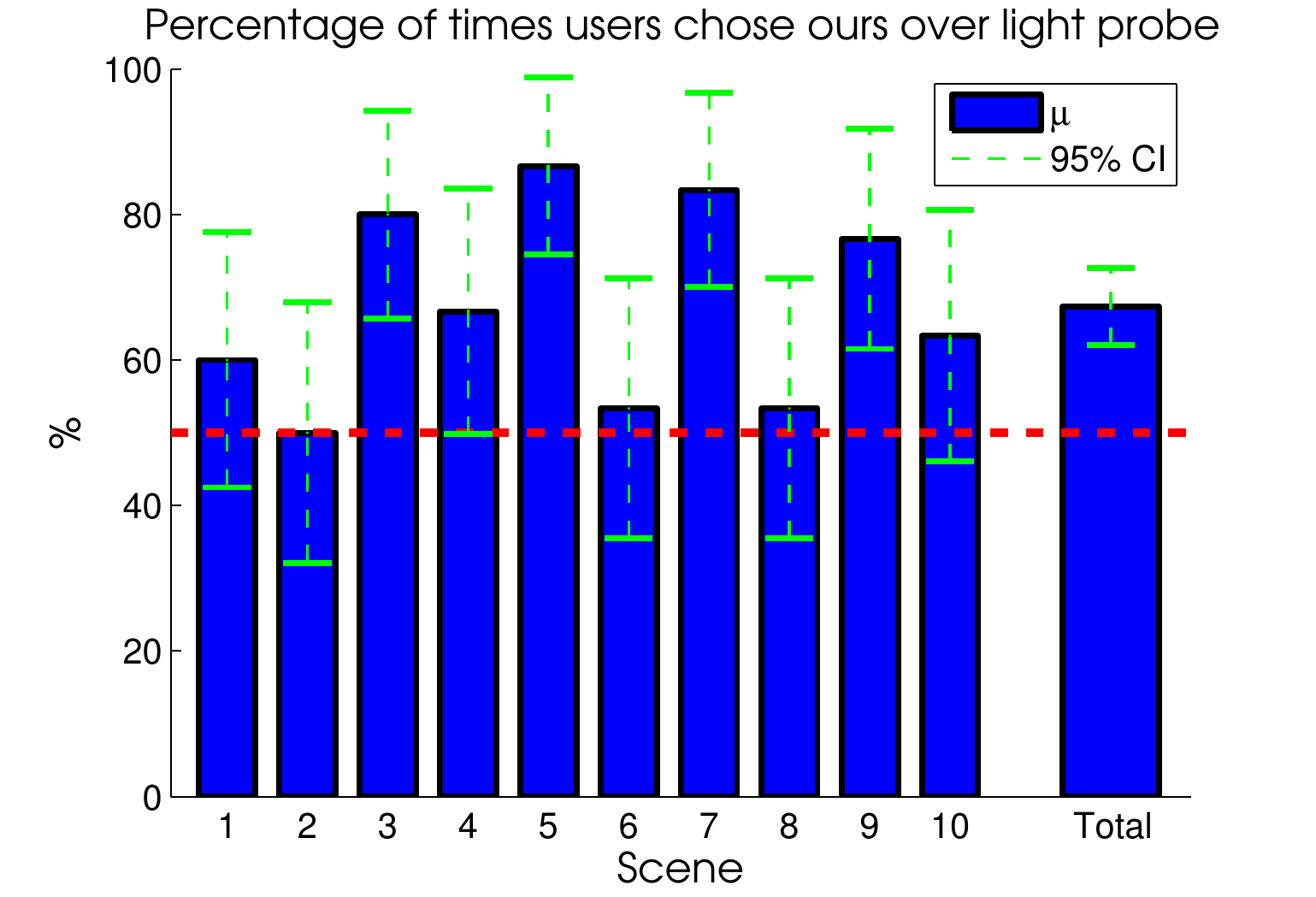}  \vspace{-3mm}  \\
\line(1,0){243} \vspace{1mm} \\
\vspace{-16mm}
\centerline{
\includegraphics[width=0.24\columnwidth]{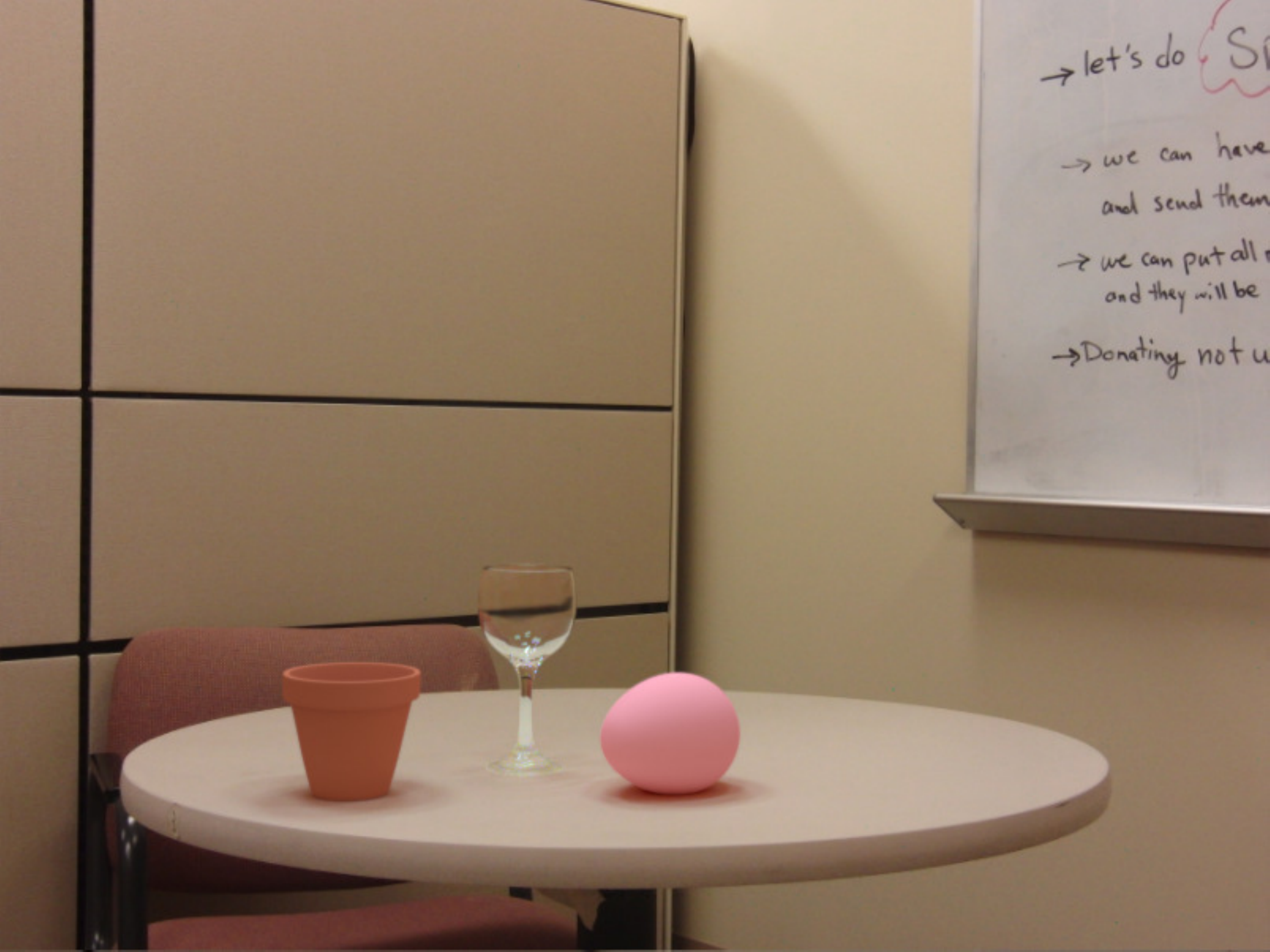}
\includegraphics[width=0.24\columnwidth]{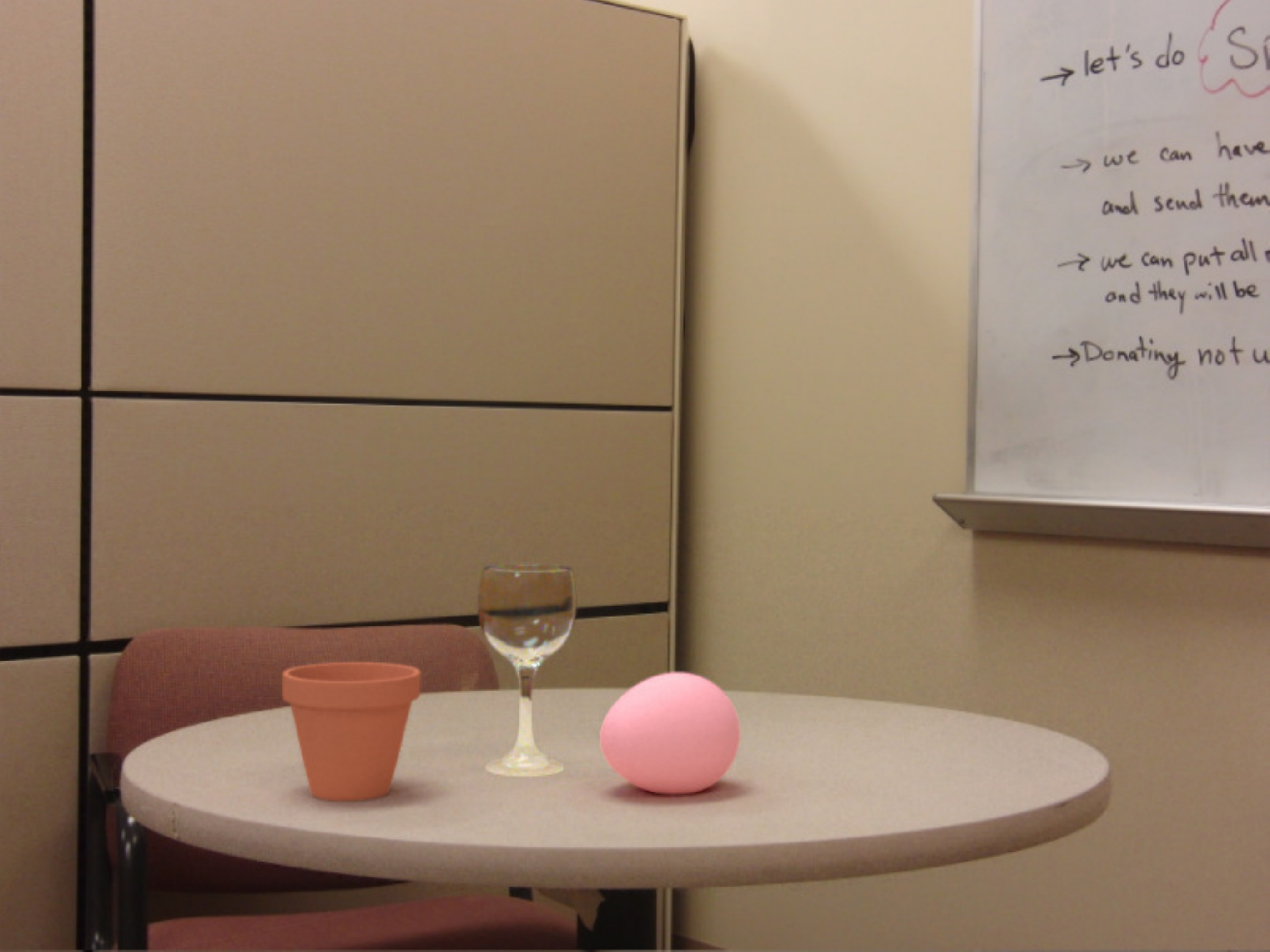}
\includegraphics[width=0.24\columnwidth]{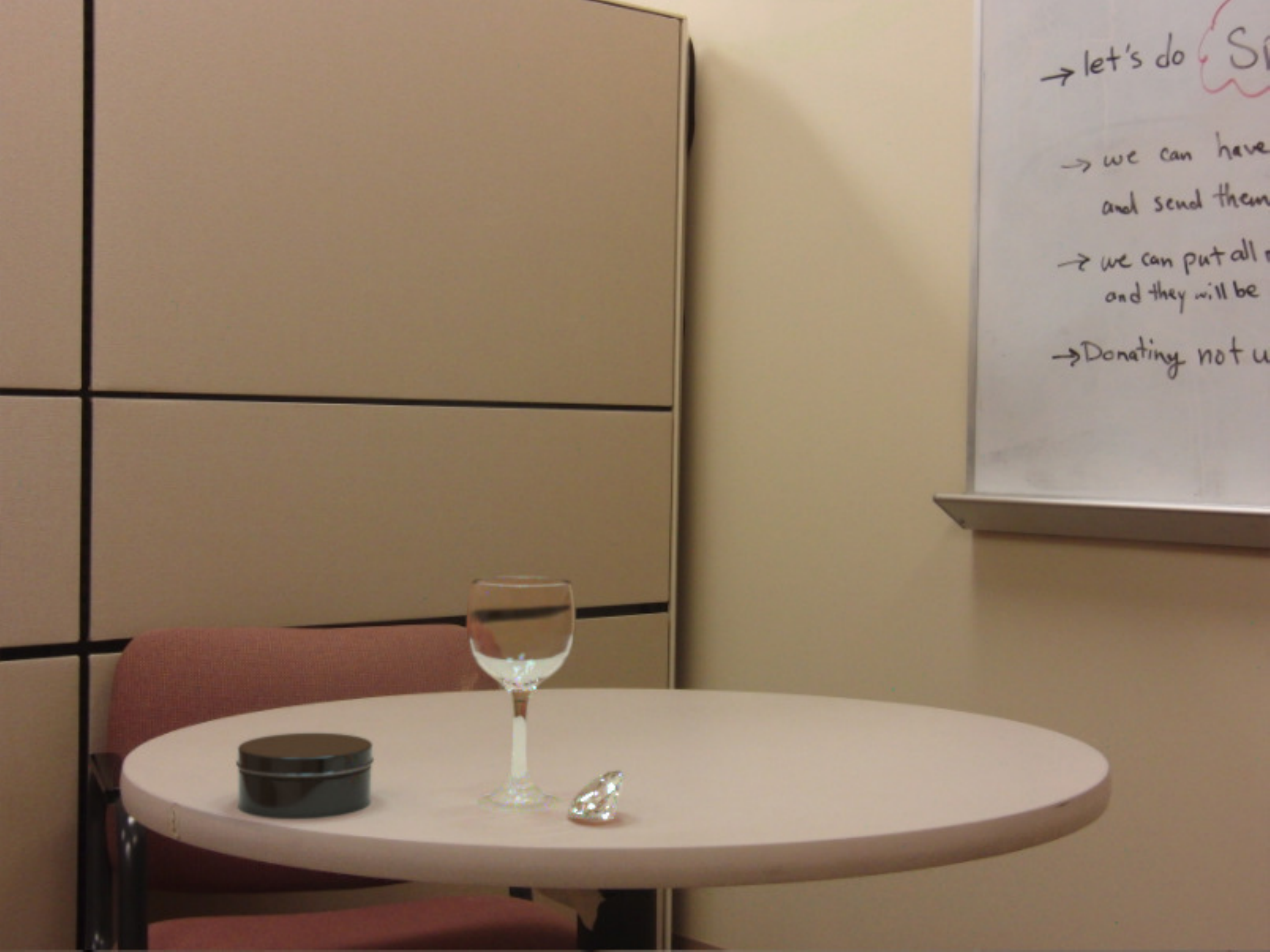}
\includegraphics[width=0.24\columnwidth]{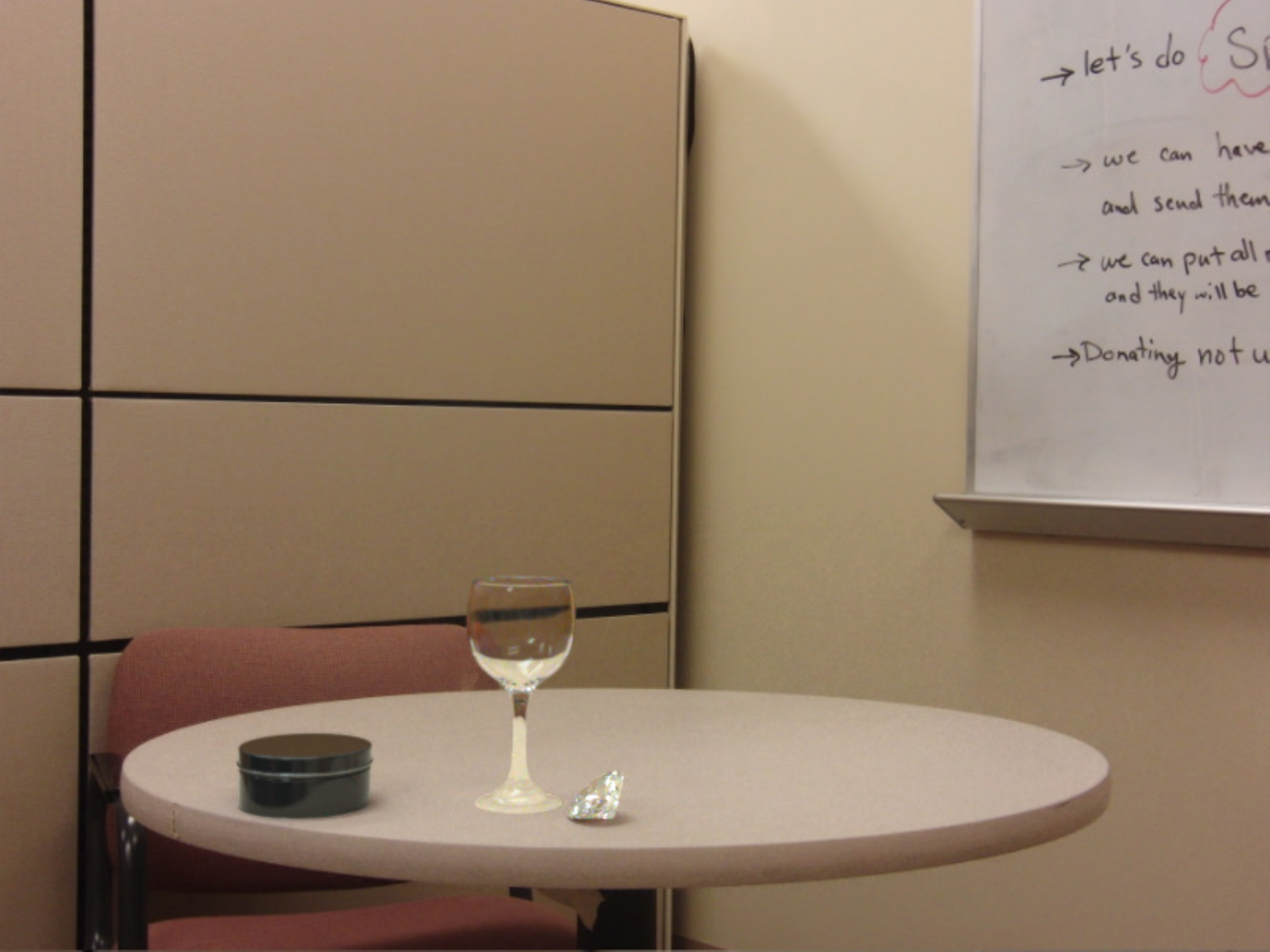}
}\\ \vspace{13mm}
\centerline{ \color{white} \small \hspace{7mm} Ours \hspace{12mm} Light probe \hspace{12mm}  Ours \hspace{12mm} Light probe \hfill}
\\
\centerline{ \hfill Scene 8 \hspace{30mm} Scene 9 \hfill }
\vspace{-2mm}
\caption{When asked to pick which method appeared more realistic, subjects chose our method over the light probe method at least 50\% of the time for each scene (67\% on average), indicating a statistically significant number of users preferred our method. The blue bars represent the mean response (30 responses per bar, 300 total), and the green lines represent the 95\% confidence interval. The horizontal red line indicates the 50\% line. The images below the graph show two scenes from the study that in total contain all objects.  Scene 8 was one of the lowest scoring scenes (53\%), while scene 9 was one of the highest scoring (77\%).
}
\label{fig:study_us_v_deb}
\end{figure}

\boldhead{Results}
We analyze the results of the different methods versus the real pictures separately from the results of our method compared to the light probe method. When describing our results, we denote $N$ as the sample size. When asked to choose which image appeared more realistic between our method and the light probe method, participants chose our image 67\% of the time (202 of 300).  Using a one-sample, one-tailed t-test, we found that users significantly preferred our method ($p\text{-value} \ll 0.001$), and on average users preferred our method more than the light probe method for all 10 scenes (see Fig~\ref{fig:study_us_v_deb}).

\begin{figure}[htp!]
\includegraphics[width=.95\columnwidth]{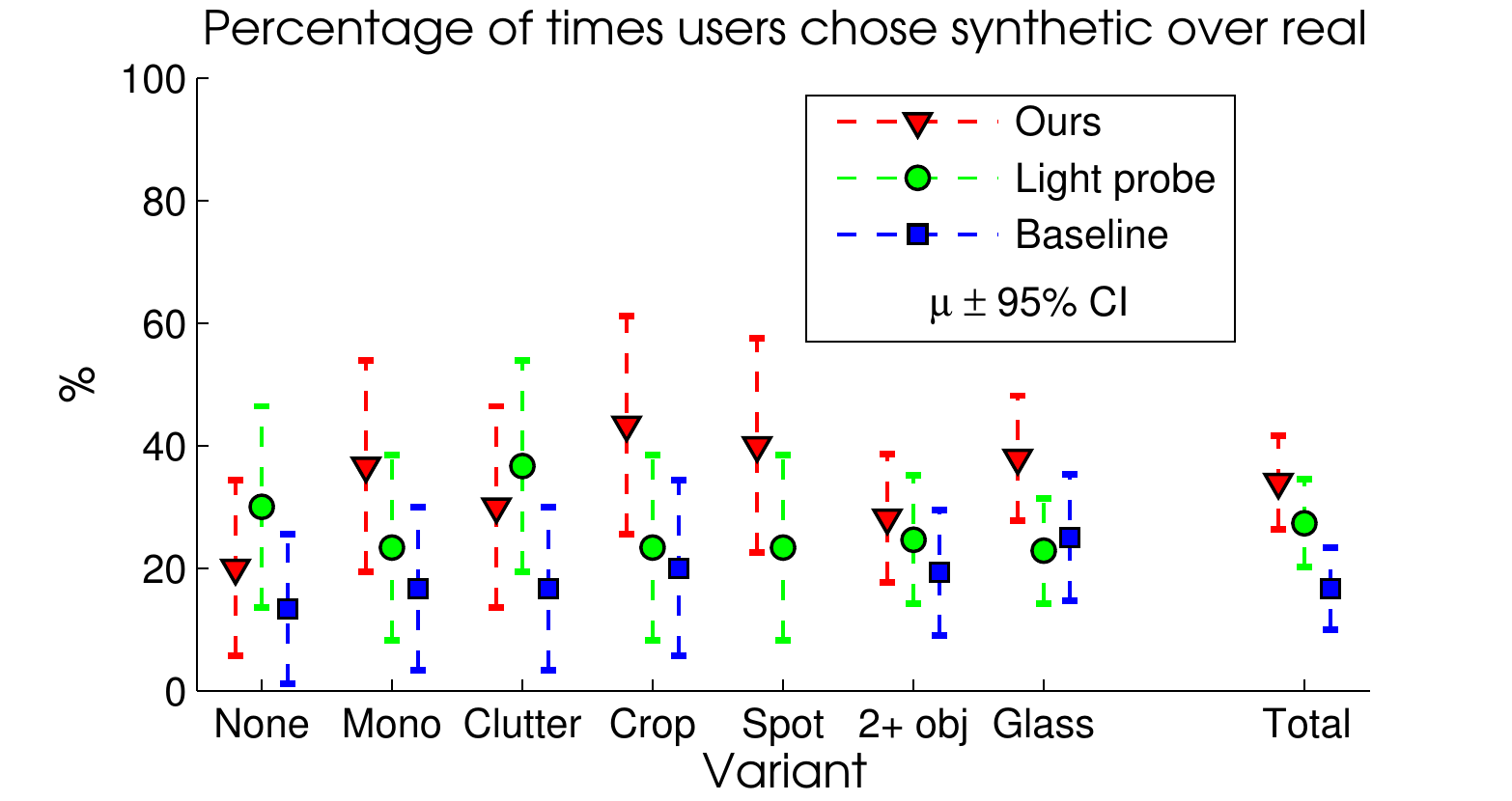}
\begin{center}
{\bf Percentage of times users chose synthetic over real}
\begin{tabular}{|c||c c c| c|} \hline
$N=30$ & ours & light probe & baseline & total \\ \hline
none   & 20 & 30 & 13.3 & 21.1  \\
monochrome & 36.7 & 23.3 & 16.7 & 26.6 \\
clutter & 30 & 36.7 & 16.7 & 27.8 \\
cropped & 43.3 & 23.3 & 20 & 28.9\\
spotlight &  40 & 23.3 & N/A & 31.7 \\ \hline
total & 34 & 27.3 & 16.7 & 26.7 \\ \hline
\end{tabular}
\end{center}
\vspace{1mm}
\begin{center}
\begin{tabular}{|c||c c c| c|} \hline
& ours & light probe & baseline & total \\ \hline
\hspace{1mm}  2+ objects \hspace{1mm}  & 28.2 & 24.6 & 19.3 & 24.4  \\
glass & 37.9 & 22.8 & 25 & 28.7 \\ \hline
\end{tabular}
\end{center}
\caption{Results for the three methods compared to a real image. In the graph, the mean response for each method is indicated by a triangle (ours), circle (light probe), and square (baseline). The vertical bars represent the 95\% binomial confidence interval. The tables indicate the average population response for each category. We also considered the effects of inserting multiple synthetic objects and synthetic objects made of glass, and these results were consistent with other variants. Both our method and the light probe method performed similarly, indicated especially by the overlapping confidence intervals, and both methods clearly outperform the baseline. Variants do appear to have a slight affect on human perception (making it harder to differentiate real from synthetic).
}
\label{fig:study_alldata}
\end{figure}

In the synthetic versus real comparison, we found overall that people incorrectly believe the synthetic photograph produced with our method is real 34\% of the time (51 of 150), 27\% of the time with the light probe method (41 of 150), and 17\% for the baseline (20 of 120). Using a two-sample, one-tailed t-test, we found that there was not a significant difference in subjects that chose our method over the light probe method ($p = 0.106$); however, there was a significant difference in subjects choosing our method over the baseline ($p = 0.001$), and in subjects choosing the light probe method over the baseline ($p=0.012$). For real versus synthetic comparisons, we also tested the variants as described above. All variants (aside from ``none'') made subjects perform worse overall in choosing the real photo, but these changes were not statistically significant. Figure~\ref{fig:study_alldata} summarizes these results.

We also surveyed four non-na\"{\i}ve users (graphics graduate students), whose results were not included in the above comparisons. Contrary to our assumption, their results {\it were} consistent with the other 30 na\"{\i}ve subjects. These four subjects selected 2, 3, 5, and 8 synthetic photographs (out of 14 real-synthetic pairs), an average of 35\%, which is actually higher than the general population average of 27\% (averaged over all methods/variants), indicating more trouble in selecting the real photo. In the comparison of our method to the light probe method, these users chose our method 5, 7, 7, and 8 times (out of 10 pairs) for an average of 68\%, consistent with the na\"{\i}ve subject average of 67\%.

\boldhead{Discussion}
From our study, we conclude that both our method and the light probe method are highly realistic, but that users can tell a real image apart from a synthetic image with probability higher than chance. However, even though users had no time restrictions, they still could not differentiate real images from both our method and the light probe method reliably. As expected, both of these synthetic methods outperform the baseline, but the baseline still did surprisingly well. Applying different variants to the pairs of images hindered subjects' ability to determine the real photograph, but this difference was not statistically significant.

When choosing between our method and the light probe method, subjects chose our method with equal or greater probability than the light probe method for each scene in the study. This trend was probably the result of our light probe method implementation, which used rough geometry and reflectance estimates produced by our algorithm, and was not performed by a visual effects or image-based lighting expert. {\it Had such an expert generated the renderings for the light probe method, the results for this method might have improved, and so led to a change in user preference for comparisons involving the light probe method. The important conclusion is that we can now achieve realistic insertions without access to the scene.}


Surprisingly, subjects tended to do a worse job identifying the real picture as the study progressed. We think that this may have been caused by people using a particular cue to guide their selection initially, but during the study decide that this cue is unreliable or incorrect, when in fact their initial intuition was accurate. If this is the case, it further demonstrates how realistic the synthetic scenes look as well as the inability of humans to pinpoint realistic cues.

\begin{figure}[htp]
\begin{center}
\includegraphics[width=0.49\columnwidth]{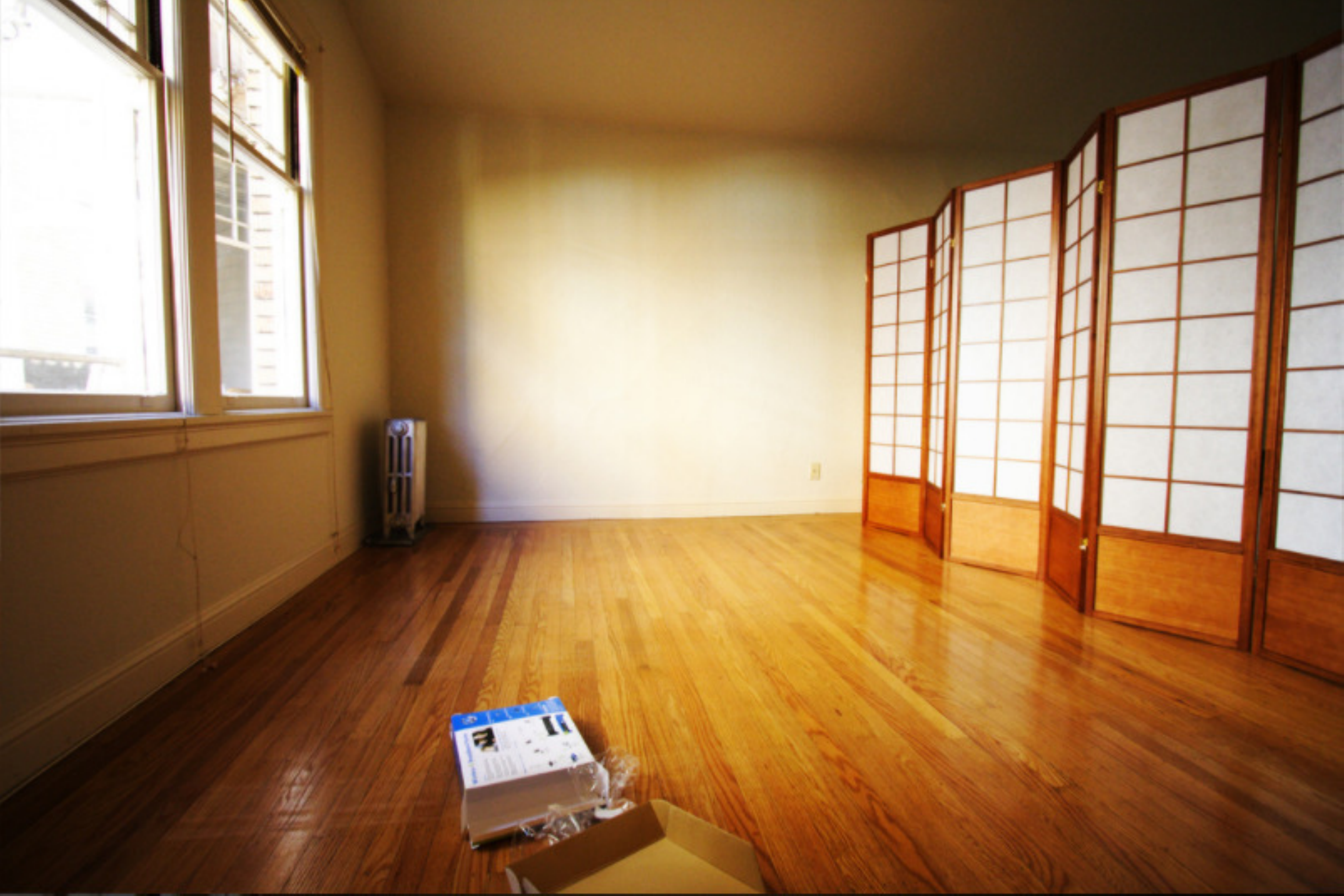}
\includegraphics[width=0.49\columnwidth]{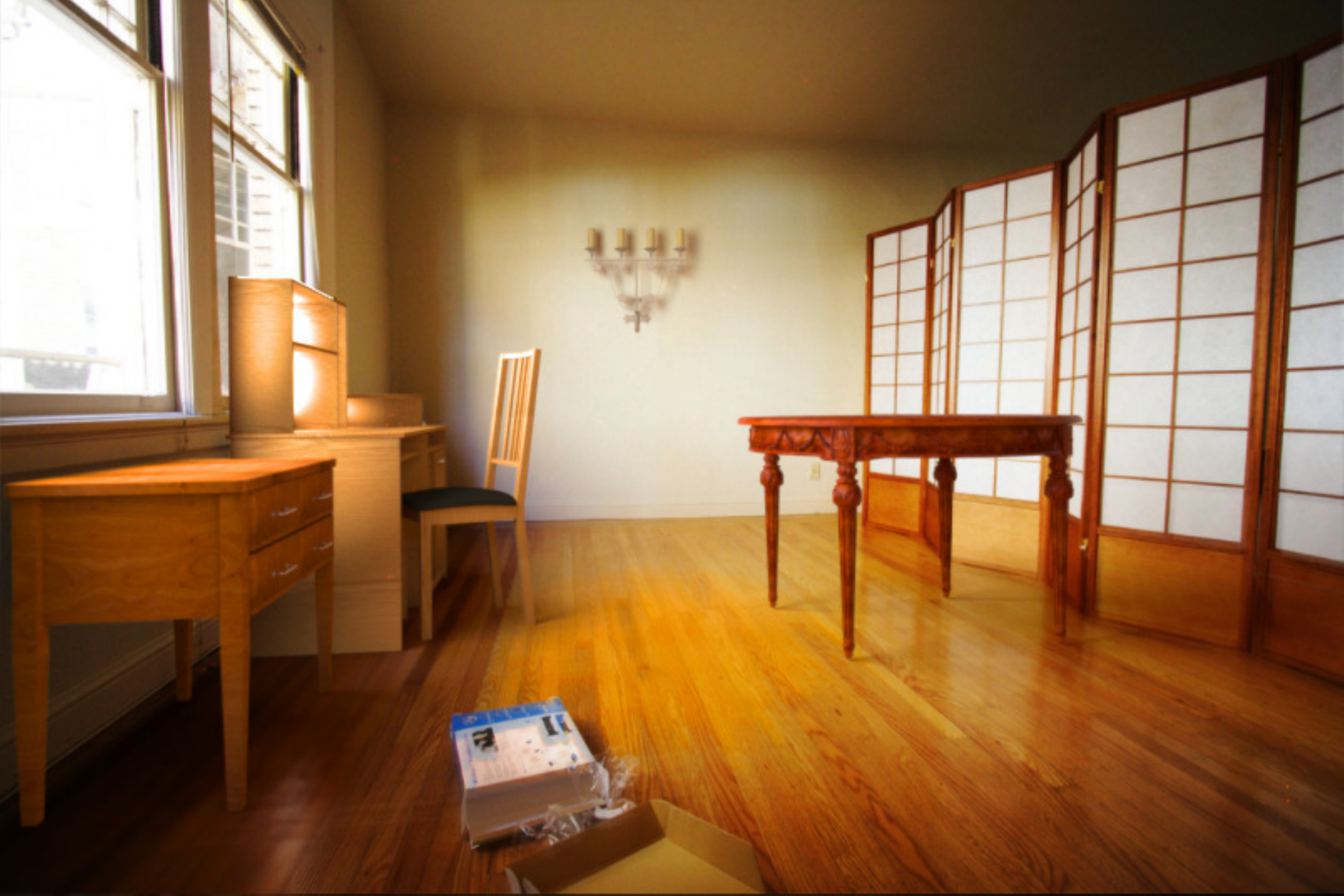}
\end{center}
\caption{Home redecorating is a natural application for our method. A user could take a picture of a room, and visualize new furniture or decorations without leaving home.
}
\label{fig:interiordecor}
\end{figure}

Many subjects commented that the task was more difficult than they thought it would be. Self assessment scores reflected these comments as self evaluations decreased for 25 of 30 subjects (i.e. a subject rated him/herself higher in the entry assessment than in the exit assessment), and in the other five subjects, the assessment remained the same. The average entry assessment was 3.9, compared to the average exit assessment of 2.8. No subject rated him/herself higher in the exit assessment than in the entry assessment.

The fact that na\"{\i}ve subjects scored comaparably to non-na\"{\i}ve subjects indicates that this test is difficult even for those familiar with computer graphics and synthetic renderings. All of these results indicate that people are not good at differentiating real from synthetic photographs, and that our method is state of the art.

\section{Results and discussion}
\begin{figure}[htp]
\includegraphics[width=0.49\columnwidth]{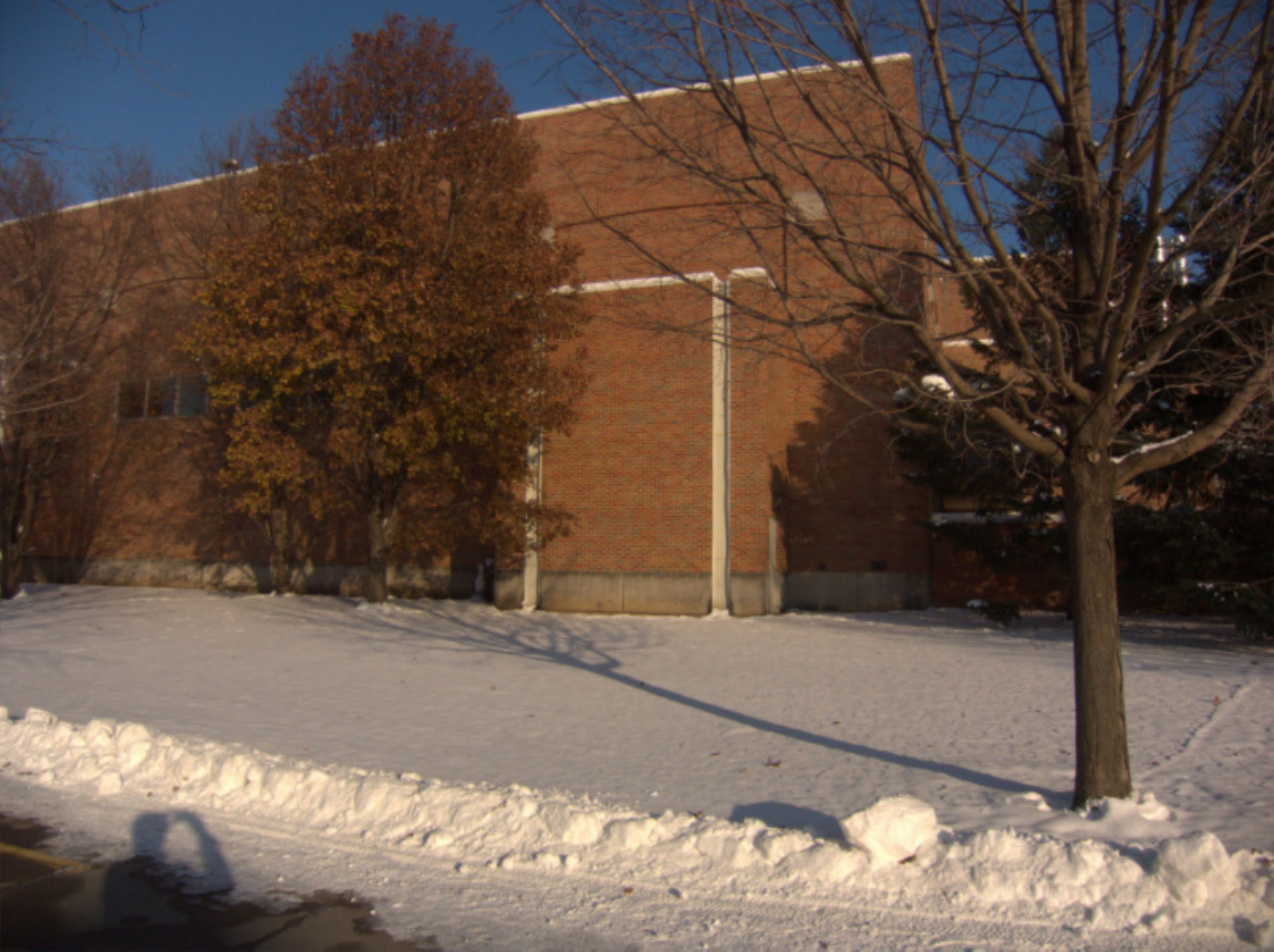}
\includegraphics[width=0.49\columnwidth]{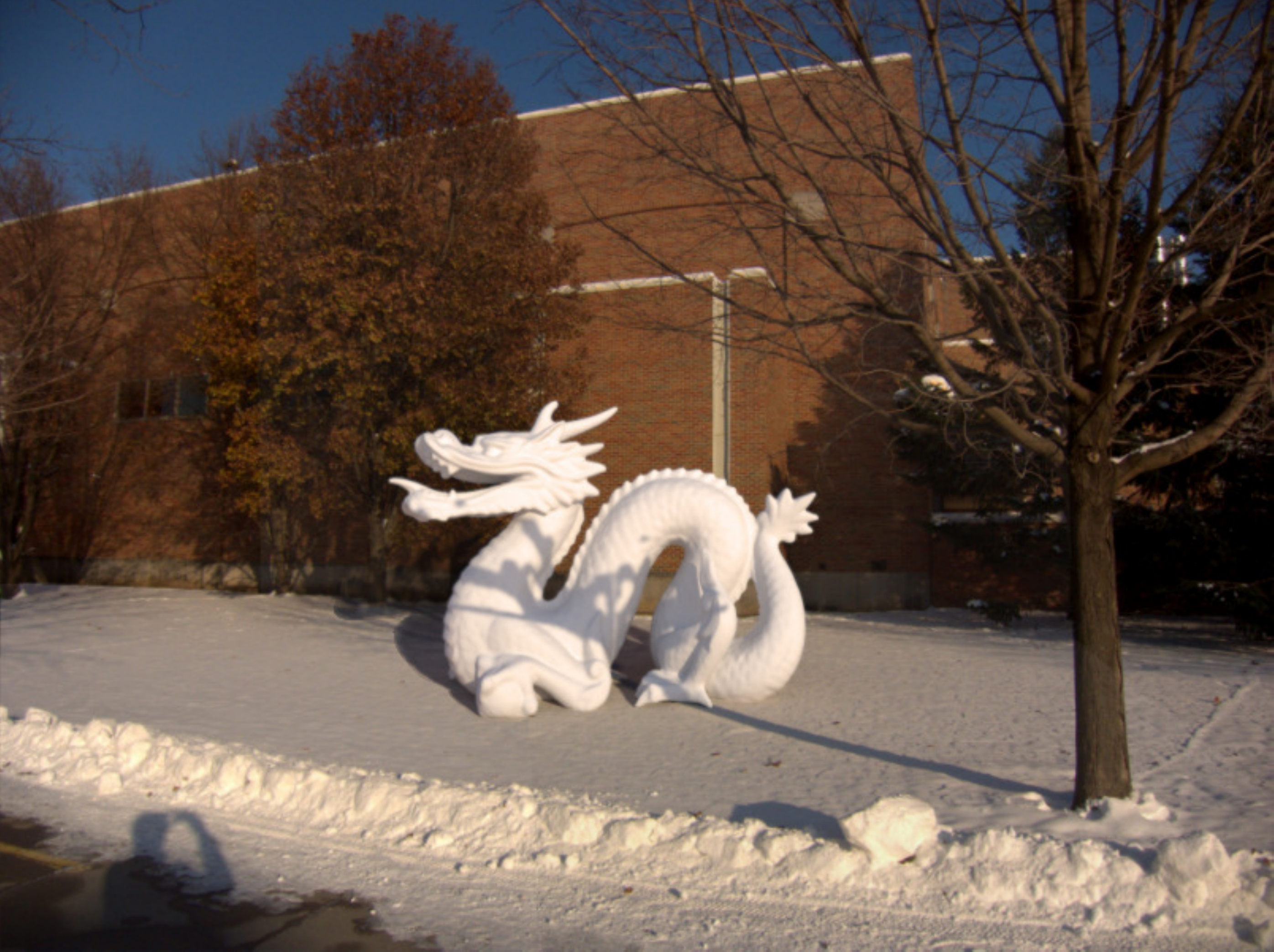}  \vspace{0.5mm} \\
\includegraphics[width=0.49\columnwidth]{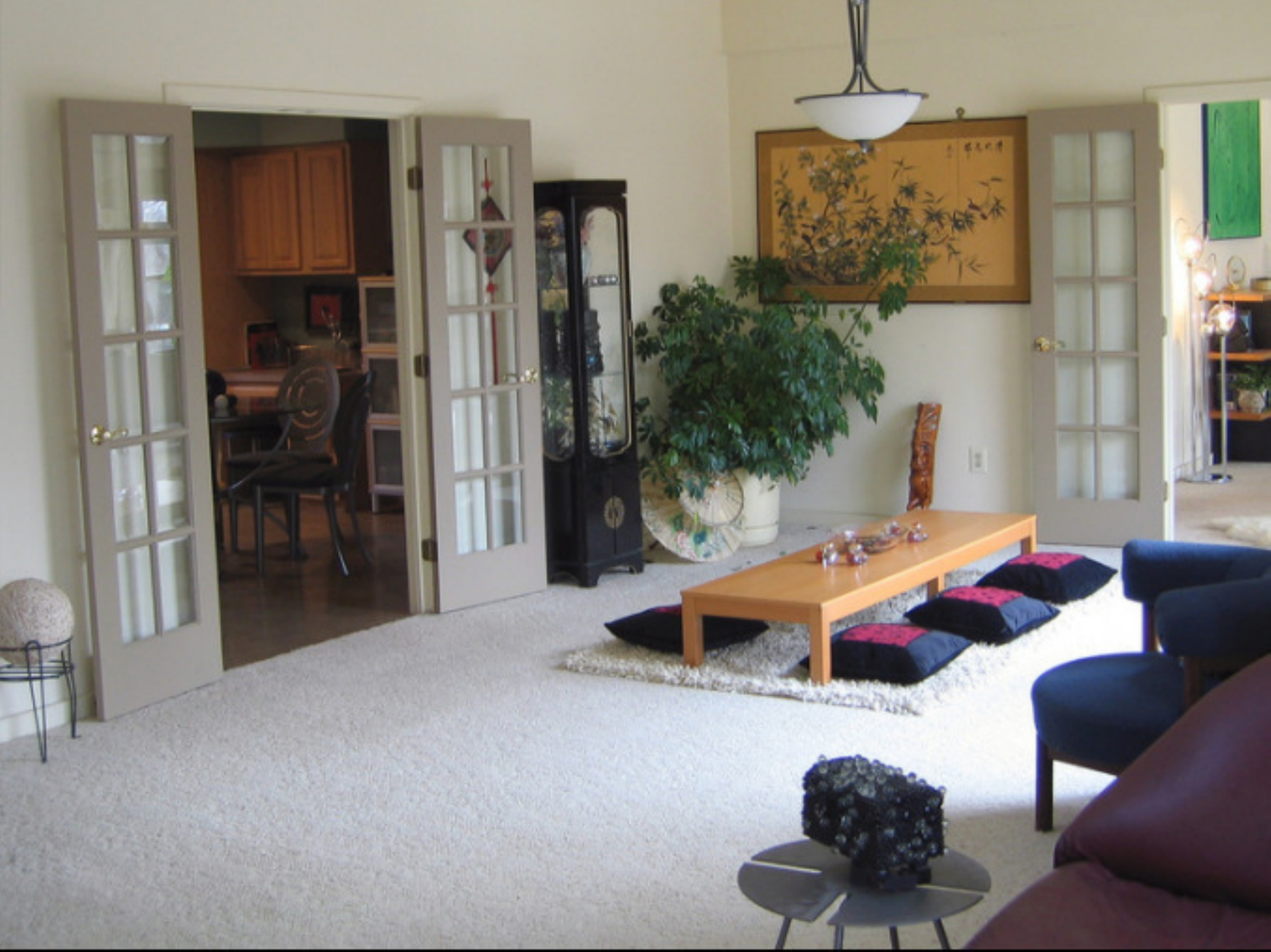}
\includegraphics[width=0.49\columnwidth]{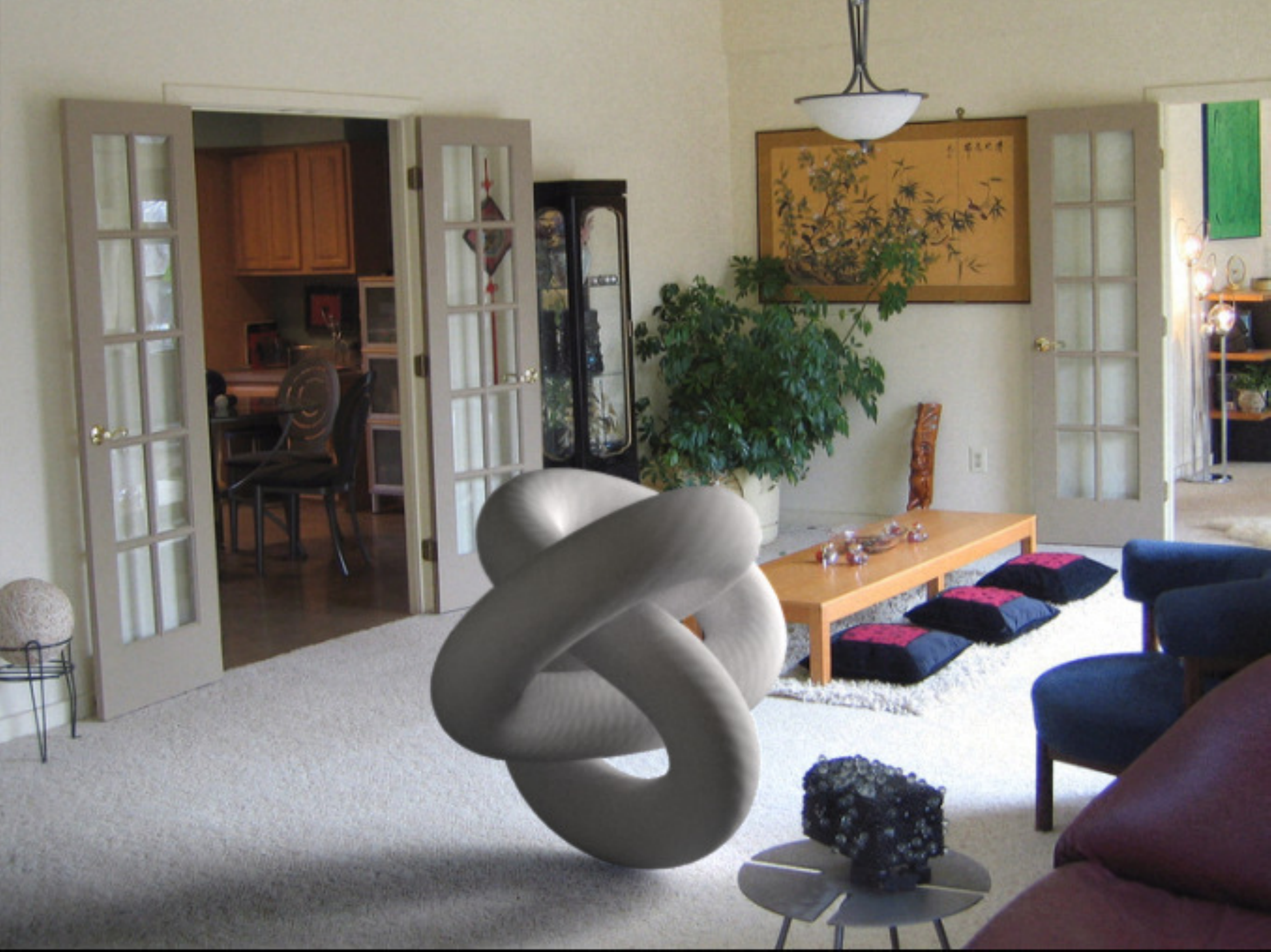}
\caption{Our algorithm can handle complex shadows \emph{(top)}, as well as out-of-view light sources \emph{(bottom)}.
\vspace{-2mm}
}
\label{fig:results1}
\end{figure}

\begin{figure}[htp]
\includegraphics[width=0.49\columnwidth]{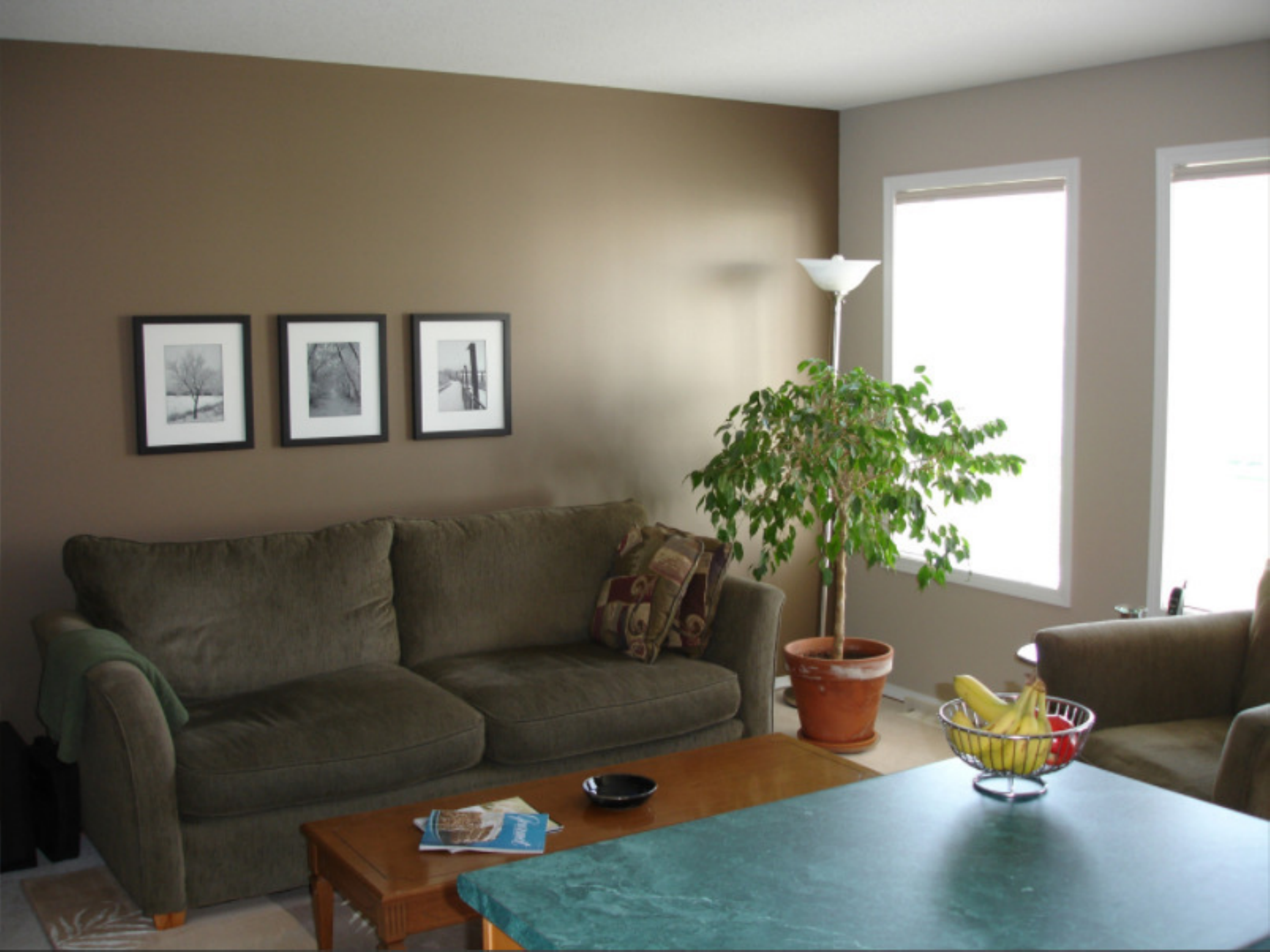}
\includegraphics[width=0.49\columnwidth]{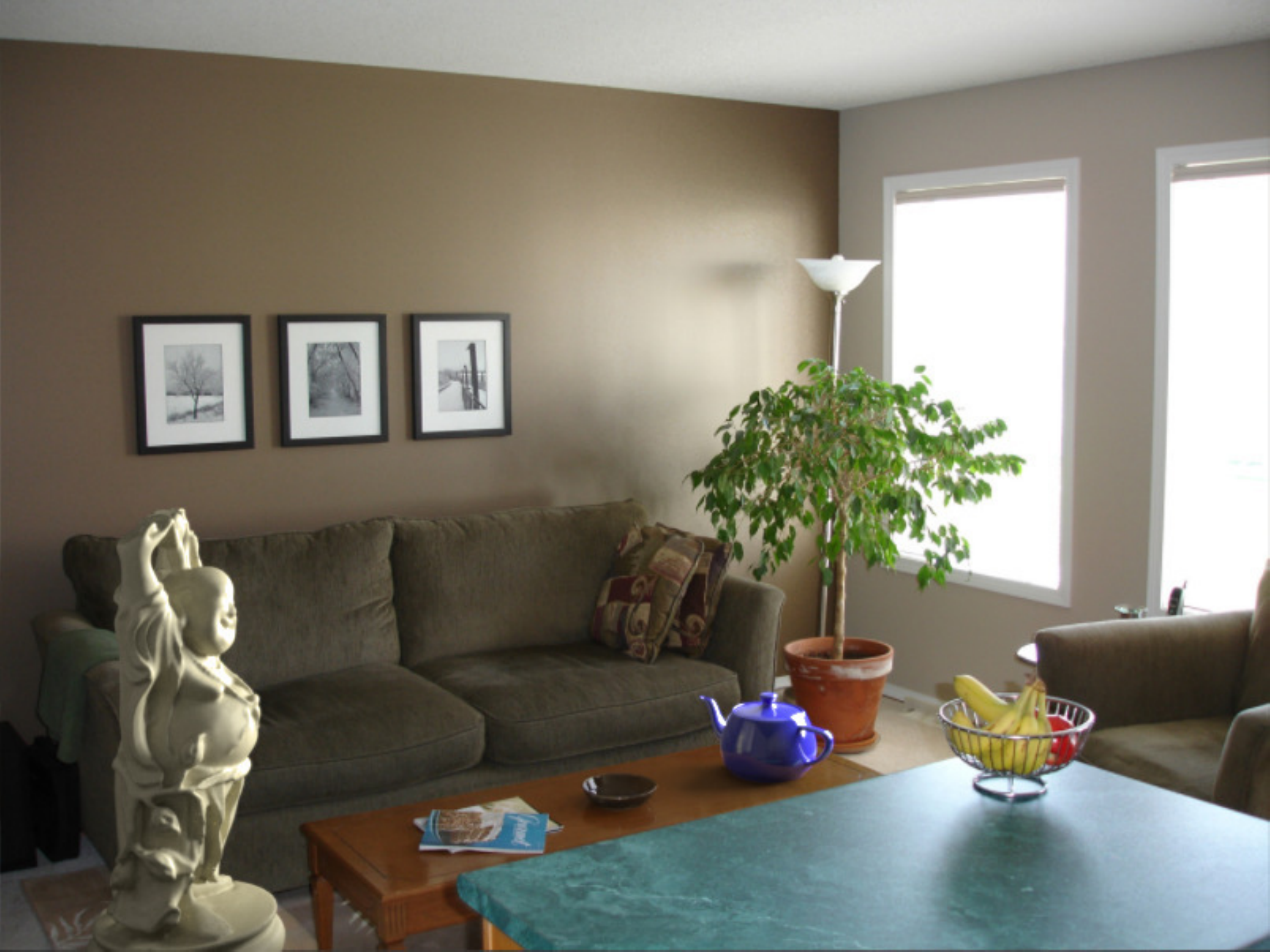} \vspace{0.5mm} \\
\includegraphics[width=0.49\columnwidth]{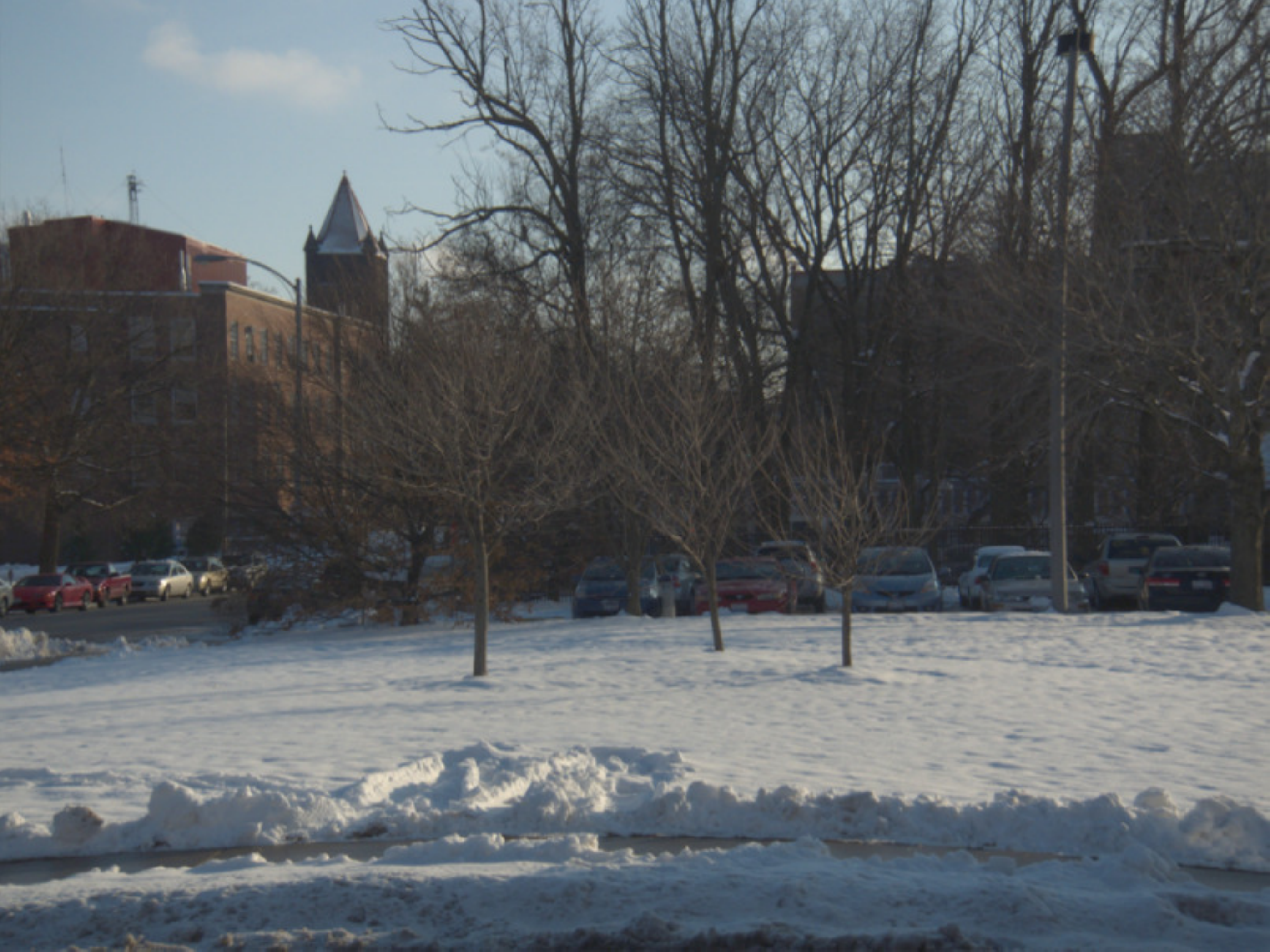}
\includegraphics[width=0.49\columnwidth]{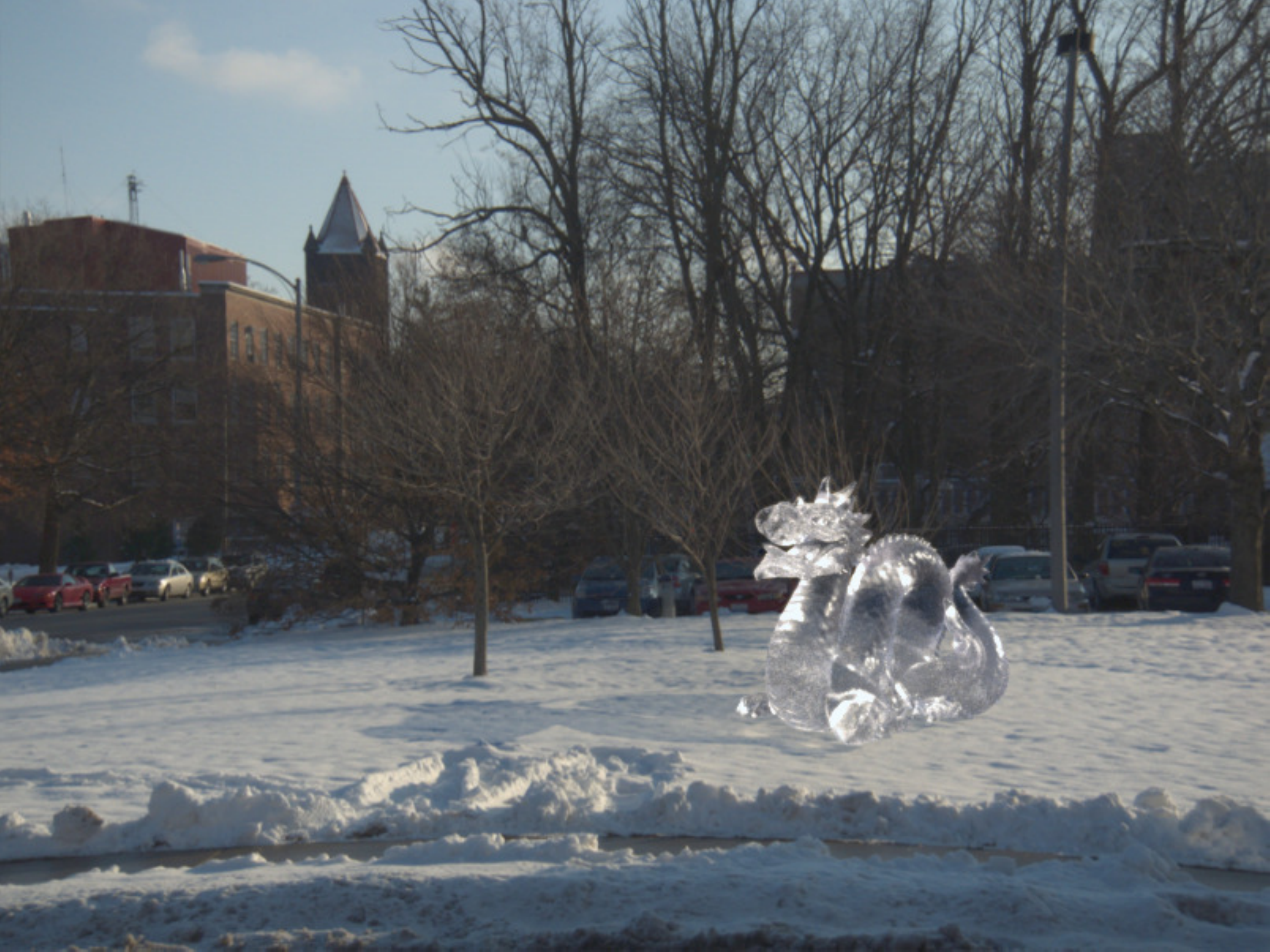}
\caption{Specular materials naturally reflect the scene \emph{(top)}, and translucent objects reflect the background realistically \emph{(bottom)}.
}
\label{fig:results2}
\end{figure}

\begin{figure}[htp]
\begin{center}
\includegraphics[width=0.49\columnwidth]{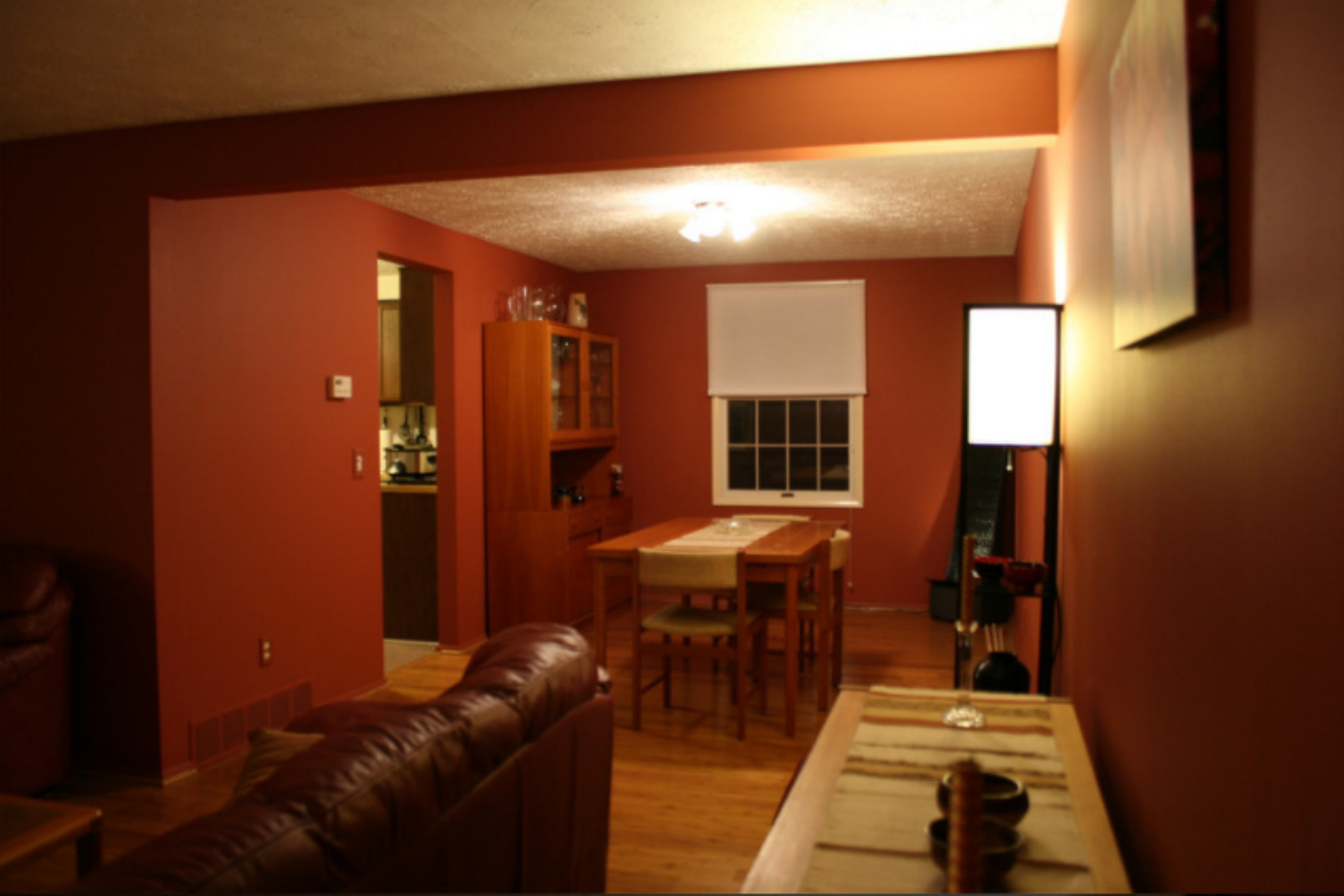}
\includegraphics[width=0.49\columnwidth]{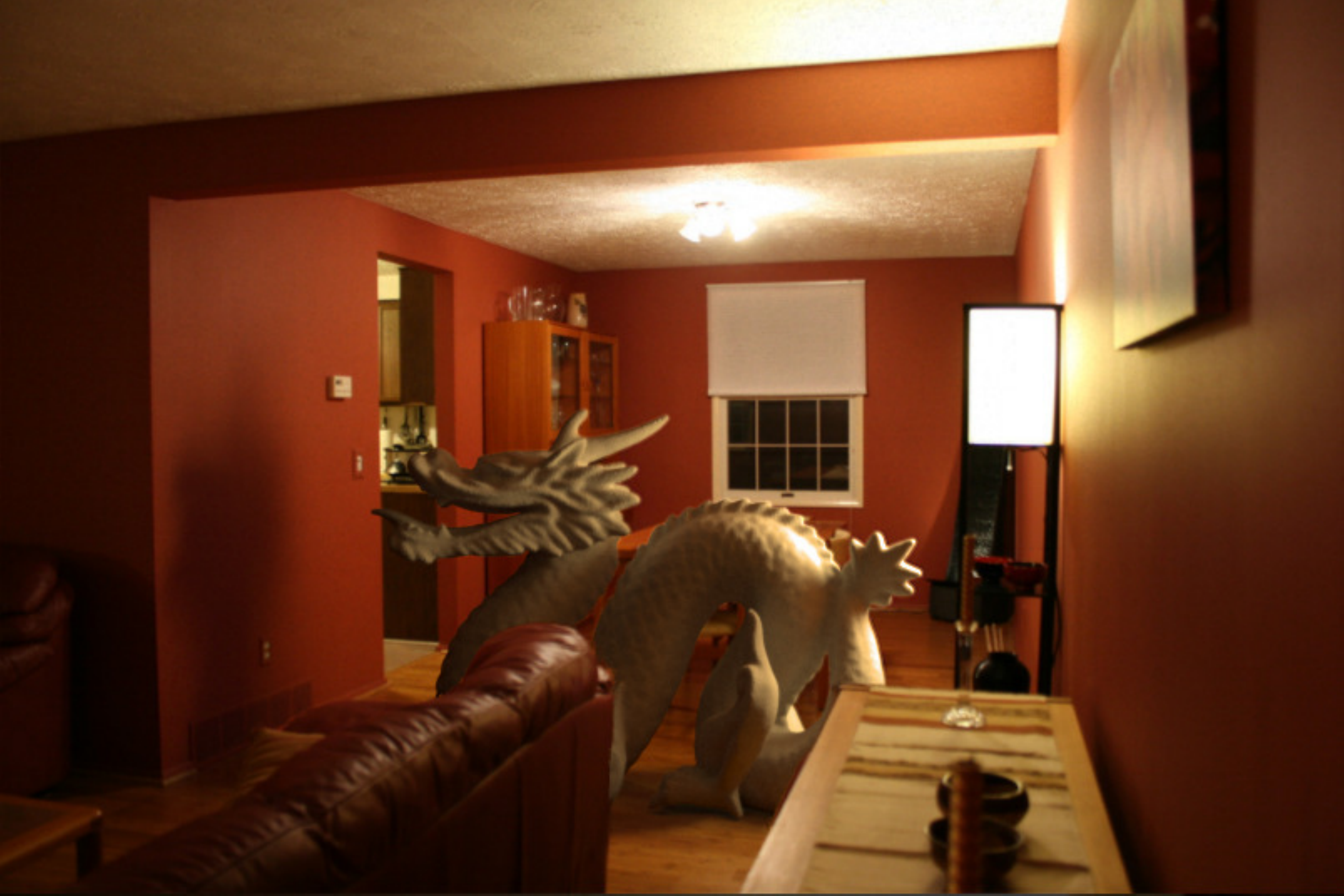}
\end{center}
\vspace{0.5mm}
\begin{center}
\includegraphics[width=0.49\columnwidth]{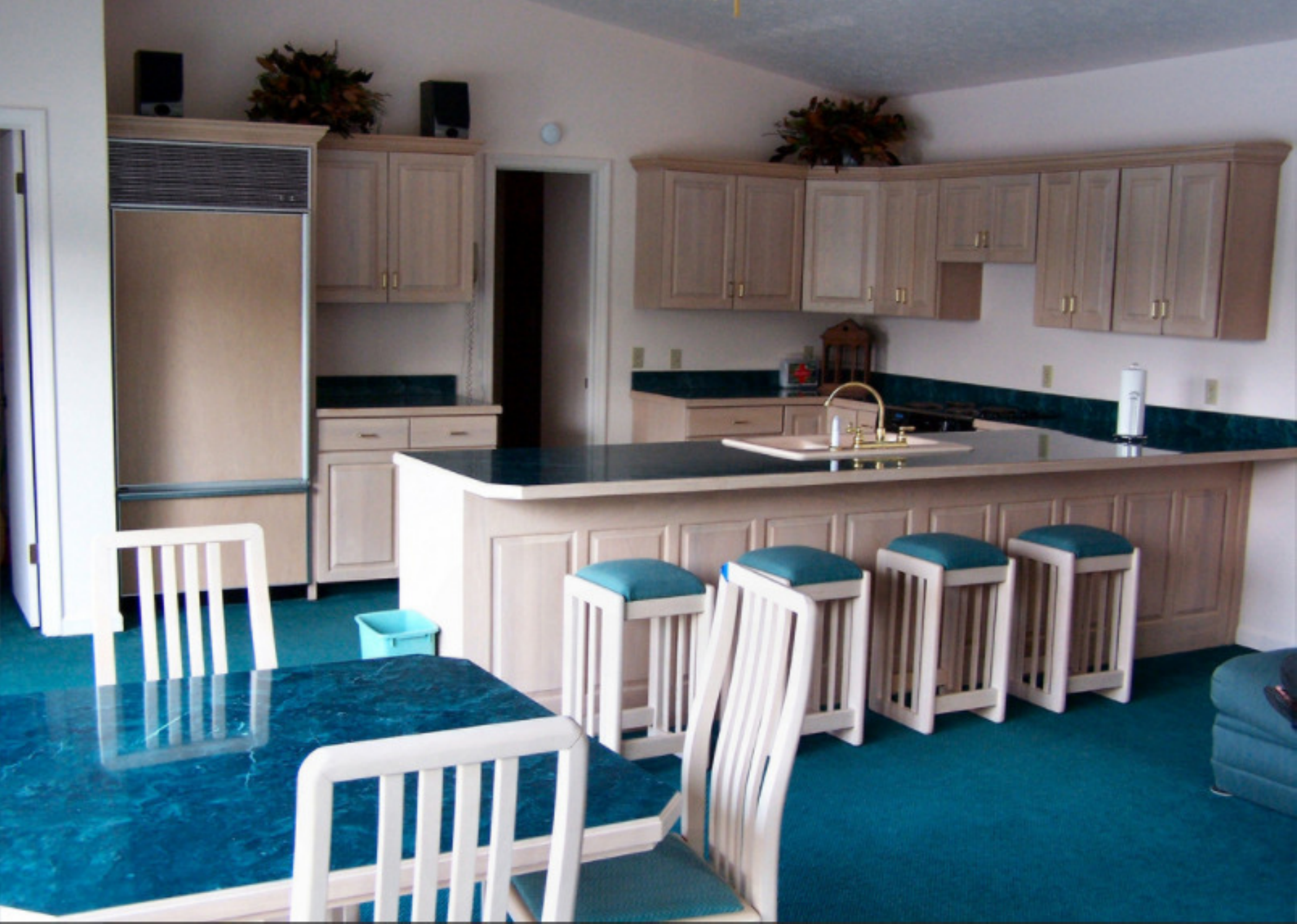}
\includegraphics[width=0.49\columnwidth]{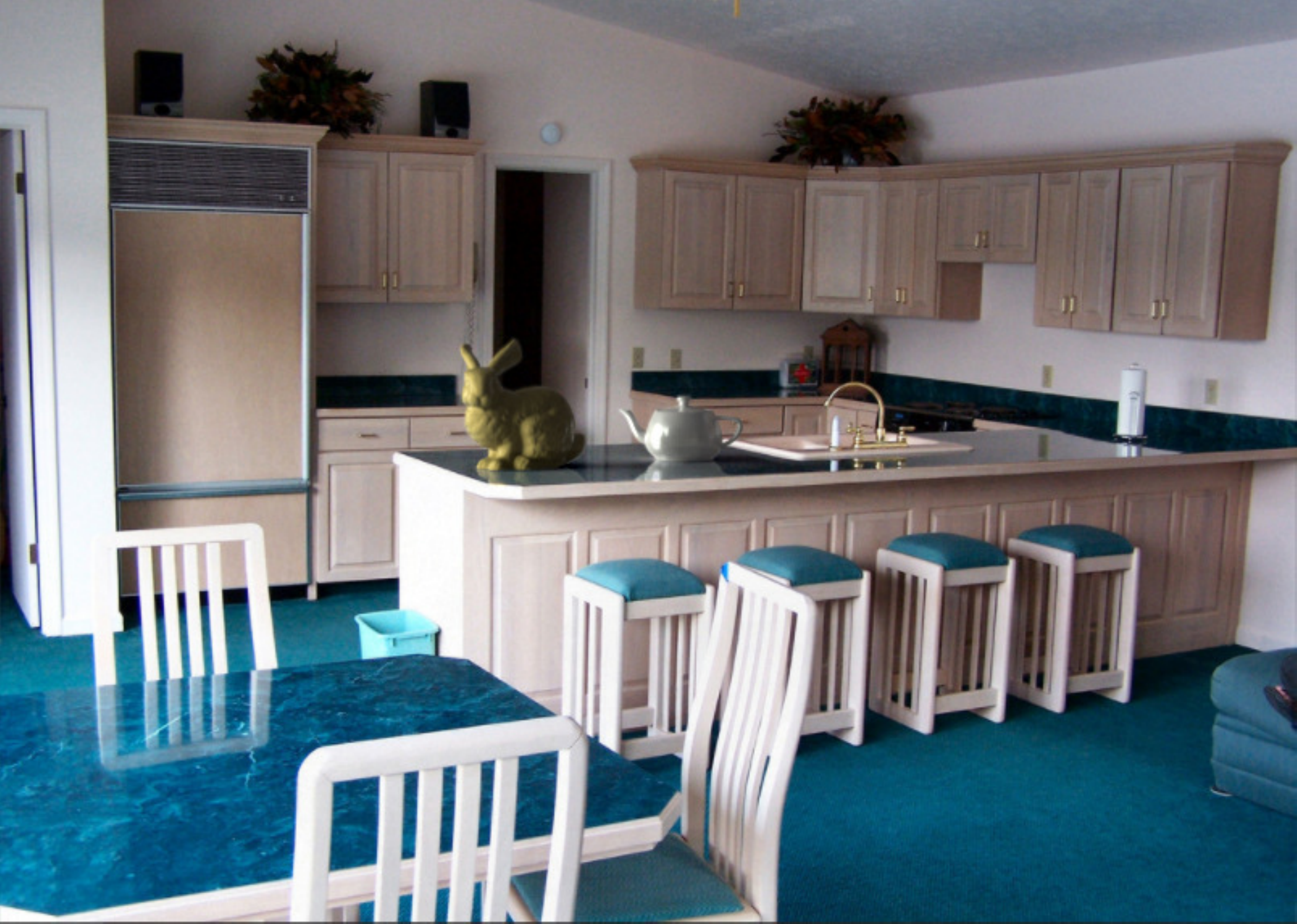}
\end{center}
\caption{Complex occluding geometry can be specified quickly via segmentation \emph{(top, couch)}, and glossy surfaces in the image reflect inserted objects \emph{(bottom, reflections under objects).}
}
\label{fig:results3}
\end{figure}

We show additional results produced with our system in Figs~\ref{fig:interiordecor}-~\ref{fig:results3}.  Lighting effects are generally compelling (even for inserted emitters, Fig~\ref{fig:mirror}), and light interplay occurs automatically (Fig~\ref{fig:interreflections}), although  result quality is dependent on inserted models/materials.  We conclude from our study that when shown to people, results produced by our method are confused with real images quite often, and compare favorably with other state-of-the-art methods.

Our interface is intuitive and easy to learn. Users unfamiliar with our system or other photo editing programs can begin inserting objects within minutes. Figure~\ref{fig:noviceresult} shows a result created by a novice user in under 10 minutes.

We have found that many scenes can be parameterized by our geometric representation. Even images without an apparent box structure (e.g. outdoor scenes) work well (see Figs~\ref{fig:results1} and~\ref{fig:results2}).

Quantitative measures of error are reassuring; our method beats natural baselines (Fig~\ref{fig:errorvis}).  Our intrinsic decomposition method incorporates a small amount of interaction and achieves significant improvement over Retinex in a physical comparison (Fig~\ref{fig:intrinsiccomp}), and the datasets we collected (Sec~\ref{sec:Evaluation}) should aid future research in lightness and material estimation. However, it is still unclear which metrics should be used to evaluate these results, and qualitative evaluation is the most important for applications such as ours.


\subsection{Limitations and future work}
For extreme camera viewpoints (closeups, etc), our system may fail due to a lack of scene information. In these cases, luminaires may not exist in the image, and may be difficult to estimate (manually or automatically). Also, camera pose and geometry estimation might be difficult, as there may not be enough information available to determine vanishing points and scene borders.

Intrinsic image extraction may fail, either because the problem is still very difficult for diffuse scenes or because surfaces are not diffuse.  For example, specular surfaces modeled as purely diffuse may cause missed reflections. Other single material estimation schemes could be used \cite{Boivin:2001:IRD:383259.383270,Debevecprobe}, but for specular surfaces and complex BRDFs, these methods will also likely require manual edits. It would be interesting to more accurately estimate complex surface materials automatically. Robust interactive techniques might also be a suitable alternative (i.e. \cite{Carroll:sg11}).


Insertion of synthetic objects into legacy videos is an attractive extension to our work, and could be aided, for example, by using multiple frames to automatically infer geometry~\cite{pmvs}, surface properties~\cite{yu99inverse}, or even light positions. Tone mapping rendered images can involve significant user interaction, and methods to help automate this process the would prove useful for applications such as ours.  Incorporating our technique within redecorating aids (e.g. \cite{Merrell:sg11,Yu:sg11}) could also provide a more realistic sense of interaction and visualization (as demonstrated by Fig~\ref{fig:interiordecor}).



\section{Conclusion}
We have demonstrated a system that allows a user to insert objects into legacy images. Our method only needs a few quick annotations, allowing novice users to create professional quality results, and does not require access to the scene or any other tools used previously to achieve this task.
The results achieved by our method appear realistic, and people tend to favor our synthetic renderings over other insertion methods.

\bibliographystyle{acmsiggraph}
\bibliography{karsch}

\begin{thebibliography}{\protect\citename{Lopez-Moreno et~al\mbox{.} }2010}

\bibitem[\protect\citename{Alnasser and Foroosh }2006]{alnasser2006}
{\sc Alnasser, M., and Foroosh, H.}
\newblock 2006.
\newblock Image-based rendering of synthetic diffuse objects in natural scenes.
\newblock In {\em ICPR}, 787--790.

\bibitem[\protect\citename{Barrow and Tenenbaum }1978]{BarTen78}
{\sc Barrow, H., and Tenenbaum, J.}
\newblock 1978.
\newblock Recovering intrinsic scene characteristics from images.
\newblock In {\em Comp. Vision Sys.}, 3--26.

\bibitem[\protect\citename{Blake }1985]{Blake}
{\sc Blake, A.}
\newblock 1985.
\newblock Boundary conditions for lightness computation in mondrian world.
\newblock {\em Computer Vision, Graphics and Image Processing 32\/}, 314--327.

\bibitem[\protect\citename{Boivin and Gagalowicz
  }2001]{Boivin:2001:IRD:383259.383270}
{\sc Boivin, S., and Gagalowicz, A.}
\newblock 2001.
\newblock {Image-based rendering of diffuse, specular and glossy surfaces from
  a single image}.
\newblock In {\em Proc. ACM SIGGRAPH}, 107--116.

\bibitem[\protect\citename{Brelstaff and Blake }1987]{BrelstaffBlake}
{\sc Brelstaff, G., and Blake, A.}
\newblock 1987.
\newblock Computing lightness.
\newblock {\em Pattern Recognition Letters 5}, 2, 129--138.

\bibitem[\protect\citename{Carroll et~al\mbox{.} }2011]{Carroll:sg11}
{\sc Carroll, R., Ramamoorthi, R., and Agrawala, M.}
\newblock 2011.
\newblock Illumination decomposition for material recoloring with consistent
  interreflections.
\newblock {\em ACM Trans. Graph. 30\/} (August), 43:1--43:10.

\bibitem[\protect\citename{Cossairt et~al\mbox{.} }2008]{cossairt2008}
{\sc Cossairt, O., Nayar, S., and Ramamoorthi, R.}
\newblock 2008.
\newblock Light field transfer: global illumination between real and synthetic
  objects.
\newblock {\em ACM Trans. Graph. 27\/} (August), 57:1--57:6.

\bibitem[\protect\citename{Criminisi et~al\mbox{.} }2000]{criminisi2000}
{\sc Criminisi, A., Reid, I., and Zisserman, A.}
\newblock 2000.
\newblock Single view metrology.
\newblock {\em Int. J. Comput. Vision 40\/} (November), 123--148.

\bibitem[\protect\citename{Debevec }1998]{Debevecprobe}
{\sc Debevec, P.}
\newblock 1998.
\newblock Rendering synthetic objects into real scenes: bridging traditional
  and image-based graphics with global illumination and high dynamic range
  photography.
\newblock In {\em Proceedings of the 25th annual conference on Computer
  graphics and interactive techniques}, SIGGRAPH '98, 189--198.

\bibitem[\protect\citename{Farenzena and Fusiello }2007]{Farenzena}
{\sc Farenzena, M., and Fusiello, A.}
\newblock 2007.
\newblock Recovering intrinsic images using an illumination invariant image.
\newblock In {\em ICIP}, 485--488.

\bibitem[\protect\citename{Fournier et~al\mbox{.} }1993]{fournier1992b}
{\sc Fournier, A., Gunawan, A.~S., and Romanzin, C.}
\newblock 1993.
\newblock Common illumination between real and computer generated scenes.
\newblock In {\em Proceedings of Graphics Interface~'93}, 254--262.

\bibitem[\protect\citename{Funt et~al\mbox{.} }1992]{Funtlightness}
{\sc Funt, B.~V., Drew, M.~S., and Brockington, M.}
\newblock 1992.
\newblock Recovering shading from color images.
\newblock In {\em ECCV}, 124--132.

\bibitem[\protect\citename{Furukawa and Ponce }2010]{pmvs}
{\sc Furukawa, Y., and Ponce, J.}
\newblock 2010.
\newblock Accurate, dense, and robust multiview stereopsis.
\newblock {\em IEEE PAMI 32\/} (August), 1362--1376.

\bibitem[\protect\citename{Greger et~al\mbox{.} }1998]{irvol}
{\sc Greger, G., Shirley, P., Hubbard, P.~M., and Greenberg, D.~P.}
\newblock 1998.
\newblock The irradiance volume.
\newblock {\em IEEE Computer Graphics and Applications 18\/}, 32--43.

\bibitem[\protect\citename{Grosse et~al\mbox{.} }2009]{grosse09intrinsic}
{\sc Grosse, R., Johnson, M.~K., Adelson, E.~H., and Freeman, W.~T.}
\newblock 2009.
\newblock Ground-truth dataset and baseline evaluations for intrinsic image
  algorithms.
\newblock In {\em ICCV}, 2335--2342.

\bibitem[\protect\citename{Guo et~al\mbox{.} }2011]{guo_cvpr11}
{\sc Guo, R., Dai, Q., and Hoiem, D.}
\newblock 2011.
\newblock Single-image shadow detection and removal using paired regions.
\newblock In {\em CVPR}, 2033--2040.

\bibitem[\protect\citename{Hartley and Zisserman }2003]{hartley2004}
{\sc Hartley, R., and Zisserman, A.}
\newblock 2003.
\newblock {\em Multiple View Geometry in Computer Vision}, 2~ed.
\newblock Cambridge University Press, New York, NY, USA.

\bibitem[\protect\citename{Hedau et~al\mbox{.} }2009]{hedau2009iccv}
{\sc Hedau, V., Hoiem, D., and Forsyth, D.}
\newblock 2009.
\newblock Recovering the spatial layout of cluttered rooms.
\newblock In {\em ICCV}, 1849--1856.

\bibitem[\protect\citename{Hoiem et~al\mbox{.} }2005]{hoiem2005siggraph}
{\sc Hoiem, D., Efros, A.~A., and Hebert, M.}
\newblock 2005.
\newblock Automatic photo pop-up.
\newblock {\em ACM Trans. Graph. 24\/} (July), 577--584.

\bibitem[\protect\citename{Horn }1974]{Horn}
{\sc Horn, B. K.~P.}
\newblock 1974.
\newblock Determining lightness from an image.
\newblock {\em Computer Vision, Graphics and Image Processing 3\/}, 277--299.

\bibitem[\protect\citename{Horry et~al\mbox{.} }1997]{TIP}
{\sc Horry, Y., Anjyo, K.-I., and Arai, K.}
\newblock 1997.
\newblock Tour into the picture: using a spidery mesh interface to make
  animation from a single image.
\newblock In {\em Proceedings of the 24th annual conference on Computer
  graphics and interactive techniques}, SIGGRAPH '97, 225--232.

\bibitem[\protect\citename{Kang et~al\mbox{.} }2001]{TIP2}
{\sc Kang, H.~W., Pyo, S.~H., Anjyo, K., and Shin, S.~Y.}
\newblock 2001.
\newblock Tour into the picture using a vanishing line and its extension to
  panoramic images.
\newblock {\em Computer Graphics Forum 20}, 3, 132--141.

\bibitem[\protect\citename{Kee and Farid }2010]{kee-farid10a}
{\sc Kee, E., and Farid, H.}
\newblock 2010.
\newblock Exposing digital forgeries from 3-d lighting environments.
\newblock In {\em WIFS}, 1--6.

\bibitem[\protect\citename{Khan et~al\mbox{.} }2006]{Khan:tog06}
{\sc Khan, E.~A., Reinhard, E., Fleming, R.~W., and B\"{u}lthoff, H.~H.}
\newblock 2006.
\newblock Image-based material editing.
\newblock {\em ACM Trans. Graph. 25\/} (July), 654--663.

\bibitem[\protect\citename{Lalonde and Efros }2007]{Lalonde_iccv07}
{\sc Lalonde, J.-F., and Efros, A.~A.}
\newblock 2007.
\newblock Using color compatibility for assessing image realism.
\newblock In {\em ICCV}, 1--8.

\bibitem[\protect\citename{Lalonde et~al\mbox{.} }2007]{lalonde2007}
{\sc Lalonde, J.-F., Hoiem, D., Efros, A.~A., Rother, C., Winn, J., and
  Criminisi, A.}
\newblock 2007.
\newblock Photo clip art.
\newblock {\em ACM Trans. Graph. 26\/} (July).

\bibitem[\protect\citename{Lalonde et~al\mbox{.} }2009]{Lalonde:sa09}
{\sc Lalonde, J.-F., Efros, A.~A., and Narasimhan, S.~G.}
\newblock 2009.
\newblock Webcam clip art: appearance and illuminant transfer from time-lapse
  sequences.
\newblock {\em ACM Trans. Graph. 28\/} (December), 131:1--131:10.

\bibitem[\protect\citename{Land and McCann }1971]{Land71}
{\sc Land, E., and McCann, J.}
\newblock 1971.
\newblock Lightness and retinex theory.
\newblock {\em J. Opt. Soc. Am. 61}, 1, 1--11.

\bibitem[\protect\citename{Lee et~al\mbox{.} }2009]{Lee:09}
{\sc Lee, D.~C., Hebert, M., and Kanade, T.}
\newblock 2009.
\newblock Geometric reasoning for single image structure recovery.
\newblock In {\em CVPR}, 2136--2143.

\bibitem[\protect\citename{Lee et~al\mbox{.} }2010]{Lee:10}
{\sc Lee, D.~C., Gupta, A., Hebert, M., and Kanade, T.}
\newblock 2010.
\newblock Estimating spatial layout of rooms using volumetric reasoning about
  objects and surfaces.
\newblock {\em Advances in Neural Information Processing Systems (NIPS) 24\/}
  (November), 1288--1296.

\bibitem[\protect\citename{Levin et~al\mbox{.} }2008]{Levin07spectralmatting}
{\sc Levin, A., Rav-Acha, A., and Lischinski, D.}
\newblock 2008.
\newblock Spectral matting.
\newblock {\em IEEE PAMI 30\/} (October), 1699--1712.

\bibitem[\protect\citename{Liebowitz et~al\mbox{.} }1999]{Liebowitz99}
{\sc Liebowitz, D., Criminisi, A., and Zisserman, A.}
\newblock 1999.
\newblock Creating architectural models from images.
\newblock In {\em Eurographics}, vol.~18, 39--50.

\bibitem[\protect\citename{Lopez-Moreno et~al\mbox{.}
  }2010]{LopezMoreno2010698}
{\sc Lopez-Moreno, J., Hadap, S., Reinhard, E., and Gutierrez, D.}
\newblock 2010.
\newblock {Compositing images through light source detection}.
\newblock {\em Computers {\&} Graphics 34}, 6, 698--707.

\bibitem[\protect\citename{Merrell et~al\mbox{.} }2011]{Merrell:sg11}
{\sc Merrell, P., Schkufza, E., Li, Z., Agrawala, M., and Koltun, V.}
\newblock 2011.
\newblock Interactive furniture layout using interior design guidelines.
\newblock {\em ACM Trans. Graph. 30\/} (August), 87:1--87:10.

\bibitem[\protect\citename{Mury et~al\mbox{.} }2009]{Mury:2009vz}
{\sc Mury, A.~A., Pont, S.~C., and Koenderink, J.~J.}
\newblock 2009.
\newblock {Representing the light field in finite three-dimensional spaces from
  sparse discrete samples}.
\newblock {\em Applied Optics 48}, 3 (Jan), 450--457.

\bibitem[\protect\citename{Oh et~al\mbox{.} }2001]{Oh01}
{\sc Oh, B.~M., Chen, M., Dorsey, J., and Durand, F.}
\newblock 2001.
\newblock Image-based modeling and photo editing.
\newblock In {\em Proceedings of the 28th annual conference on Computer
  graphics and interactive techniques}, SIGGRAPH '01, 433--442.

\bibitem[\protect\citename{Rother }2002]{rother2002}
{\sc Rother, C.}
\newblock 2002.
\newblock A new approach to vanishing point detection in architectural
  environments.
\newblock {\em IVC 20}, 9-10 (August), 647--655.

\bibitem[\protect\citename{Sato et~al\mbox{.} }2003]{satoikeuchi}
{\sc Sato, I., Sato, Y., and Ikeuchi, K.}
\newblock 2003.
\newblock Illumination from shadows.
\newblock {\em IEEE PAMI 25}, 3, 290--300.

\bibitem[\protect\citename{Saxena et~al\mbox{.} }2008]{Make3D}
{\sc Saxena, A., Sun, M., and Ng, A.~Y.}
\newblock 2008.
\newblock Make3d: depth perception from a single still image.
\newblock In {\em Proceedings of the 23rd national conference on Artificial
  intelligence - Volume 3}, AAAI Press, 1571--1576.

\bibitem[\protect\citename{Sinha et~al\mbox{.} }2008]{1409112}
{\sc Sinha, S.~N., Steedly, D., Szeliski, R., Agrawala, M., and Pollefeys, M.}
\newblock 2008.
\newblock Interactive 3d architectural modeling from unordered photo
  collections.
\newblock {\em ACM Trans. Graph. 27\/} (December), 159:1--159:10.

\bibitem[\protect\citename{Tappen et~al\mbox{.} }2005]{Tappenpami}
{\sc Tappen, M.~F., Freeman, W.~T., and Adelson, E.~H.}
\newblock 2005.
\newblock Recovering intrinsic images from a single image.
\newblock {\em IEEE Trans. Pattern Anal. Mach. Intell. 27\/} (September),
  1459--1472.

\bibitem[\protect\citename{Tappen et~al\mbox{.} }2006]{TappenCVPR06}
{\sc Tappen, M.~F., Adelson, E.~H., and Freeman, W.~T.}
\newblock 2006.
\newblock Estimating intrinsic component images using non-linear regression.
\newblock In {\em CVPR}, vol.~2, 1992--1999.

\bibitem[\protect\citename{Wang and Samaras }2003]{wangsamaras}
{\sc Wang, Y., and Samaras, D.}
\newblock 2003.
\newblock Estimation of multiple directional light sources for synthesis of
  augmented reality images.
\newblock {\em Graphical Models 65}, 4, 185--205.

\bibitem[\protect\citename{Weiss }2001]{WeissII}
{\sc Weiss, Y.}
\newblock 2001.
\newblock Deriving intrinsic images from image sequences.
\newblock In {\em ICCV}, II: 68--75.

\bibitem[\protect\citename{Yeung et~al\mbox{.}
  }2011]{Yeung:2011:MCT:1899404.1899406}
{\sc Yeung, S.-K., Tang, C.-K., Brown, M.~S., and Kang, S.~B.}
\newblock 2011.
\newblock Matting and compositing of transparent and refractive objects.
\newblock {\em ACM Trans. Graph. 30\/} (February), 2:1--2:13.

\bibitem[\protect\citename{Yu et~al\mbox{.} }1999]{yu99inverse}
{\sc Yu, Y., Debevec, P., Malik, J., and Hawkins, T.}
\newblock 1999.
\newblock Inverse global illumination: recovering reflectance models of real
  scenes from photographs.
\newblock In {\em Proceedings of the 26th annual conference on Computer
  graphics and interactive techniques}, SIGGRAPH '99, 215--224.

\bibitem[\protect\citename{Yu et~al\mbox{.} }2011]{Yu:sg11}
{\sc Yu, L.-F., Yeung, S.-K., Tang, C.-K., Terzopoulos, D., Chan, T.~F., and
  Osher, S.~J.}
\newblock 2011.
\newblock Make it home: automatic optimization of furniture arrangement.
\newblock {\em ACM Trans. Graph. 30\/} (August), 86:1--86:12.

\bibitem[\protect\citename{Zhang et~al\mbox{.} }2001]{Zhang01}
{\sc Zhang, L., Dugas-Phocion, G., Samson, J., and Seitz, S.}
\newblock 2001.
\newblock Single view modeling of free-form scenes.
\newblock In {\em CVPR}, 990--997.

\end{thebibliography}
\end{document}